\documentclass[letterpaper, 11pt, onecolumn]{IEEEtran}
\usepackage[margin=1.0in]{geometry}
\usepackage[pdftex]{graphicx}
\graphicspath{{Figures_Glucose-Measurement-Survey/}}
\usepackage{xcolor}
\usepackage{multirow}
\usepackage{algorithm}
\usepackage{balance}
\usepackage{placeins}
\usepackage{array}
\usepackage{blindtext}
\usepackage{subfigure}
\usepackage{caption}
\usepackage{capt-of}
\usepackage{booktabs}
\usepackage{supertabular}

\begin{document}

%\begin{flushleft}
%\title{A Survey of Approaches for Noninvasive Glucose Measurement}
%\title{Everything You Wanted to Know About Noninvasive Glucose Measurement}
%
%\title{Everything You Wanted to Know About Noninvasive Glucose Monitoring}
%\title{Everything You Wanted to Know About Bloodless Noninvasive Glucose Monitoring}
%\title{Everything You Wanted to Know About Noninvasive Bloodless Glucose Monitoring}

\title{Everything You Wanted to Know About Noninvasive Glucose Measurement and Control}

%\end{flushleft}
%

%\title{iGLU: An Edge-Device for Non-Invasive Precision Blood Glucose Monitoring in Smart Healthcare}
	
%\title{iGLU: A NIR based Machine-Learning Integrated Edge-Device for Non-Invasive Precision Blood Glucose Monitoring in Smart Healthcare}

\author{
%		\begin{flushleft}
\begin{tabular}{ccc}
				\\
 Prateek Jain &	 Amit M. Joshi   & Saraju P. Mohanty \\
Dept. of ECE &  Dept. of ECE  & Computer Science and Engineering \\
MNIT, Jaipur, India  &  MNIT, Jaipur, India  & University of North Texas, USA\\
prtk.ieju@gmail.com  &  amjoshi.ece@mnit.ac.in   &  smohanty@ieee.org
			\end{tabular}
%		\end{flushleft}
		%\begin{tabular}{ccc}
		%Prateek Jain & Amit M. Joshi & Saraju P. Mohanty \\
		%Electronics \& Commu. Eng. & Electronics \& Commu. Eng. & Computer Science and Engineering   \\
		%MNIT, Jaipur, India. & 	MNIT, Jaipur, India. & University of North Texas, USA. \\
		%prtk.ieju@gmail.com &amjoshi.ece@mnit.ac.in & saraju.mohanty@unt.edu
		%\end{tabular}	
		%Prateek~Jain,~\IEEEmembership{Graduate Student Member,~IEEE};
		%Amit M.~Joshi,~\IEEEmembership{Member,~IEEE};\\
		%and Saraju P. Mohanty,~\IEEEmembership{Senior Member,~IEEE}% <-this % stops a space
		%\thanks {P. Jain is with the Department of Electronics and Communications Engineering (ECE), Malaviya National Institute of Technology, Jaipur, Email: prtk.ieju@gmail.com.} %\protect\\
		%\thanks{A. M.~Joshi is with the Department of Electronics and Communications Engineering (ECE), Malaviya National Institute of Technology, Jaipur, E-mail: amjoshi.ece@mnit.ac.in.} %\protect\\
		%\thanks{S. P. Mohanty is with the Department of Computer Science and Engineering, University of North Texas, E-mail: saraju.mohanty@unt.edu.}%\protect\\
}

\maketitle

\begin{abstract}
Diabetes is a chronicle disease where the body of a human is irregular to dissolve the blood glucose properly. The diabetes is due to lack of insulin in human body. The continuous monitoring of blood glucose is main important aspect for health care. Most of the successful glucose monitoring devices is based on methodology of pricking of blood. However, such kind of approach may not be advisable for frequent  measurement. The paper presents the extensive review of glucose measurement techniques. The paper covers various non-invasive glucose methods and its control with smart healthcare technology.  
To fulfill the imperatives for non-invasive blood glucose monitoring system, there is a need to configure an accurate measurement device. Noninvasive glucose-level monitoring device for clinical test overcomes the problem of frequent pricking for blood samples. There is requirement to develop the Internet-Medical-Things (IoMT) integrated Healthcare Cyber-Physical System (H-CPS) based Smart Healthcare framework for glucose measurement with purpose of continuous health monitoring.  The paper also covers selective consumer products along with selected state of art glucose measurement approaches.  The paper has also listed several challenges and open problems for glucose measurement. 
\end{abstract}


%\end{frontmatter}

% \linenumbers

%%%%%%%%%%%%%%%%%%%%%%%%%%%%%%%%%%%%%%%%%%%
\section{Introduction}

\begin{figure}[htbp]
	\centering
	\includegraphics[width=0.80\textwidth]{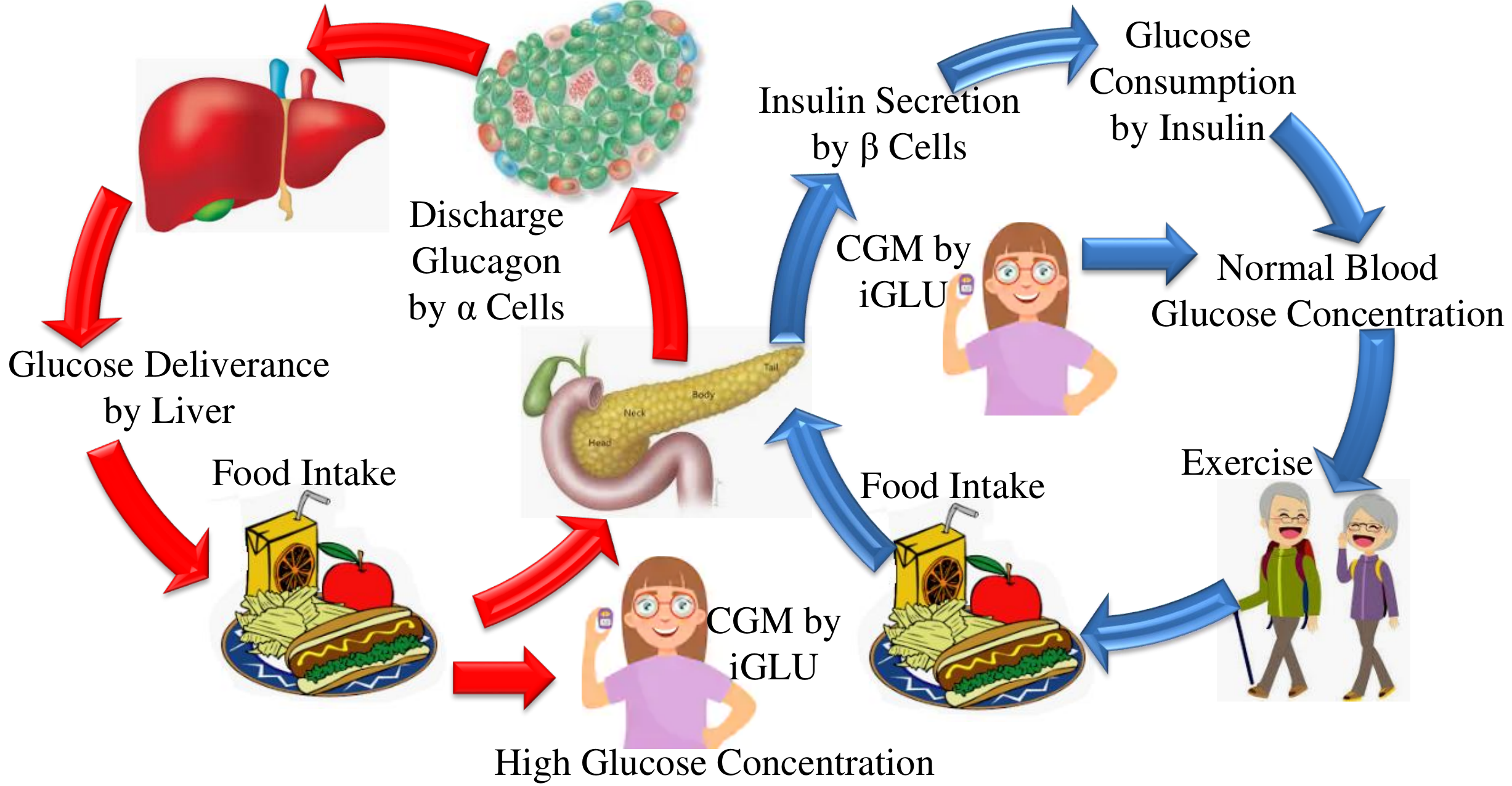}
\caption{Illustration of the closed loop form of glucose generation and consumption \cite{jain2020iglu}.}
	\label{FIG:Closed_Loop_Glucose_Generation_and_Consumption}
\end{figure}

The glucose is considered as important source of energy for the human body. The body requires blood glucose  of normal range (80 to 150 mg/dl) in order to perform the daily activities \cite{Diabetestalk_URL_2018}. However, the higher or lower value of glucose would lead to various complication inside the body. At the same time, insulin is also crucial hormone generated inside the body from the food intake. The glucose is produced from the food digestion which enters the blood cell to supply the energy and also helps in the growth. In case, the insulin is not properly generated then blood would accumulate the high glucose concentration. Fig. \ref{FIG:Closed_Loop_Glucose_Generation_and_Consumption} illustrates the closed-loop of glucose generation and consumption in human body \cite{jain2020iglu}. A consistently high blood
glucose concentration is possible if the generation of $\alpha$ cells is larger as
compared to that of the $\beta$ cells. Because of this condition, enough insulin is not
secreted in the body for glucose consumption. This condition
refers to as the Diabetes Mellitus. Diabetes is termed as chronic disease which defines high blood glucose levels inside the human body. The unbalanced glycemic profile is main reason for the cause of diabetic condition. The rate of prevalence for Non Communicable Diseases (NCD)/Chronic Disease has increased with many fold from last several years. There are around 20 million death reported yearly through cardiovascular disease, for which high blood glucose is significant predisposing factors. 
Moreover, people with diabetes are more affected during the viral pandemic outbreaks \cite{9174644, joshi2020smartA, alsamman2020transcriptomic}.

%The diabetes has also been considered as one of the major comorbidity links with the novel corona virus \cite{9174644}. The Coronavirus Disease 2019 (COVID-19) has affected more than 465 millions people around the world and around 20 \% to 50 \% people had pre-existing diabetes disease. However, the diabetes is also linked with previous viral pandemics. The diabetes was also defined as one of the prominent risk which was associated with Severe Acute Respiratory Syndrome (SARS-CoV-1) during 2002. Similarly, It was main disease which was observed among hospitalised patients at the time of Influenza A (H1N1) in 2009 \cite{alsamman2020transcriptomic}.  
%During the prevalence of Middle East Respiratory Syndrome Coronavirus (MERS-CoV), there were around 50 \% of the diabetic patients  in comparison with entire population and the chances of the infection would be more for underneath diabetes. 
%The rate of death was almost 35 \% for diabetes during MERS outbreak  \cite{joshi2020smartA}.

%\subsection{Diabetes Trends}

\begin{figure}[htbp]
	\centering
	\includegraphics[width=0.60\textwidth]{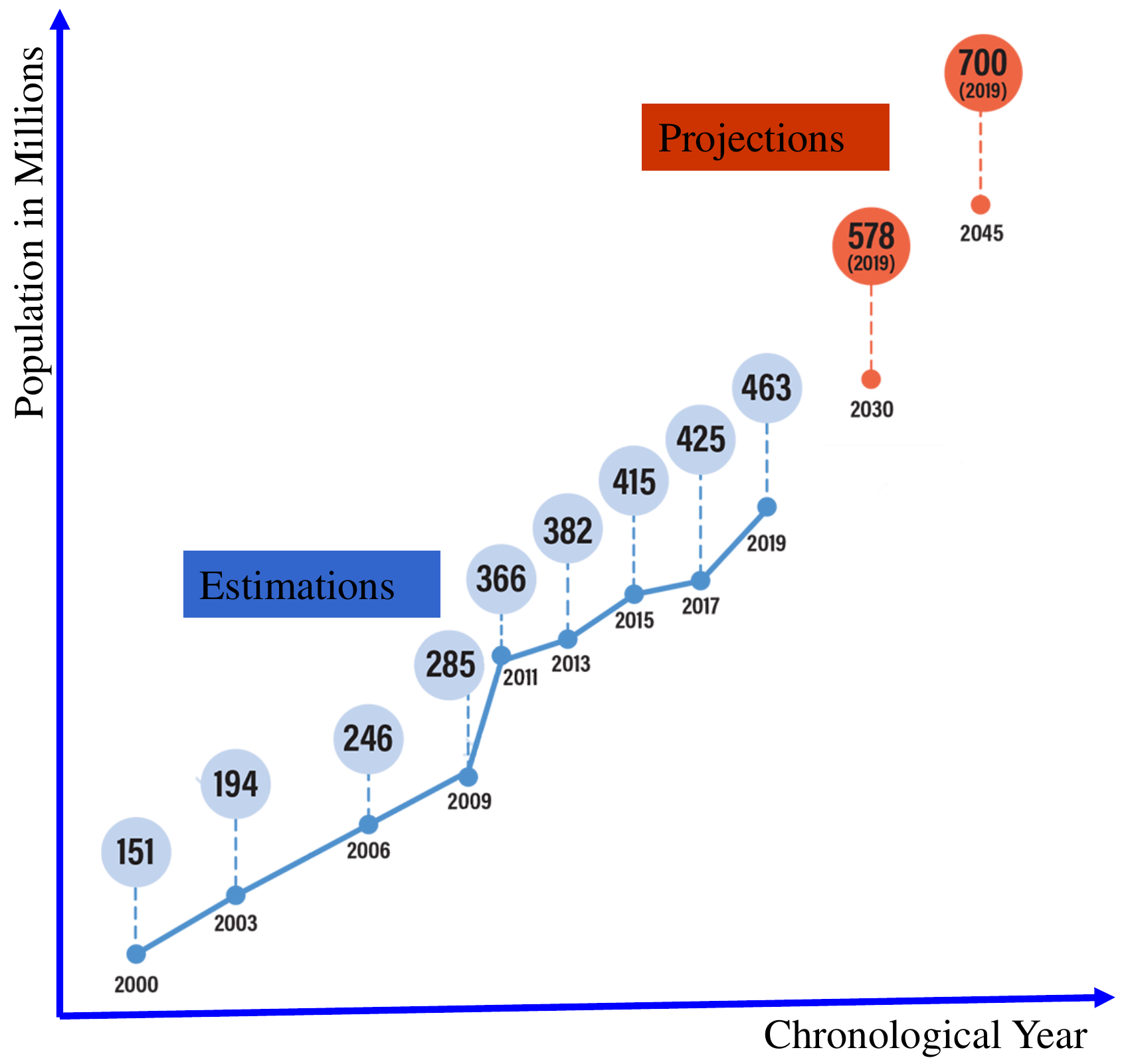}
	\caption{Global trend of Diabetes, Adopted from \cite{diabetesatlas_URL_2020}.}
	\label{FIG:Diabetes_Projection_Trend}
\end{figure}

There has been exponential growth of diabetes patients over past few years because of obesity, unhealthy diet plan, old-age population, and inactive lifestyle. Diabetes is considered as one of the fastest growing health challenges, with the number of adults living with diabetes having more than tripled over the past 2 decades (Refer Fig. \ref{FIG:Diabetes_Projection_Trend}) \cite{diabetesatlas_URL_2020}. The prevalence of diabetes around the world was 9.3\%  during 2019  with approximate 463 million people. It is expected to rise to 578 million by 2030 with 10.2\% prevalence rate and the same would be 10.9\% with 700 million population by 2045.  It has been observed that prevalence is quite higher in urban to 10.8\% whereas 7.2\% in rural region. Almost half of the diabetes patients unaware about their situation due to lack of knowledge.  
The diabetes has indeed global outbreak which has affected presently almost 1 in 10 people around the world. It is projected that more than
0.5 billion adults would suffer from the diabetes in the next decade \cite{Cho_DRCP_2018-Apr}. As per the report from International Diabetes
Federation (IDF), the death from diabetes has large number than combined death from Malaria (0.6mio), HIV/AIDS (1.5mio) and tuberculosis (1.5mio)  \cite{saeedi2019global}. There are around 8 million new patients are being added to diabetic community every year. This has grown the demands immensely for the 
effective diabetic management. 
It is important to monitor the blood glucose over time to time for avoiding late-stage complication from diabetes. This has necessitate the design of various reliable and robust solutions for efficient diabetes management.
The market of diabetes devices has also grown rapidly with significant requisite for frequent glucose measurement for better glycemic profile control.

%\subsection{Negative Impacts of Diabetes}

Diabetes is one of the major chronic disease  which has long-term impact of the well-being life of a person. Diabetes Mellitus (DM) is considered as physiological dysfunctions with high blood
glucose because of insufficient insulin, insulin resistance,  or excess generation of
glucagon \cite{clevelandclinic_URL_2020}. It is the critical health issue of $21^{st}$ century. Type 2 Diabetes (T2DM) has shown rapid growth around the world from past few years. Any form of diabetes may lead to complications in various body parts which increase the possibility of premature death. The higher value of blood glucose  known as hyperglycemia,
would lead to thickening of blood vessels which could resulted in kidneys damage
and loss of sight and some times even to these organs failure. Diabetes is also associated with limb amputation, peripheral vascular diseases and myocardial. 
Contrary, the low blood glucose defined as hypoglycemia may occur in Type 1 Diabetes Patients (T1DM) for excessive insulin dosage \cite{drugs_URL_2020}. 
The most common symptoms for hypoglycemia pateints are dizziness,
sweating and fatigue  and  in the worst case it can lead to coma and death.
%In type 1 diabetes, the immune system of the body attacks and destroys the cells of the pancreas which produce insulin. This results in the affected person who will be unable to generate insulin naturally. Type 2 diabetes is the most common diabetic stage which is most commonly seen in the people over the world. In this type of diabetes, the pancreas will be able to generate some amount of insulin. Gestational diabetes occurs in women in the later stages of pregnancy. 
The diabetic patients would have several common symptoms such as thirsty, tiredness, changes in vision, consistently hungriness, unexpected weight loss and the excretion of urine within short durations \cite{endocrineweb_URL_2019}. If the diabetes remain untreated over the period of time, it may cause blindness, heart stroke, kidney disease, lower limb amputation and blindness. It would lead to increase the probability of death almost 50\% higher in comparison of the patients without diabetes. The diabetes also brings the additional financial burden for the treatment and point of care. The diabetic patients could also result in loss of productivity at workplace and may lead to disability. There are several health issues which may also arise from diabetes like depression, digestive problem, anxiety disorders, mood disorder and eating habits change. 
%Hence, the diabetes carries an unsustainable economic burden regarding persistently high healthcare costs; greater care and financial support desired from family members/carers; loss of workplace productivity; and disability.
The diabetes could be controlled with proper diet plan, through some physical exercise, insulin dosage and medicines. The early stage of diabetes is possible to control with oral medicines. The  diabetes control also helps to reduce the associated risk of high blood pressure, cardiovascular and amputation.

The rest of the article is organized in the following manner: 
Section \ref{SEC:Diabetes_and_Glucose-Measurement} briefly presents different types of diabetes while making case for the need of glucose level monitoring.
Section \ref{SEC:Glucose-Level_Measurement_Overview} presents overview of various types of glucose-level measurement mechanisms.
Section \ref{SEC:Glucose-Level_Noninvasive_Approaches} provides details of available approaches for noninvasive glucose-level monitoring.
Section \ref{SEC:Postprocessing_and_Calibration_Techniques} has discussions on various post-processing and calibration techniques for noninvasive glucose-level monitoring.
Section \ref{SEC:Consumer_Products_for_Measurement} briefly discusses various consumer products for noninvasive glucose level measurement.
Section \ref{SEC:Glucose_Controls_Approaches_and_Consumer-Products} presents the approaches for glucose-level control and corresponding consumer products.
Section \ref{SEC:IoMT_Perspectives} provides the Internet-of-Medical-Things (IoMT) perspectives of glucose level measurements and control in healthcare Cyber-Physical Systems (H-CPS) that makes smart healthcare possible.
Section \ref{SEC:Shortcomings_and_Open-Problems} outlines the shortcomings and open problems of glucose-level measurements and control.
Section \ref{SEC:Conclusions} summarizes the learning of this comprehensive review work.

%%%%%%%%%%%%%%%%%%%%%%%%%%%%%%%%%%%%%%%%%%
\section{The Health Issue of Diabetes and Need for Glucose-Level Measurement}
\label{SEC:Diabetes_and_Glucose-Measurement}

This Section presents details of different types of diabetes, the health issues arise due to diabetes, while making case for the need of glucose level monitoring.

\subsection{Types of Diabetes}
\begin{figure}[htbp]
	\centering
	\includegraphics[width=0.65\textwidth]{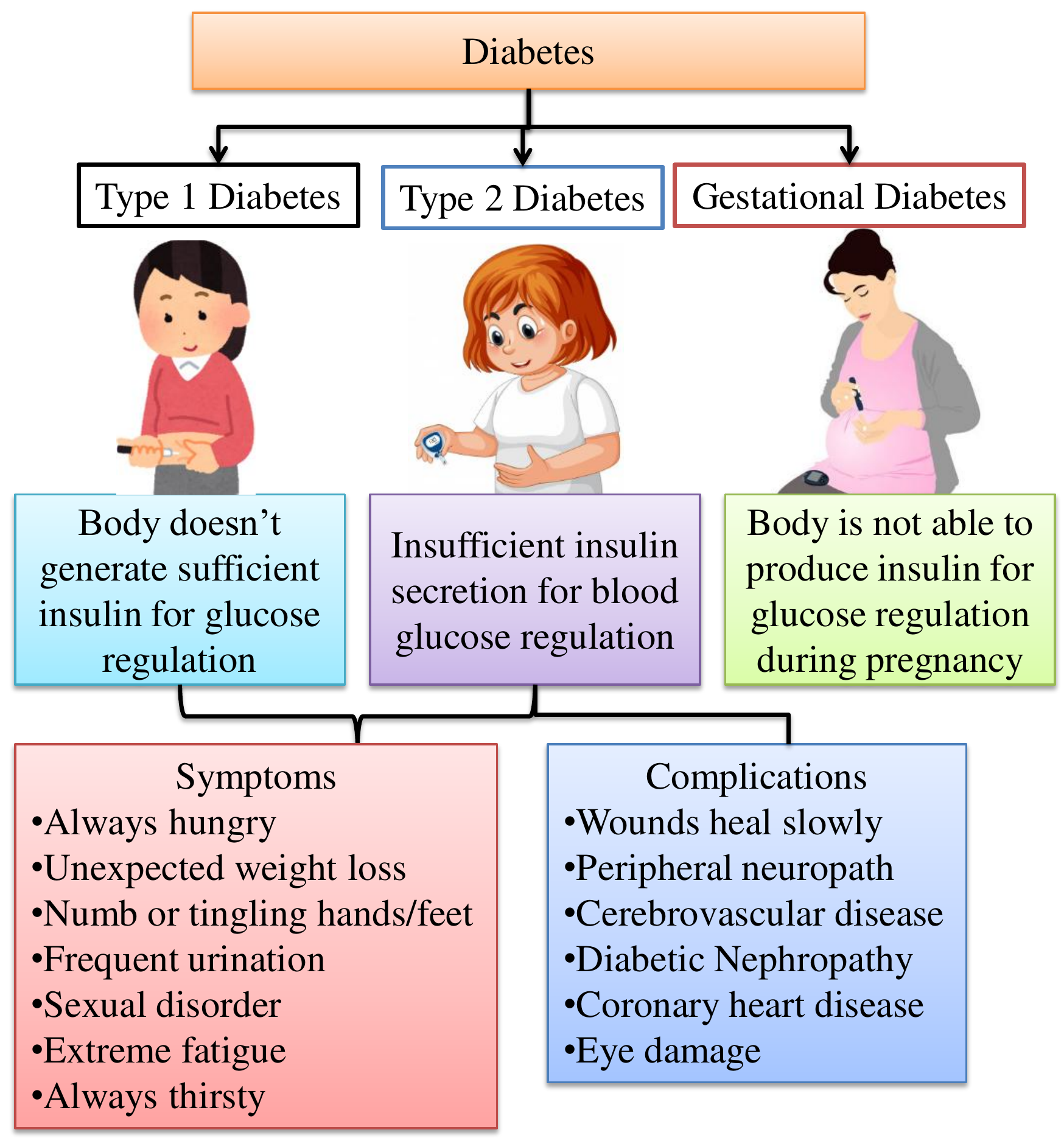}
	\caption{Different types of diabetes and their symptoms.}
	\label{FIG:Diabetes_Symptoms_Complication}
\end{figure}

The diabetes occurs because of insufficient insulin with respect to glucose generated inside the body. The insulin from body is either insufficient or not any which is generated from beta cells of the pancreas. In case of diabetes, the cells of liver, muscles and fat unable to balance glucose insulin effectively. The diabetes are classified mainly in three categories: Type 1 diabetes, Type 2 diabetes and gestational diabetes (Refer Fig. \ref{FIG:Diabetes_Symptoms_Complication}) \cite{healthline_URL_2018}. 

For diabetes of type-1, the pancreas does not produce insulin inside the body which is resulted in a weak immune system. This results in a person who is unable to generate insulin naturally  \cite{jain2020iglu, Fowler42}. In case of type 2 diabetes, the amount of insulin from pancreas  is not sufficient to maintain glycemic profile of the body. Gestational diabetes usually occurs in a pregnant woman at later stage of the delivery. 
%The diabetic patients may have one or more symptoms of thirsty, tiredness, weight loss, vision issue, frequent hungriness, excretion of urine within short durations, etc. 
in the year 2020, total 2 billion adults around the globe suffers from overweight, and 300 million of them are obese. In addition, a minimum of 155 million children in the world is overweight or obese. It is projected that the prevalence of hyperglycemia is 8.0\% and expected to increase to 10\% by 2025 \cite{Cho_DRCP_2018-Apr}.
There has been concern for diabetic people specially in developing countries due to increase in Type 2 Diabetes cases rapidly at earlier age which have overweight children even before puberty. Whereas for developed countries, most of people have high blood glucose at age around 60 years. Most frequently affected are at middle aged between 35 and 64 in developed countries \cite{diabetesatlas_URL_2020}. In 2019, 69.2 millions population in India had Type-2 diabetes. Approximately 2.35 million adults have Type-1 diabetes. 
In general, there are around 5\% adults have been considered for Type-1 diabetic patients while the others 90-95\% are of Type-2 diabetic patients. Type-1 diabetic patient must have insulin to control the blood glucose level. Type-2 diabetic patients can control their glucose level by following an optimized diet with medication and a regular physical exercise schedule. 
 %The non-invasive device for continuous glucose measurement of diabetes patients 
%can help for insulin dosage or diet plan. The certain meal is also needed to control diabetes.

\subsection{The Health Crisis due to Diabetes}
The diabetes mainly occurs due to unbalanced  glucose insulin level of the body where insulin is  demolished and muscles and cells are not able to generate insulin properly \cite{Yin_TETC_2019-2958946, Fowler42}.
%The prolong impact of diabetes may result in nerve damage, heart failure, kidney disease, and blindness.
The probability of death would also increase upto 50\% in comparison to non-diabetes case. The control action of the diabetes would be possible using proper precautionary measure after frequent glucose measurements.
%The control of glycemic profile for diabetes persons lead to reduce in other associated risk factors such as high systolic, diastolic blood pressure and cardiovascular disease such as lipid profile.
% It mainly affects the people of Type-1 Diabetes patient  because they find difficulty to control the blood sugar due to inadequate generation of insulin. Moreover, most common people are of type 2 diabetes cases where the limited insulin is produced by body. The diabetes may lead to failure of heart \& kidney and in worst case the blindness due  not being controlled at proper time. 
Therefore, there is a real need for smart healthcare solution which would provide instant self measurement of blood glucose with high accuracy.

\begin{figure}[htbp]
	\centering
	\includegraphics[width=0.80\textwidth]{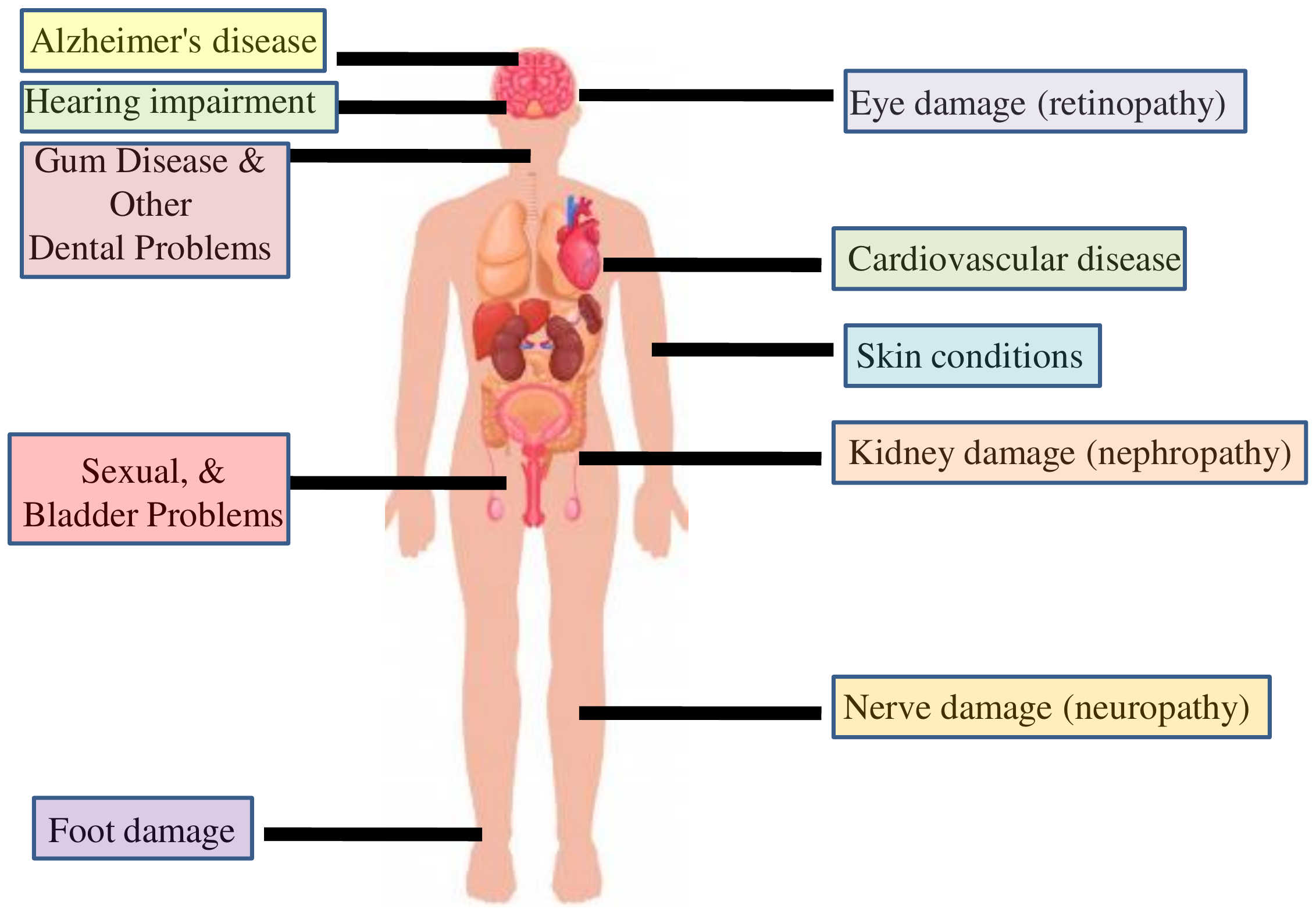}
	\caption{Diseases in human body due to diabetes}
	\label{FIG:Diabetes_Health_Crisis}
\end{figure}

Hyperglycemia is the major issue which has been considered by several health organizations at worldwide level \cite{Zhang2011, Venkataraman2011}. There are several attempts which have been used for glucose measurement \cite{Wild2569}. There have been substitutional work using various techniques  to make the device more familiar with clinicians and patients \cite{WHITING2011311}. Diabetes is possible in the age group 18 to 80 years usually \cite{Siegel2015, Alavi2001}. The normal range of glucose is in the range of 70-150 mg/dL and pathophysiological would be from 40 mg/dL to 550 mg/dL \cite{Li2015}. One of the emerging issues is to design the glucose measurement device for continuous health care monitoring \cite{Pai2017}. The devices for monitoring the glucose level are available for last two decades \cite{Reddy2017}.

\subsection{Glucose Measurement: A Brief History}
\begin{figure}[htbp]
	\centering
	\includegraphics[width=0.75\textwidth]{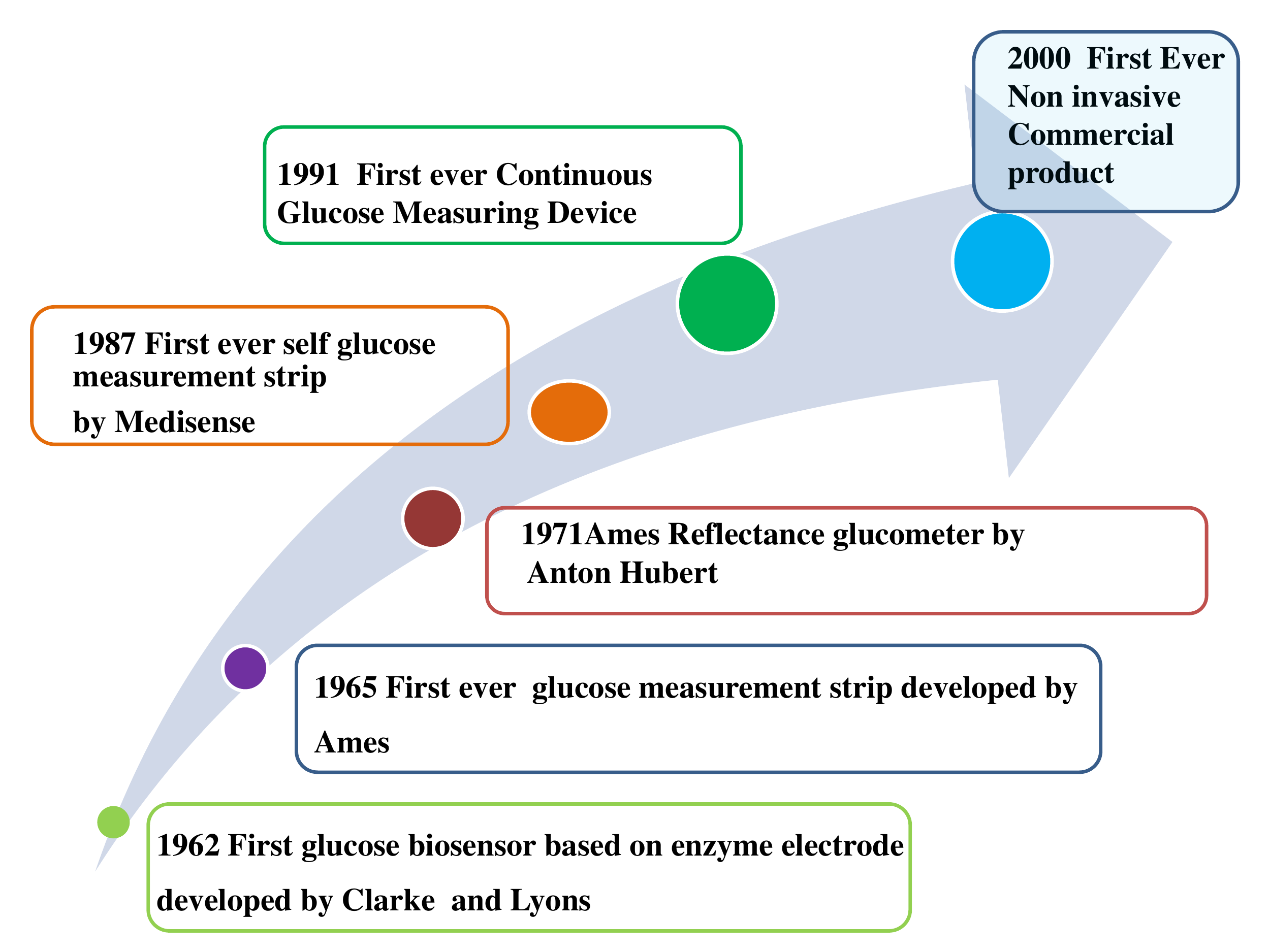}
	\caption{History of Glucose Measurement.}
\label{FIG:Glucose_Measurement_History}
\end{figure}

The glucose meter (aka glucometer) is a portable medical device for predicting the glucose level concentration in the blood \cite{Jain2020, Joshi_TCE_iGLU2_TCE.2020.3011966}. It may also be a strip based dipped into any substance and determined the glucose profile. It is a prime device for blood glucose measurement by people with diabetes mellitus or hypoglycemia. With the objective of glucose monitoring device advancement, the concept of the biosensor has been proposed earlier in 1962 by Lyons and Clark from Cincinnati. Clark is known as the ``father of biosensors'', and modern-day glucose sensor which is used daily by millions of diabetics. This glucose biosensor had been composed with an inner oxygen semipermeable membrane, a thin layer of GOx, an outer dialysis membrane and an oxygen electrode. Enzymes could be gravitated at an electrochemical detector to form an enzyme electrode \cite{Mohanty_POT_2006-Mar}. However, the main disadvantage of first-generation glucose biosensors was that there was the requirement of high operation potential of hydrogen peroxide amperometric measurement for high selectivity. The first-generation glucose biosensors were replaced by mediated glucose biosensors (second-generation glucose sensors). The proposed biosensors till present scenario represent the advancements in terms of portability of device and precision in measurement. But, due to some environmental and measurement limitations; these biosensors were not taken for real-time diagnosis.  The history of glucose measurement is shown in Fig. \ref{FIG:Glucose_Measurement_History} \cite{Salam2016TheEO}.

\begin{figure}[htbp]
	\centering
	\includegraphics[width=0.75\textwidth]{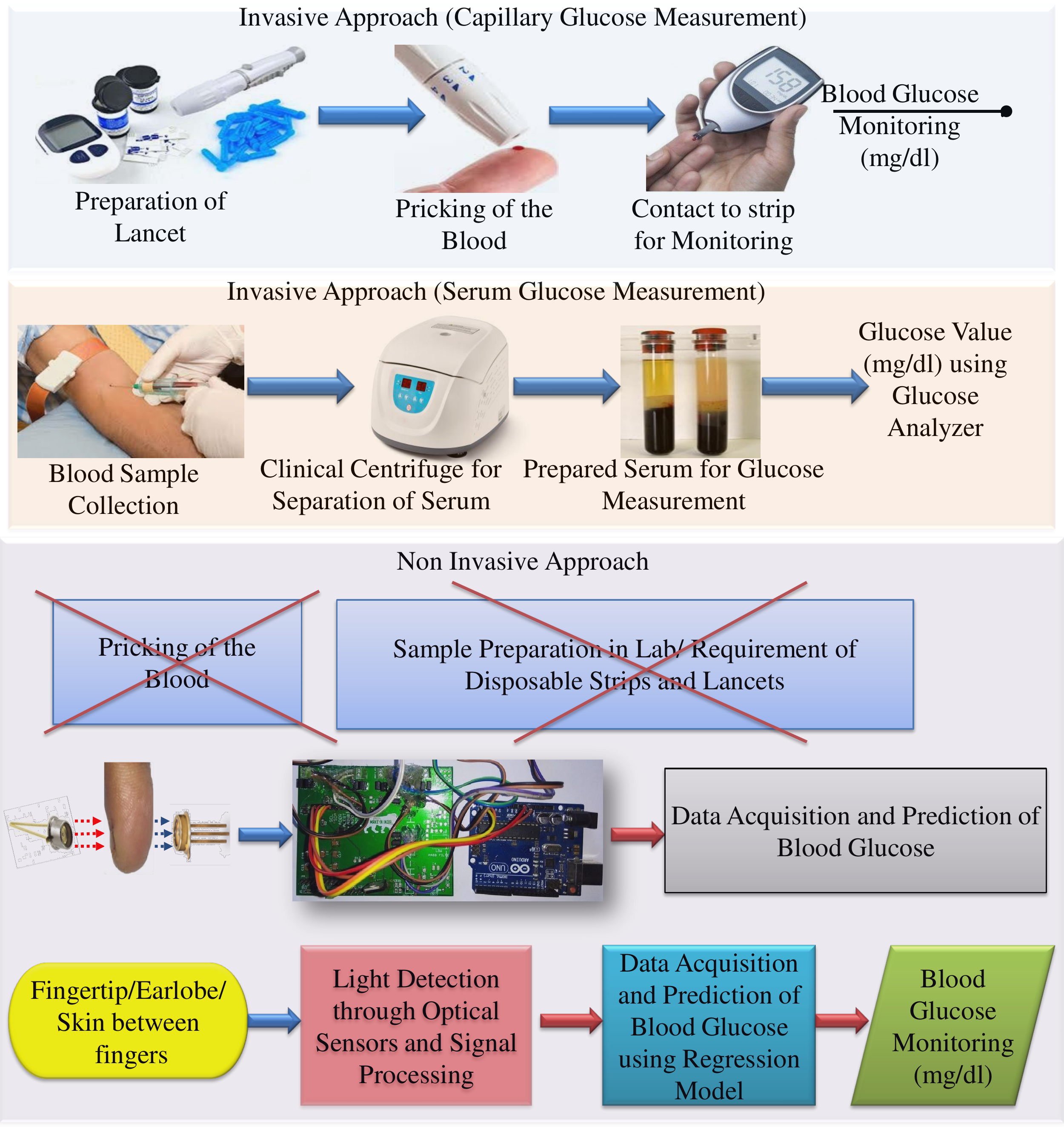}
	\caption{Invasive versus Noninavsive Glucose Measurement.}
	\label{FIG:Invasive_versus_Noninavsive_Glucose_Measurement}
\end{figure}

%\begin{figure*}[htbp]
%	\centering
%	\includegraphics[width=0.995\textwidth]{Glucose_Measurement_Options_Taxonomy}
%	\caption{An overview of the Glucose Measurement Options.}
%	\label{FIG:Glucose_Measurement_Options_Taxonomy}
%\end{figure*}

\subsection{Glucose Measurement Technique}

Presently, the glucose monitoring is carried out either laboratory based technique or home based monitoring. These both approaches are invasive in nature which provides discomfort by blood pricking and it only helps to measure the glucose measurement at that point of time. It is also not very convenient  for the user to take out blood samples multiple times in a day and many patients are reluctant to opt such type of solution. Therefore, significant changes of glycemic profile may go unnoticed because of unanticipated side effects and low compliance from the patients. This could impact on improper insulin dosage and unknown food ingredient.  However, they are reliable solution due to their good sensitivity and higher accuracy for glucose measurement \cite{Jain_arXiv_2019-Nov30-1911-04471_iGLU1, Mohanty_arXiv_2020-Jan-28-2001-09182_iGLU2}.

The novel approach for glucose measurement has been explored from past several years which is based on the principle of physical detection than conventional chemical based principle. Such non-invasive based method does not require the blood sample but uses the interstitial fluid (ISF) for glucose molecule detection. There are several attempts in the same direction for glucose measurement through sweat, saliva, tears and skin surface \cite{acsnano7b06823}. However, the main challenge is to have precise measurement, good sensitivity and reliability from such measurement. Such approach could be suitable for Continuous Glucose Measurement (CGM) and self monitoring purpose. Such CGM techniques would provide the frequent measurement in a day which would helpful for better glucose control and also for the necessary preventive actions for hyperglycemia and hypoglycemia patients. Such kind of techniques would also support for the dietician and healthcare provider to prepare proper diet plan according to glucose fluctuation for the patient.

\subsection{The Need for Continuous Glucose Measurement (CGM)}

The measurement of glucose could be done through non-invasive, semi (or minimal) invasive and invasive approach. The frequent measurement may not be possible using invasive method which can cause trauma. The semi-invasive and non-invasive could be useful for Continuous Glucose Measurement (CGM) without any pricking of the blood. However, the non-invasive glucose measurement is most suitable technique which helps to measure the blood glucose painlessly \cite{Siegel2016}.

%Blood glucose measurement is possible using invasive, minimally invasive and non-invasive methods. Frequent pricking, as needed in invasive methods, for glucose measurement causes trauma.
% Therefore, the semi-invasive approach has the advantage of continuous glucose monitoring without multiple times pricking. 
% However, non-invasive methods can completely eliminate pricking which opens door to painless and continuous glucose monitoring (CGM) .

\begin{figure}[htbp]
	\centering
	\includegraphics[width=0.750\textwidth]{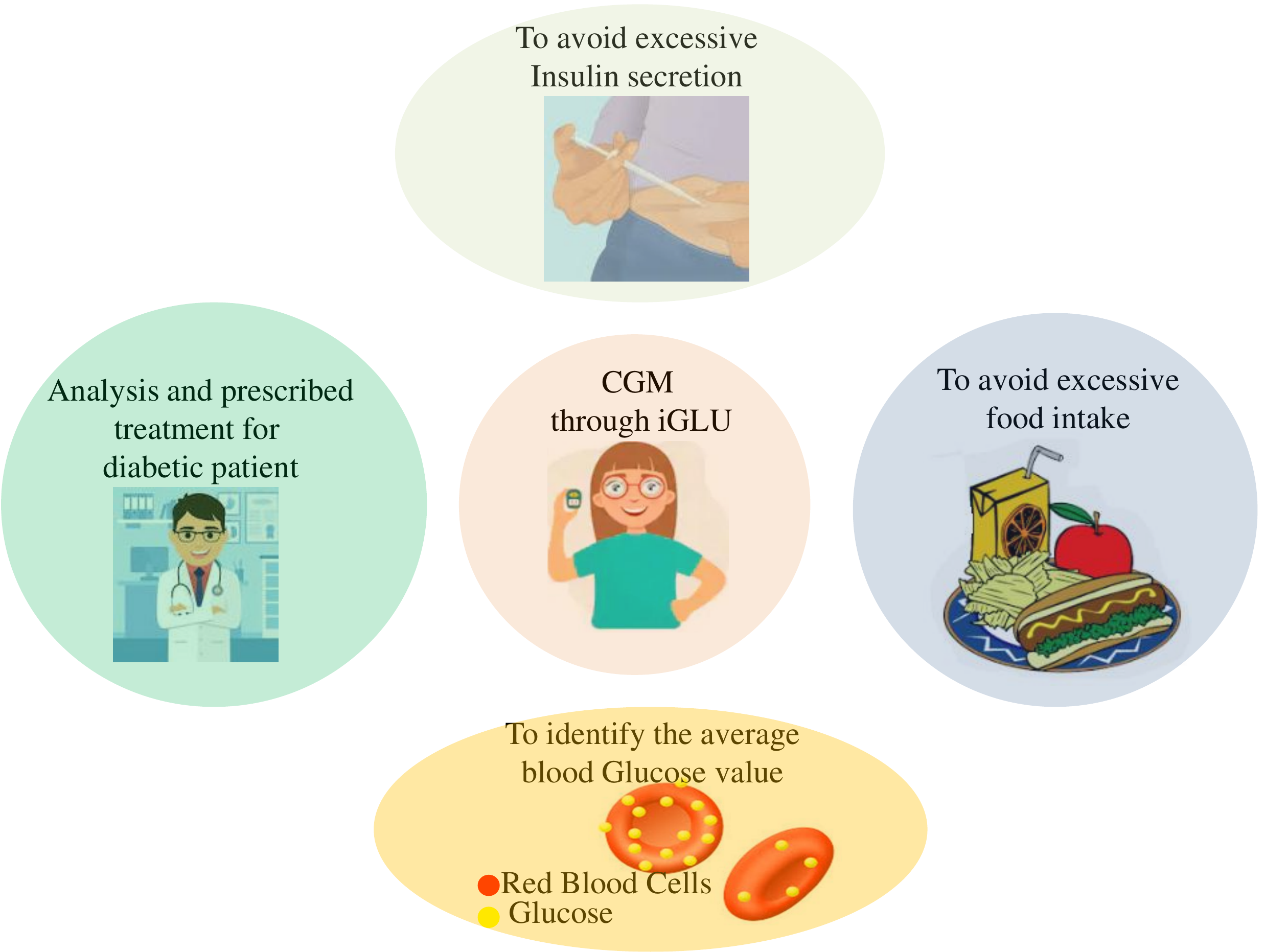}
	\caption{The objectives of continuous glucose monitoring.}
	\label{FIG:CGM_Need}
\end{figure}

CGM assist to have proper blood glucose level analysis at each prandial mode. It helps to measure  glucose insulin level after insulin secretion,hysical exercise or subsequent to medication. The frequent glucose reading also helpful to endocrinologist for providing the proper prescription.  It mainly helps for type 1 diabetic patients to take care of their insulin dosage over the period of time. The proper diet management could be possible with help of recurrent glucose monitoring and flow diagram of CGM is shown as Fig. \ref{FIG:CGM_Need} \cite{Jain_arXiv_2019-Nov30-1911-04471_iGLU1, Mohanty_arXiv_2020-Jan-28-2001-09182_iGLU2}. The CGM is useful for the patients for frequent glucose measurement over the period of time. This would helpful to identify the average blood glucose value for the last 90 days, by which glycated haemoglobin (HbA1c) can be determined.

%%%%%%%%%%%%%%%%%%%%%%%%%%%%%%%%%%%%%%%%%%%%%%%%%%%%
%\section{What are Possible Approaches Glucose-Level Measurement?}
%%%%%%%%%%%%%%%%%%%%%%%%%%%%%%%%%%%%%%%%%%%%%%%%%%%%
\section{Approaches for Glucose-Level Measurement: A Broad Overview}
\label{SEC:Glucose-Level_Measurement_Overview}

\begin{figure*}[htbp]
	\centering
	\includegraphics[width=1.35\textwidth, angle=90]{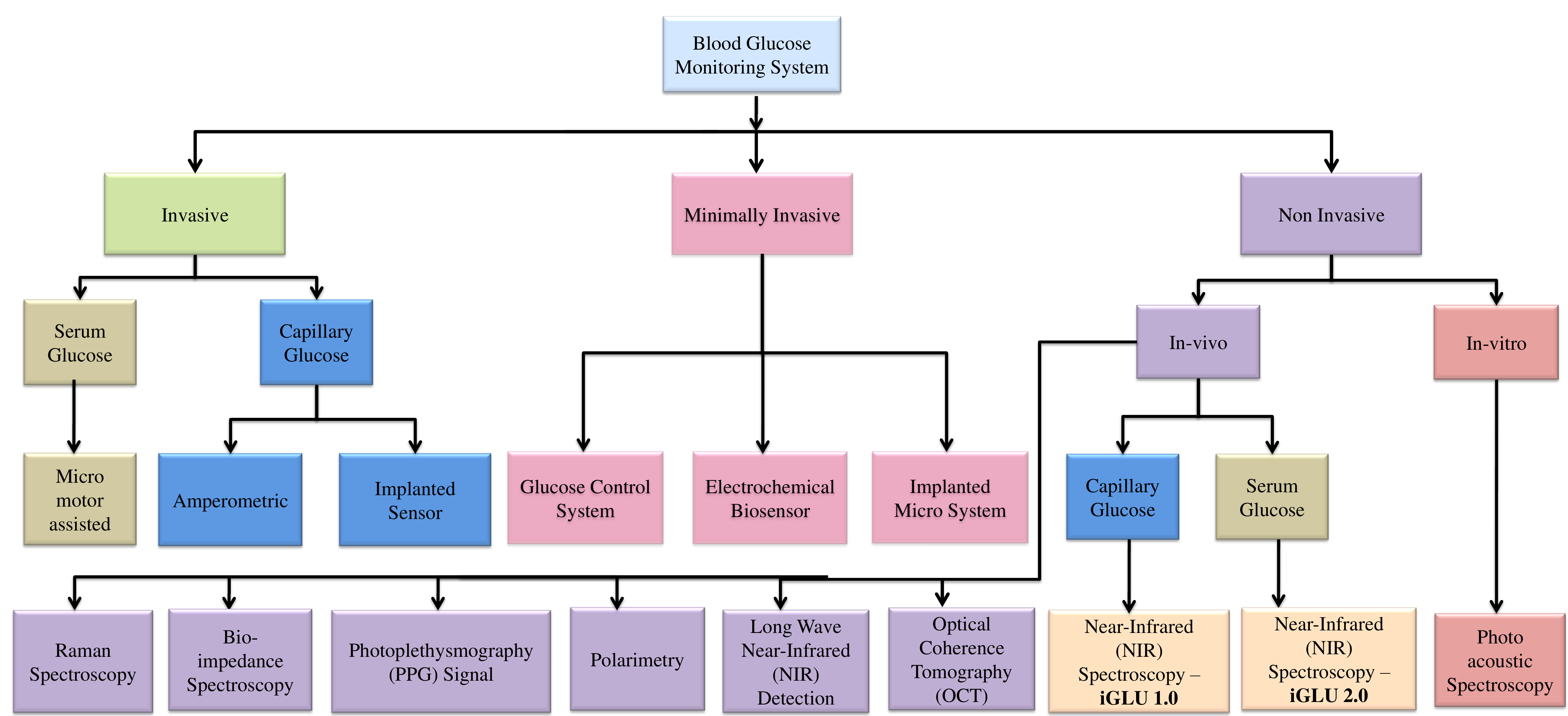}
\caption{An overview of the Glucose Measurement Options \cite{Jain_IEEE-MCE_2020-Jan_iGLU1, Jain_arXiv_2019-Nov30-1911-04471_iGLU1, Mohanty_arXiv_2020-Jan-28-2001-09182_iGLU2, Joshi_TCE_iGLU2_TCE.2020.3011966}.}
	\label{FIG:Glucose_Measurement_Options_Taxonomy}
\end{figure*}

This Section discusses an overview of various types of glucose-level measurement mechanisms.
In the past, many works has done for the glucose measurement. They can be invasive, non-invasive, or minimally invasive.
A lot of works has been completed based on the non-invasive technique. They are technically based on optical and non-optical methods.
Some of the optical techniques used methods based on Raman Spectroscopy, NIR spectroscopy, and PPG method. A taxonomy of the different methods is provided in Fig. \ref{FIG:Glucose_Measurement_Options_Taxonomy} \cite{Joshi_TCE_iGLU2_TCE.2020.3011966, Jain_arXiv_2019-Nov30-1911-04471_iGLU1, Mohanty_arXiv_2020-Jan-28-2001-09182_iGLU2,  Jain_IEEE-MCE_2020-Jan_iGLU1}.

\subsection{Invasive Methods}

Many commercial continuous blood glucose measurement devices use cost-effective electrochemical sensors \cite{Zhilo2017}. They are available to respond quickly for glucose detection in blood \cite{Gusev2017}. Lancets (for pricking the blood) is used at the primary stage for blood glucose monitoring for various commercial devices available in the market \cite{Sari2016}. The frequent measurement through the process is so much panic due to picking the blood sample from the fingertip more than 3-4 times in a day for frequent monitoring\cite{Lekha2015}.
The low invasive biosensor for glucose monitoring has been developed with glucose oxidase that require around 1mm penetration inside the skin for measurement \cite{li2016fine}. The technique of photometric was attempted to detect glucose with help of small blood volumes \cite{demitri2017measuring}.

\subsection{Minimally Invasive Methods}

The minimally invasive method using prototype sensor was developed to have frequent monitoring of glucose tissue \cite{lucisano2017glucose}. The  sensor is wearable and is implanted on membrane 
which contains the immobilized glucose oxidase.
The glucose monitoring through implantable devices were developed \cite{7576627}. The semi or minimal invasive method using biosensors designed for diabetes patient \cite{5291722}. The wearable micro system explored for frequent measurement of glucose \cite{7933990}. Similarly, there was an attempt of continuous glucose monitoring with help of microfabricated biosensor through transponder chip \cite{4956982}. The signal coming out of transponder chip was used for the calibration for semi invasive approach of Dexcom sensor \cite{acciaroli2018reduction}. The diabetes control explored by glucose sensor with artificial pancreas system \cite{6778812}. The minimal invasive approaches have limitations mainly accuracy and may have shorter life span for monitoring.

%\subsection{Wearable Microsystem for Blood Glucose Monitoring}

This is a \textit{wearable microsystem} for the continuous monitoring of the blood glucose. It's a minimally invasive method for the glucose monitoring. The main idea behind this is that it uses micro-actuator which consists the shape memory alloy (SMA) for the extraction of the blood sample from human skin \cite{wang2017wearable}. An upgraded version of SMA is used for the implementation of PCB. Because of it's feasibility and performance, it can be considered as the first wearable device for the glucose monitoring but it is large in size which makes it inconvenient.

\subsection{Non-invasive Methods}

Non-invasive measurement would mitigate all the previous issues and would provide painless and accurate solutions \cite{Kossowski2016, Pavlovich2013}. The non-invasive glucose measurement solution for smart healthcare had developed through portable measurement \cite{Siegel2016}. A lot of approaches have been introduced for glucose measurement \cite{Liu2016b}. The non-invasive measurement are more convenient for continuous glucose measurement in comparison to invasive method and semi invasive \cite{Kossowski2016}, \cite{Pavlovich2013}. The glucose measurement with help of optical method has observed more reliable and precise in the literature \cite{Sharma2012}. The popular optical methods include non-invasive measurement such as Raman spectroscopy, near infer-red spectroscopy,polarimetric,scattering spectroscopy \cite{Zhao2016}, photoacoustic spectroscopy \cite{Tanaka2016} etc. For the development of a non-invasive measurement device, it is considered by the researcher that the device would be much convenient for the user’s perspective \cite{Gouzouasis2016, Nikawa2001}. Calibration of the blood glucose to interstitial glucose dynamics have been considered for the accuracy of continuous glucose monitoring system \cite{Shao2016,Siegel2014}. Several calibration algorithms have been developed and implemented for portable setup \cite{Wang2014}.
% Sometimes, accuracy is not considered as a serious issue as per reliability and error detection \cite{Buda2014}.
% But, reliability has been approved for main requirements and tried to improve it \cite{Abdalsalam2013}. 
There has been several concious efforts  towards  the development of the self-monitoring system \cite{Bayasi2013}. 
 %A lot of work has also been done on fault detection for continuous monitoring \cite{Menon2013}, \cite{Liu2016b}.

\begin{figure}[htbp]
	\centering
	\includegraphics[width=0.65\textwidth]{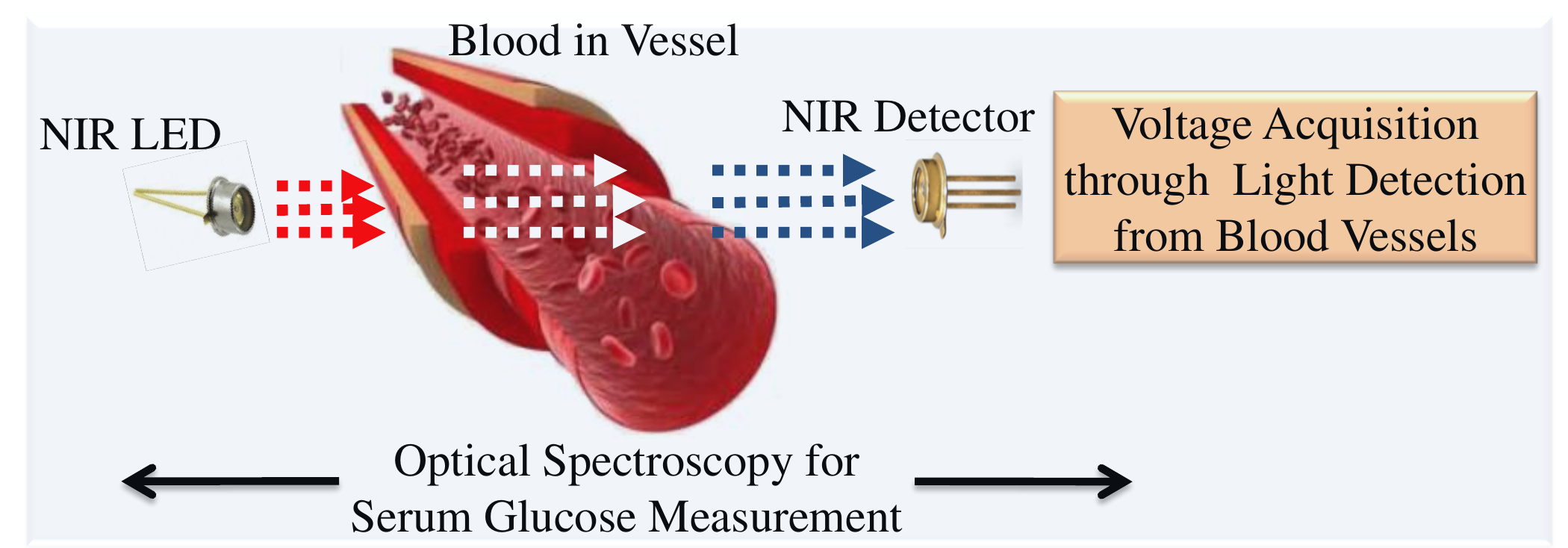}
	\caption{NIR Spectroscopy Mechanism of Serum Glucose Measurement.}
	\label{FIG:NIR-Spectroscopy_Serum-Glucose_Measurement}
\end{figure}

\subsection{Invasive Versus Non-invasive Glucose Measurements: The Trade-Offs}

\begin{figure}[htbp]
	\centering
	\includegraphics[width=0.90\textwidth]{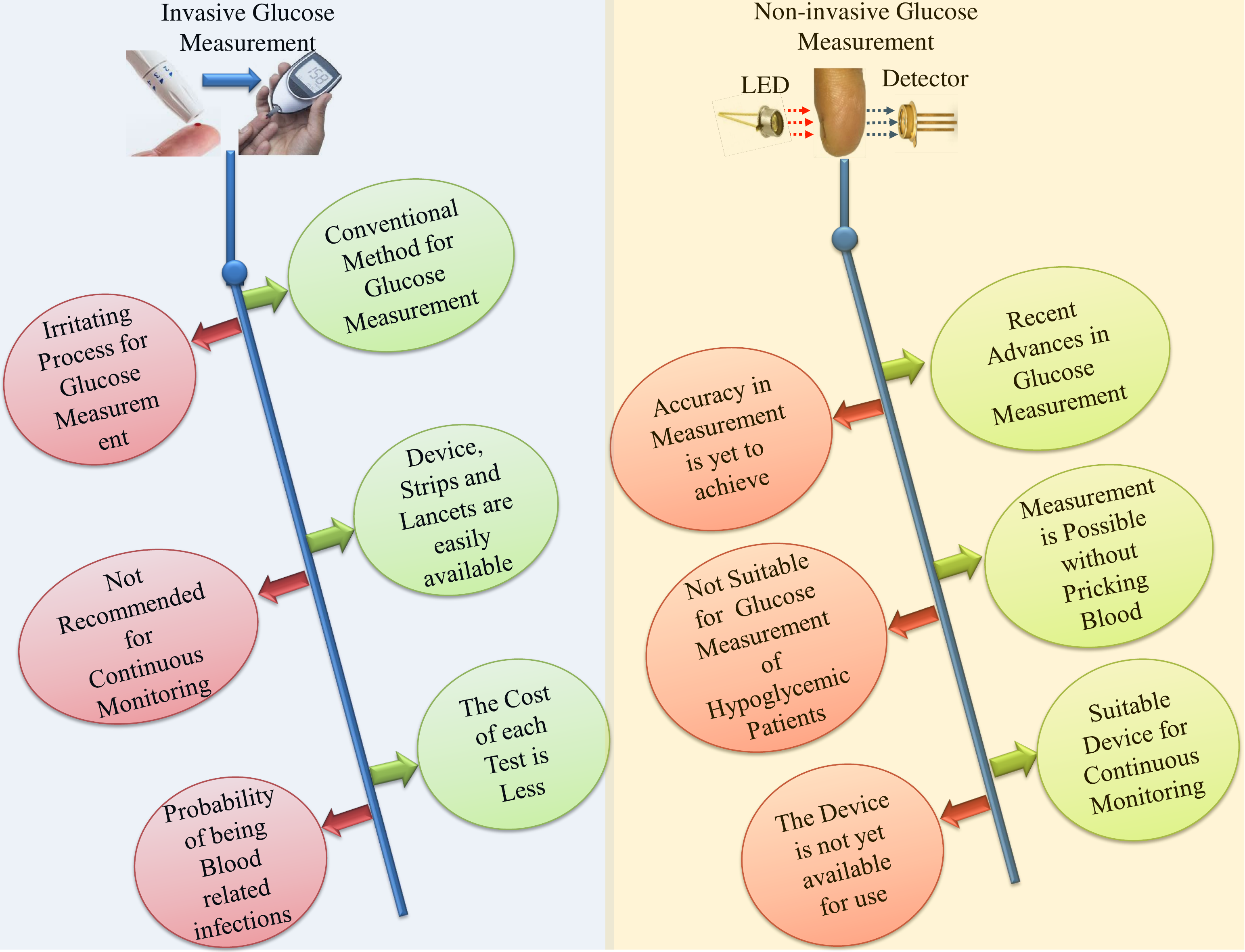}
	\caption{Representation of Tradeoffs between Invasive and Non-invasive Glucose Measurement.}
	\label{FIG:invasive_vs_noninvasive_tradeoffs}
\end{figure}

Recent glucose measurement methods for the ever-increasing the diabetic patients over the world are invasive, time-consuming, painful and a bunch of the disposable items which constantly burden for the household budget. The non-invasive glucose measurement technique overcomes such limitations, for which this has become significantly researched era. Although, there is tradeoff between these two methods which is represented in Fig. \ref{FIG:invasive_vs_noninvasive_tradeoffs}.

\subsection{Capillary Glucose versus Serum Glucose for Noninvasive Measurement}

The serum glucose value is precise which is always close to actual blood glucose measurement with compare to 
capillary glucose level. Traditional approaches able to  measure capillary glucose instantly but the serum glucose measurement identification is difficult. It is observed that the glucose level of capillary is always higher than serum glucose. The accurate measurement of blood glucose would help for appropriate control actions. Therefore, it is really important to measure the serum glucose than the capillary glucose which is more reliable for medication.
Capillary blood glucose measurement has been used widely than serum glucose estimation for medication purpose. 
%Serum glucose estimation is not viable for multiple measurements in a day. 
The serum glucose is not possible for continuous glucose measurement or frequent measurement for diabetes. The blood glucose is controlled in much better way if one can measure serum glucose at regular interval. Laboratory analysis of glycosylated haemoglobin (HbA1c) which provides 6-8 weeks blood glucose measurement is also being done through the serum blood only.
For the non-invasive measurement point of view, serum and capillary glucose are being measured through the optical spectroscopy. The mechanism of blood glucose measurement is based on received IR light after absorptions and scattering from glucose molecules which flow in blood vessels. The methodology is quite similar for both types of glucose measurement except the post-processing computation models which are necessary for blood glucose estimation.

\subsection{Non-invasive Method for Glucose Level Estimation by Saliva}

As the most convenient method to estimate glucose level is via saliva \cite{agrawal2013noninvasive} and is used for children and adults.
This saliva has specific type of parts which can be defined as: 
(1)  gland-specific saliva and (2) whole saliva.
The collection of the Gland-specific saliva is done by individual glands like parotid, Sub mandibular, sublingual, and minor salivary glands. This diagnosis is done by the history 
of the patient in terms of associated risk factors, family history, age, sex, duration of diabetes,and any associated illness. 
Other Glucose measuring methods consist of measurement using photo-metric glucometers requiring very small sample volumes \cite{demitri2016measuring}.  The basic approach is based on the reaction of the chemical test strip that reacts with sample. Measurement is done by capturing the reflections of the test area and then glucose level is estimated. It requires validation in large number of patients. 

%\begin{figure}[h]
%\centering
%\includegraphics[width=14cm]{Fig1}
%\caption{ General Flow of Model }\label{comp2d-cd-crwop}
%\end{figure}

%%%%%%%%%%%%%%%%%%%%%%%%%%%%%%%%%%%%%%%%%%%%%%%%%%%%%
%\section{What are Different Non-invasive Methods?}
%%%%%%%%%%%%%%%%%%%%%%%%%%%
\section{Approaches for Noninvasive Glucose-Level Measurement}
\label{SEC:Glucose-Level_Noninvasive_Approaches}

\begin{figure}[htbp]
	\begin{center}
		\includegraphics[width=0.75\textwidth]{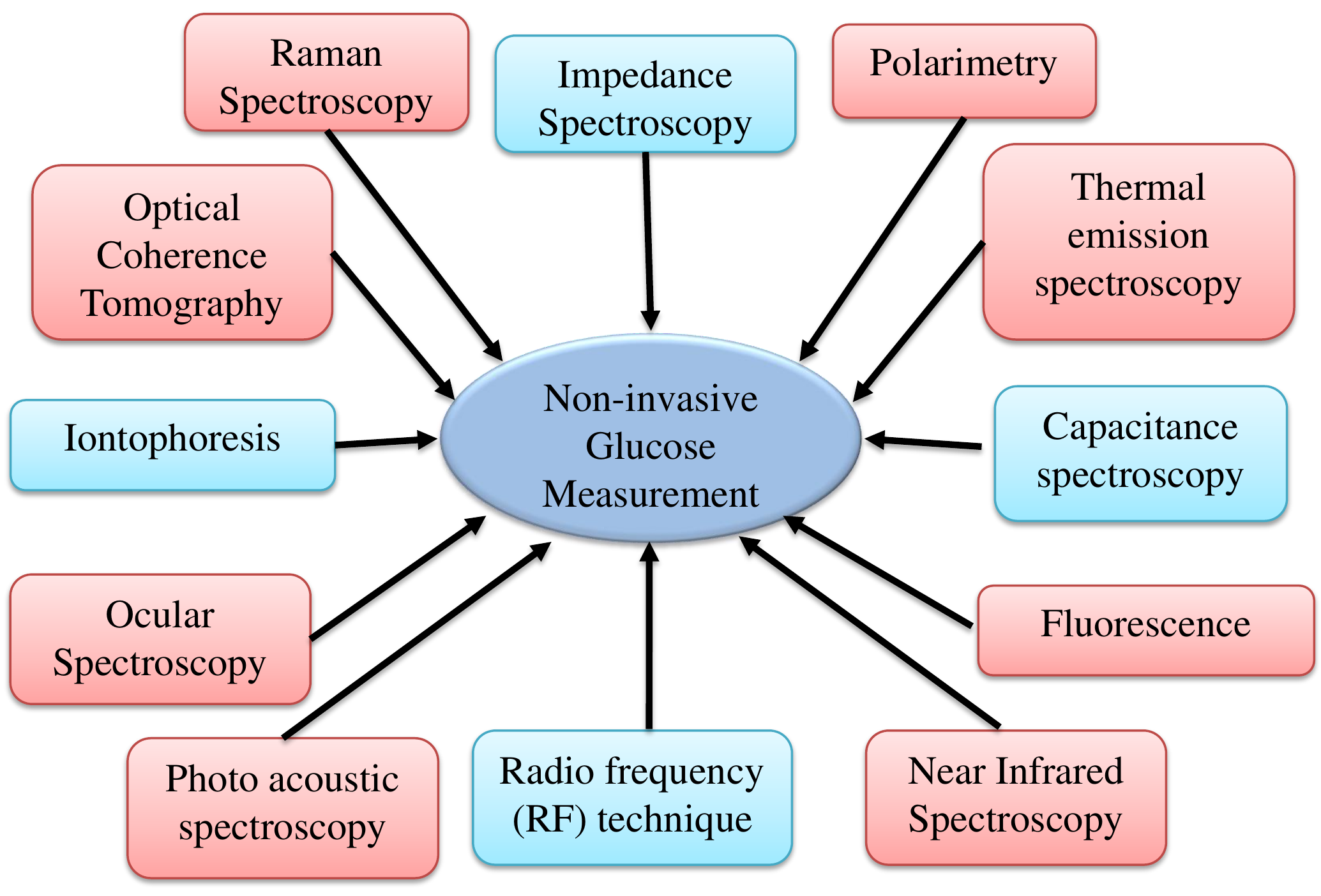}
\caption{Various spectroscopy techniques for noninvasive glucose measurement.}
		\label{FIG:Noninvasive_Glucose_Measurement_Methods}
	\end{center}
\end{figure}

\begin{figure}[htbp]
	\begin{center}
		\includegraphics[width=0.85\textwidth]{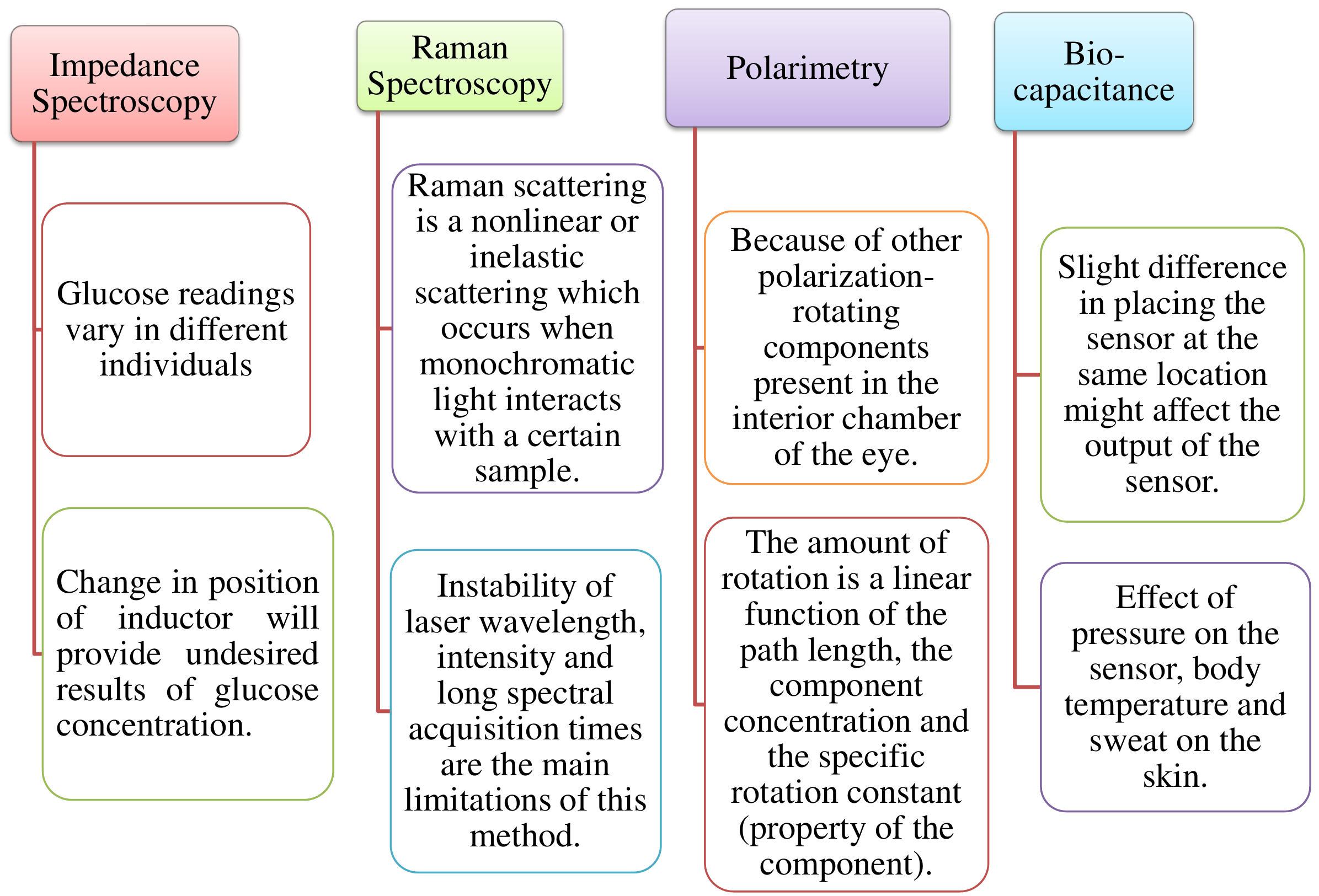}
\caption{Comparative Perspective of Various popular spectroscopy techniques for noninvasive glucose measurement.}
		\label{FIG:Noninvasive_Glucose_Measurement_Comparative}
	\end{center}
\end{figure}

This Section presents detailed discussions of various
available approaches for noninvasive glucose-level monitoring.
%Non-invasive approaches of measurement are more advanced compared to the current invasive method to make the painless device. The portable system of measurement (SoM) of the non-invasive glucose measurement device is desirable for smart healthcare system. Non-invasive approaches of measurement are more advanced compared to the current invasive method. The optical method is more reliable, cost-efficient for glucose measurement according to the analysis.
There have been several efforts for noninvasive glucose measurement using optical techniques \cite{Salam2016TheEO, Mohanty_arXiv_2020-Jan-28-2001-09182_iGLU2, 10.3390/s20051251, 10.1007/s00216-018-1395-x}. These techniques are mainly based on various spectroscopy based methods. For the development of a non-invasive measurement device, it is considered by the researcher that the device would be much convenient for the users perspective. 
%This Section provides a summary of various optimized techniques for glucose detection and previously developed devices for non-invasive measurement.
Fig. \ref{FIG:Noninvasive_Glucose_Measurement_Methods} presents summary of various types of noninvasive glucose measurement techniques, whereas their comparative perspectives are presented in Fig. \ref{FIG:Noninvasive_Glucose_Measurement_Comparative}. A qualitative comparative perspective of various noninvasive methods is summarized in Table \ref{TBL:Noninvasive_Comparions}.

%%%%%%%%%%%%%%%%%%%%%%%%%%%%%%%%%%
\noindent
\begin{center}
%\begin{longtable}{p{3.0cm}p{6.5cm}p{6.5cm}}
%\label{TBL:Noninvasive_Comparions}
	% \begin{table}
	%\begin{tabular}{p{3.0cm}p{6.5cm}p{6.5cm}}
	% \begin{tabular}{| c | c | c |}
%	\hline 
\tablehead{\hline	\textbf{Technique} & \textbf{Advantages} & \textbf{Disadvantages} \\ \hline \hline
}
\tabletail{\hline}
\tablecaption{Qualitative comparison of various noninvasive glucose-level monitoring methods.}
\label{TBL:Noninvasive_Comparions}
%	\hline
\begin{supertabular}{p{3.0cm}p{6.5cm}p{6.5cm}}
	Near Infra-Red (NIR) & \begin{itemize}
		\item The signal intensity is directly proportional to glucose molecule
		\item The glucose detection concept would work with other interfacing substance such as plastic or glass
	\end{itemize}  &
	\begin{itemize}
		\item The glucose signal weak comparatively so complex machine learning model is required for interpretation
		\item High scattering level
	\end{itemize}  \\
	\hline
	Mid Infra-Red (MIR) & \begin{itemize}
		\item The glucose molecule absorption stronger
		\item Low scattering
	\end{itemize}  &
	\begin{itemize}
		\item The light has limited penetration with tissue
		\item Noise is present in the signal so water and other non-glucose metabolites would be detected.
	\end{itemize}  \\
	\hline
	Far Infra-Red (FIR)/Thermal emission spectroscopy & \begin{itemize}
		\item Frequent Calibration is not required
		\item  Least sensitive towards scattering 
	\end{itemize}  &
	\begin{itemize}
		\item The radiation intensity depends on temprature and substance thickness
		\item Strong absorption with water so it is difficult to have precise glucose measurement
	\end{itemize}  \\
	\hline
	Raman Spectroscopy & \begin{itemize}
		\item Less sensitivity towards temperature and water
		\item High specificity
	\end{itemize}  &
	\begin{itemize}
		\item Requirement of the laser radiation source hence it can dangerous cell for CGM
		\item Susceptible towards noise interference so low SNR
	\end{itemize}  \\
	\hline
	Photo acoustic & \begin{itemize}
		\item Simple and compact sensor design
		\item Optical radiation will not harmful for the tissue
	\end{itemize}  &
	\begin{itemize}
		\item Signal is vulnerable towards acoustic noise, temperature,motion etc. 
		\item It carries some noise from some non-glucose blood
		components 
	\end{itemize}  \\
	\hline
	Polarimetry & \begin{itemize}
		\item The laser intensity
		variation will not change much the glucose prediction
	\end{itemize}  &
	\begin{itemize}
		\item Requirement external laser source and requires proper alignment with eye
		\item sensitive for the change in PH and temperature
	\end{itemize}  \\
	\hline
	Reverse Iontophoresis & \begin{itemize}
		\item Based on simple enzyme based electrode system
		\item  Highly accurate as it measure glucose from interstitial fluid 
	\end{itemize}  &
	\begin{itemize}
		\item Difficult to have proper calibration 
		\item Not so user-friendly approach due to passing of the current through skin 
	\end{itemize}  \\
	\hline 
	Fluorescence &
	\begin{itemize}
		\item Highly sensitive for glucose molecule detection due to immune for light scattering
		\item  Good sensitivity because of distinctive optical properties
	\end{itemize}  &
	\begin{itemize}
		\item Very much sensitive for local pH and/or oxygen, 
		\item  Suffers from foreign body reaction
	\end{itemize}  \\
	\hline 
	%  Fluorescence &
	% \begin{itemize}
	% \item Highly sensitive for glucose molecule detection due to immune for light scattering
	% \item  Good sensitivity because of distinctive optical properties
	%  \end{itemize}  &
	%  \begin{itemize}
	%  \item Very much sensitive for local pH and/or oxygen, 
	% \item  Suffers from foreign body reaction
	% \end{itemize}   \\
	%  \hline 
	Bio impedance spectroscopy  & 
	\begin{itemize}
		\item Comparatively less extensive 
		\item Easy for CGM 
	\end{itemize}  &
	\begin{itemize}
		\item Prone towards sweating, motion and temperature 
		\item Require large calibration period 
	\end{itemize} \\
	\hline 
	Millimetre and Microwave sensing  & 
	\begin{itemize}
		\item Deep penetration depth for precise glucose  measurement
		\item No risk for ionization
	\end{itemize}  &
	\begin{itemize}
		\item Poor selectivity 
		\item  Very much sensitive for physiological parameters such as sweating, breathing and cardiac activity
	\end{itemize}  \\
	\hline 
	Optical Coherence Tomography  & 
	\begin{itemize}
		\item High resolution and good SNR
		\item Not vulnerable for blood pressure and cardiac activity
	\end{itemize} & 
	\begin{itemize}
		\item Glucose value may change as per skin and motion
		\item Suffers from tissue inhomogeneity
	\end{itemize} \\
	\hline 
	Surface Plasma Resonance & 
	\begin{itemize}
		\item Small glucose molecule can be detected due to high sensitivity
	\end{itemize} & 
	\begin{itemize}
		\item Long calibration process and size is bulky
		\item Glucose value changes with variation in temperature,sweat and motion
	\end{itemize}  \\
	\hline 
	Time of flight and THz Time domain Spectroscopy & 
	\begin{itemize}
		\item strong absorption and dispersion for glucose molecule
	\end{itemize} &
	\begin{itemize}
		\item Lesser depth resolution and longer time for measurement
	\end{itemize}  \\
	\hline 
	Metabolic Heat Conformation  & 
	\begin{itemize}
		\item Uses the concept of well-known various physiological parameters for glucose prediction
	\end{itemize} &
	\begin{itemize}
		\item Sensitive towards variation in temperature and sweat
	\end{itemize} \\
	\hline 
	Electromagnetic sensing & 
	\begin{itemize}
		\item   low-cost and can be easily miniaturized
		\item   No risk of ionization
	\end{itemize} &
	\begin{itemize}
		\item Lack of selectivity due to 
		dielectric constant is mainly affected with other blood components 
		\item  More sensitive for the slight change of temperature
	\end{itemize} \\
	\hline 
	Ultrasound Technology & 
	\begin{itemize}
		\item   Well established technology with not much harm to tissue cell
		\item   Long penetration below the skin or tissue
	\end{itemize} &
	\begin{itemize}
		\item  Limited accuracy with ultrasound only hence mostly used with NIR as multi-model
		\item  costly technology for measurement and not useful for CGM
	\end{itemize} \\
	\hline 
Sonophoresis & 
	\begin{itemize}
		\item   Favourable technology  as  there is no side-effect to skin
		\item   Based on well known enzymatic method
	\end{itemize} &
	\begin{itemize}
		\item  Error prone due to environmental parameters
	\end{itemize} \\
	\hline 
	% \end{tabular}
	% \end{table}
%\end{longtable}
\end{supertabular}

\end{center}

\subsection{Near-Infrared (NIR) Spectroscopy}

\begin{figure}[t]
	\begin{center}
		\includegraphics[width=0.80\textwidth]{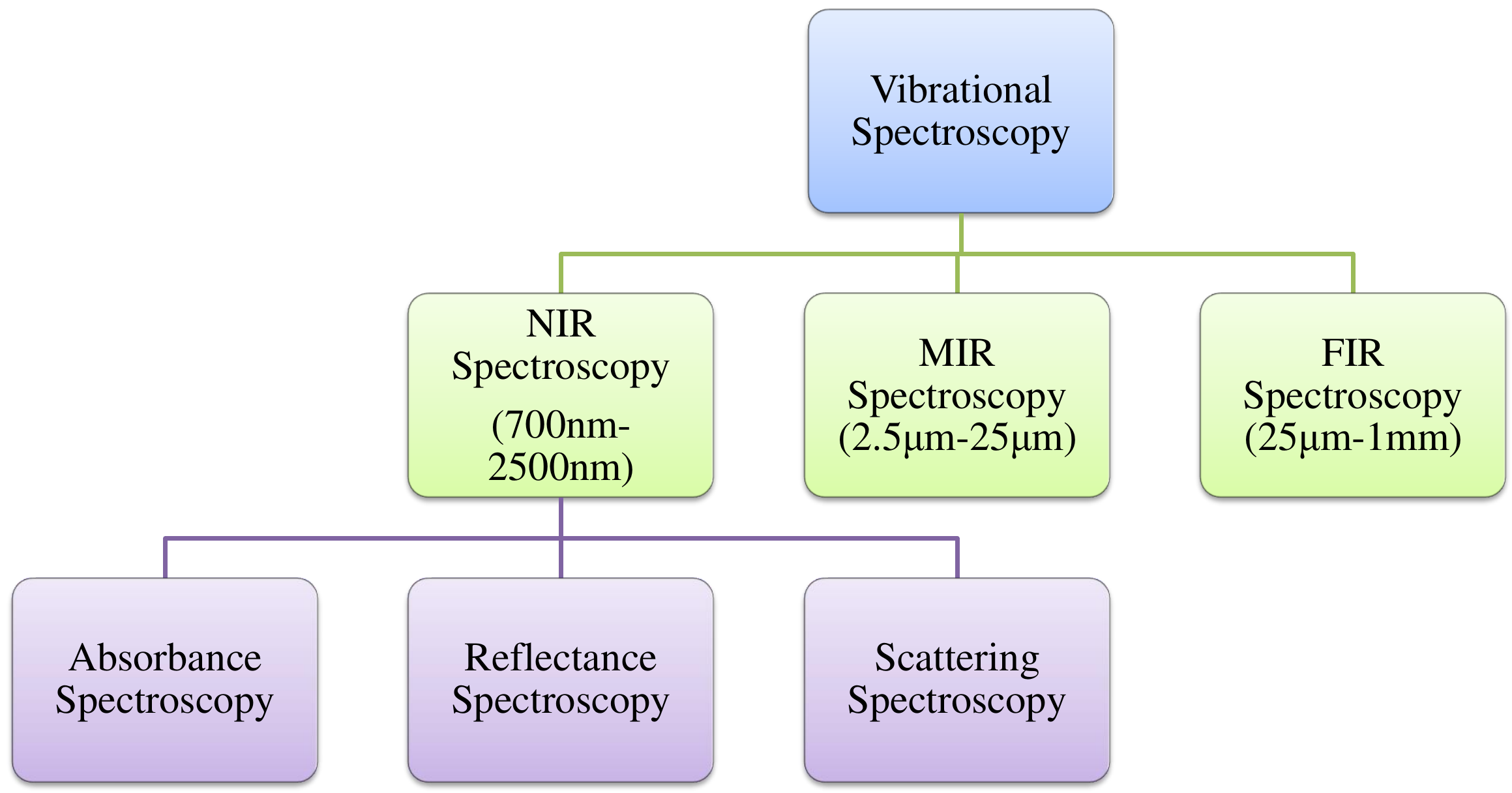}
		\caption{Classification of vibrational spectroscopy \cite{menon2013voltage}.}
		\label{FIG:Vibrational_Spectroscopy_Taxonomy}
	\end{center}
\end{figure}

It is well known as Infrared spectroscopy (IR spectroscopy) or vibration spectroscopy where radiation of infrared type are incident on the matter \cite{Madzhi2014, Aziz2014}. Various types of IR spectroscopy is shown in  Fig. \ref{FIG:Vibrational_Spectroscopy_Taxonomy}. In general, IR spectroscopy includes reflection, scattering and absorption spectroscopy \cite{azolifesciences_URL_2020}. The wave from IR absorption cause the molecular vibration and generate the spectrum band with wavelength number in $cm^{-1}$ \cite{Muley2014}.

In this case, the light in the wavelength range of 700nm to 2500nm for Near-infrared region is applied at the object (may be finger or ear lobe)  \cite{Lai2016}.
The light may interact with blood components and it may scattered, absorbed and reflected \cite{Tamilselvi2015, Lawand2015}. The intensity of received light varies as per glucose concentration as per Beer-Lambert law \cite{Yadav2014, Abidin2013}
The receiver would help to measure the presence glucose molecule from the blood vessel \cite{Nikawa2006}. 

%In this technique, the light in the near-infrared range (700nm to 2500nm) is passed through the object (ear lobe or finger) \cite{Lai2016}.
% The passed light through the finger or ear lobe interacted with the components of blood and gets reflected, absorbed and scattered \cite{Tamilselvi2015}. The penetration depth will be varied with a change in wavelength \cite{Lawand2015}. According to Beer-Lambert law, the attenuation of light in tissue or vessel relates the intensity of light, reflection, scattering coefficient and path length of light through tissue or vessel \cite{Yadav2014}. Attenuation occurs due to absorption of scattering of light \cite{Abidin2013}. The value of absorption coefficient depends upon the change in glucose concentration \cite{Anas2012}. The value of glucose concentration in blood vessel could be indicated due to change in intensity of transferred light through the vessel \cite{Menon2013a}. The change in glucose concentration is measured through light detector \cite{Nikawa2006}.

\begin{figure}[htbp]
	\begin{center}
		\includegraphics[width=0.75\textwidth]{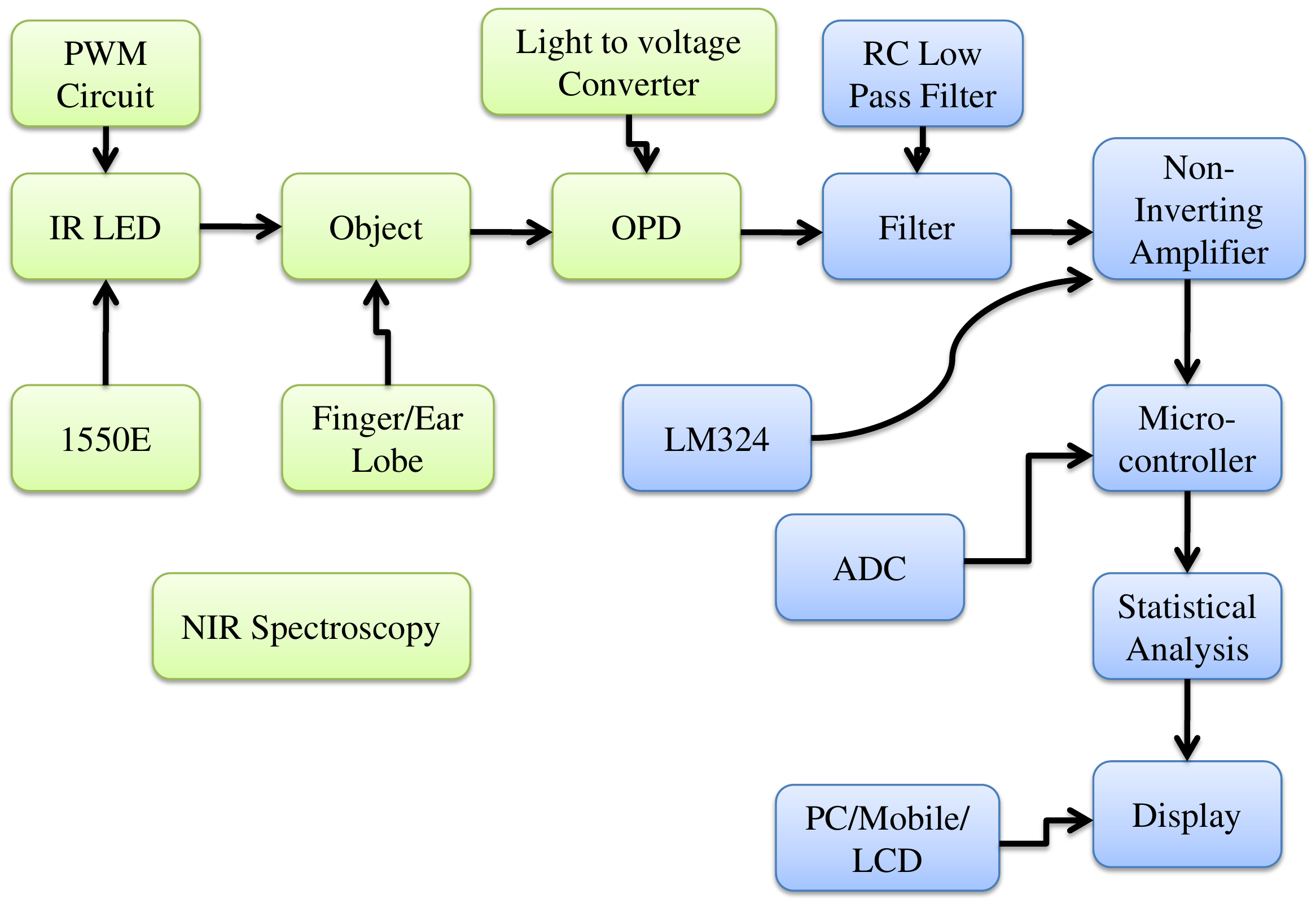}
		\caption{Block diagram representation of IR spectroscopy.}
		\label{FIG:IR_Spectroscopy}
	\end{center}
\end{figure}

\subsubsection{Long-Wave versus Short-Wave NIR Spectroscopy}

\begin{figure}[htbp]
	\centering
	\includegraphics[width=0.650\textwidth]{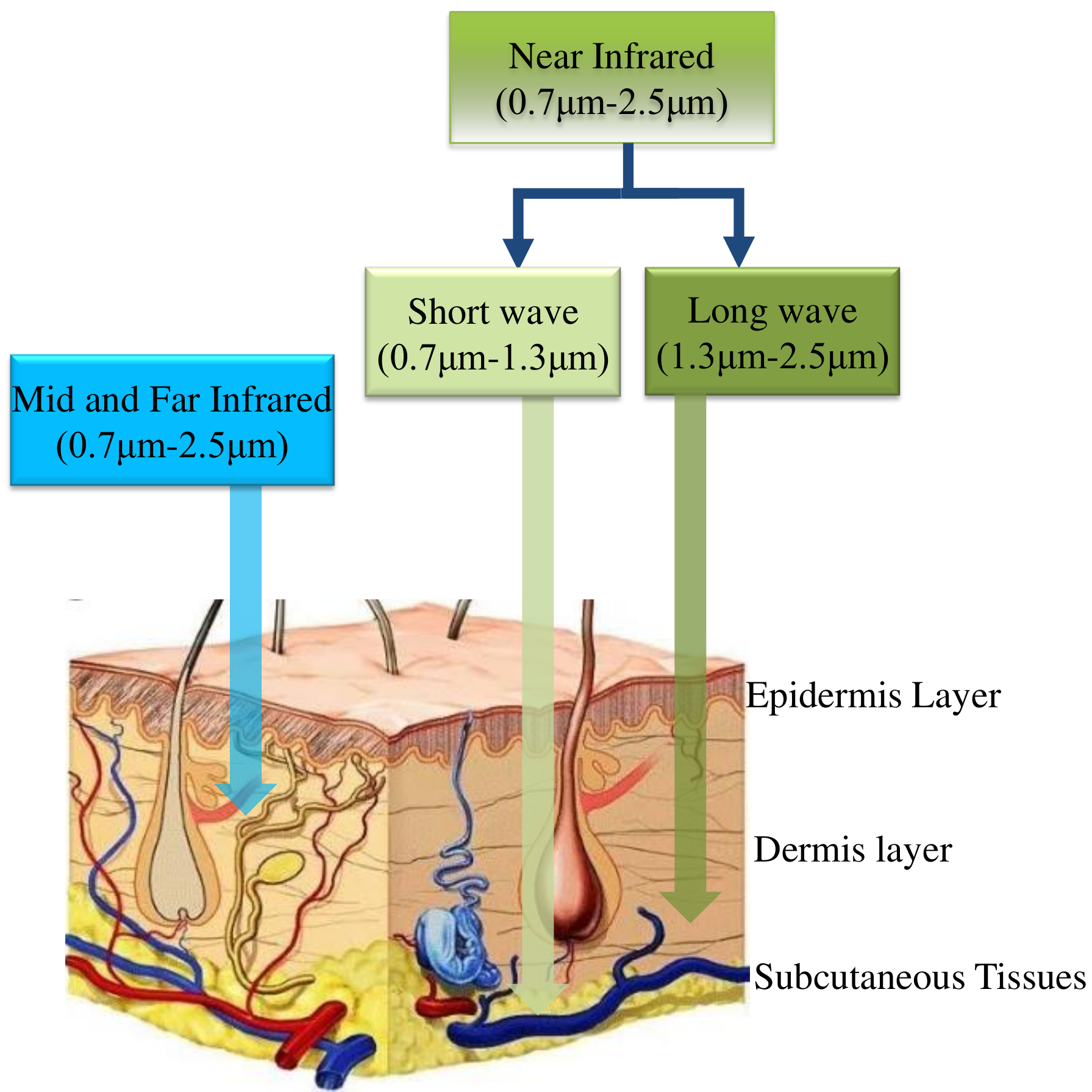}
\caption{Penetration depth various Infrared Signals in Human Skin \cite{Joshi_TCE_iGLU2_TCE.2020.3011966, Jain_IEEE-MCE_2020-Jan_iGLU1}.}
	\label{FIG:IR_Skin_Penetration_Depth}
\end{figure}

The optical detection is useful approach to have precise glucose measurement. FIR (Far infra-red) based optical technique help to get the resonance between OH and CH for first overtone. However, long wave NIR has good performance in vitro testing. In similar way, the fiber-optic sensor is used along with laser based mid-infrared spectroscopy for vitro based glucose measurement. The continuous glucose measurement has been achieved with multivariate calibration model for error analysis \cite{goodarzi2016selection}. The FIR approach has limitation of shallow penetration in comparison with short wave NIR. The short NIR would help to detect the glucose molecule more accurately \cite{sharma2013efficient}. The concept of NIR spectroscopy for glucose detection is shown in Fig. \ref{FIG:IR_Skin_Penetration_Depth}. The specific wavelength of NIR spectroscopy has already been applied earlier for precise glucose measurement using non-invasive measurement \cite{uwadaira2010factors}. Some specific wavelength such as 940 nm has been considered for the detection of glucose \cite{haxha2016optical}. The vibration of CH molecule has been observed at 920 nm with NIR spectroscopy \cite{uwadaira2010factors}. In some other works, the glucose absorption has been validated for the range 1300 to 1350 nm and stretching of glucose has been identified in NIR region \cite{zhang2013discussion, golic2003short}.  The presence of glucose component has been measured at 1300 nm in the work \cite{Song2015}. 

\subsubsection{NIR Spectroscopy Based Methods}

A method to estimate the non-invasive blood glucose with NIR spectroscopy using PPG has been proposed in literature \cite{ramasahayam2015noninvasive}. This method is performed using NIR LED and photo detector with an optode pair. At NIR wavelengths(935nm, 950nm, 1070nm), PPG signal is obtained by implementation of analog front end system. The glucose levels has been estimated using Artificial Neural Network (ANN) running in FPGA.
A microcontroller is used, for the painless and autonomous blood extraction \cite{heller2005integrated}. The ideal system Blood Glucose Measurement (BGM) in which the microcontroller is used to display the blood glucose and for the transmission of blood glucose. A remote device is used for the tracking of the insulin pump which is needed for diabetes management. This type of measurement \cite{pai2015nir} method uses change in the pressure of the sensitive body part, because it generates the sound waves. The response of the photo acoustic signals will be stronger when glucose concentration is higher. In order to improve SNR and for the reduction of noise to transfer the signal to the computer for further processing,the signal is then amplified. Feature extraction and glucose estimation is estimated by photo acoustic amplitude. In order to gather the photo acoustic signals, two pulsed laser diodes and piezoelectric transducer is used. Utilization of the LASER makes the setup costly and bulky.

\subsubsection{Non-invasive Blood Glucose Measurement Device (iGLU)} 
In this approach, ``Intelligent Glucose Meter (iGLU)'' \cite{jain2019iglu} has been utilized for the acquisition of data. This device works on a combination of NIR spectroscopy and machine learning. This device has been implemented using three channels. It uses an Internet-of-Medical-Things platform for storage and remote monitoring of data.
In the proposed device, an NIR Spectroscopy is used with multiple short wavelengths  \cite{jain2019precise}. It uses three channels for data collection. Each channel has its own emitter and detector for optical detection. Then the data collection processed by a 16 bit ADC with the sampling rate of 128 samples/second. Regression techniques is used to calibrate and validate the data and analyse the optimized model. The data that is stored on cloud can be used and monitored by the patients and the doctors. Treatment can be given based on the stored data values. This is a low cost device with more than 90 \% accuracy but it does not give real time results.

%Figure 3 gives an overview of measurement techniques for blood glucose. 

%\begin{figure}[h]
%\centering
%\includegraphics[width=14cm]{overview of blood glucose measurement method}
%\caption{An overview of Blood Glucose measurement techniques }\label{comp2d-cd-crwop}
%\end{figure}

\subsubsection{Why NIR is Preferred Over other Noninvasive Approaches?}

Glucose measurement has been done using various non-invasive approaches such as impedance spectroscopy, NIR light spectroscopy, PPG signal analysis and so on. But, apart from optical detection, other techniques have not be able to provide the precise measurement. PPG is one of the promising alternative but the PPG signal varies according to blood concentration \cite{monte2011non, habbu2019estimation}. 
It may not be useful to have precise prediction of the blood glucose.
The saliva and sweat properties vary from one person to another person. Therefore, it could not be reliable glucose measurement method.
%Desirable accuracy is not possible from measurement through sweat and saliva as properties always vary for each person. PPG signal analysis is based on extracted features of logged signal which is not based on principle of glucose molecular detection.
%  The methodology of glucose measurement through PPG signal and NIRS are represented in Figure \ref{ppgvsnirs}. 
The other spectroscopy have been also applied for
glucose measurement. However, they are not able to provide portable, cost effective and accurate prediction of body glucose.The glucose measurement using optical detection using long NIR wave which is not capable to detect the glucose molecules beneath the skin as it has shallow penetration \cite{sharma2013efficient}. Therefore, small NIR wave has been considered as potential solution for real-time glucose detection \cite{haxha2016optical, Ali2017}.

\subsection{Mid Infrared (MIR) Spectroscopy}

The bending and stretching of glucose  molecules would be observed very well with Mid Infrared (MIR) spectroscopy \cite{C0AN00537A}. The depth of skin penetration is very less because it tends to have larger absorption of water. This technique helps to have ISF glucose value in vivo measurement. There are some attempts for precise glucose measurement through saliva and palm samples.

\subsection{Blood Glucose Level Measurement using PPG}
%\subsection{Glucose Measurement using PPG Signal}

\begin{figure}[htbp]
	\centering
	\includegraphics[width=0.90\textwidth]{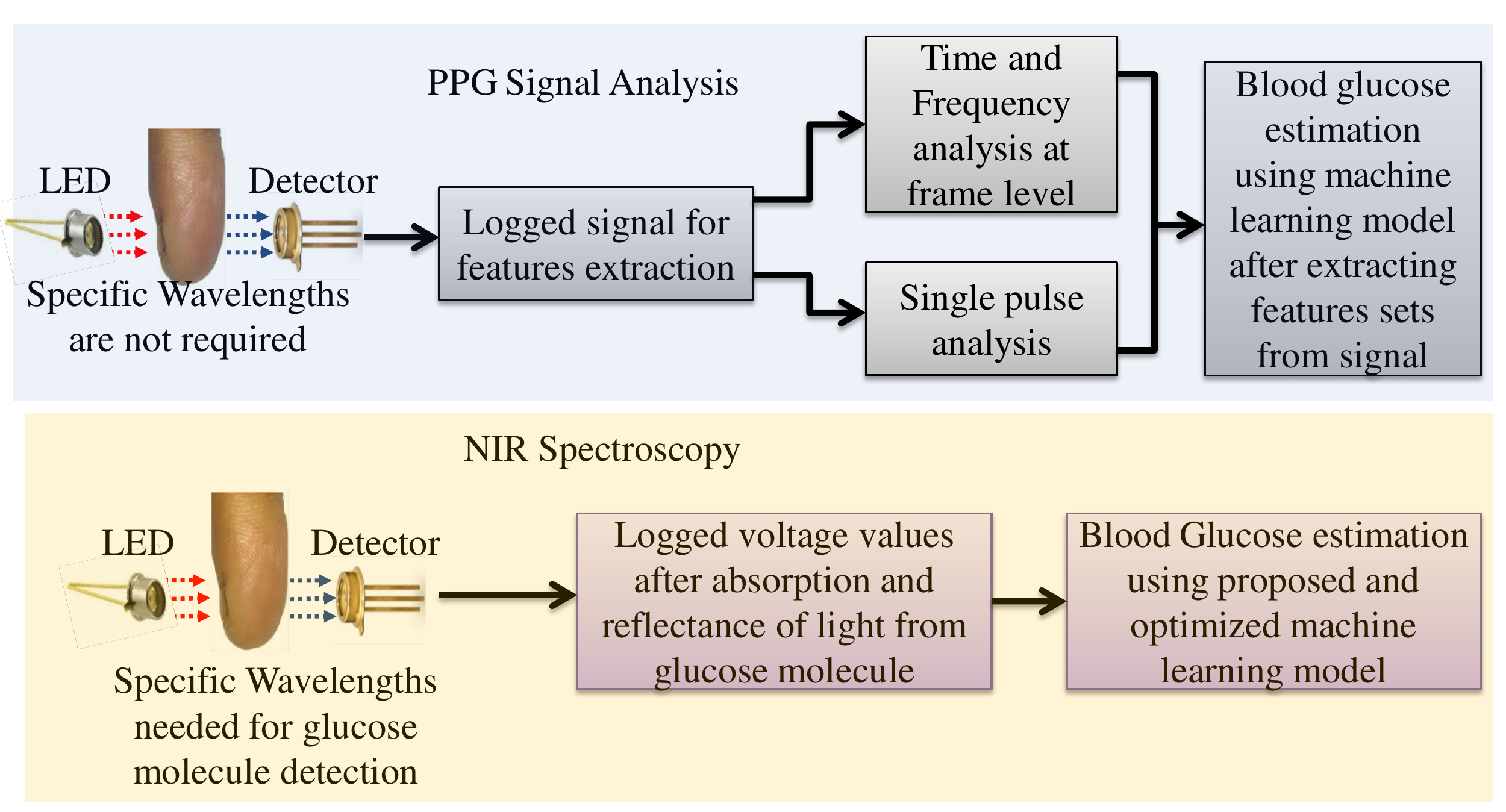}
	\caption{PPG Versus NIR for Non-invasive Glucose Measurement \cite{Joshi_TCE_iGLU2_TCE.2020.3011966, Jain_IEEE-MCE_2020-Jan_iGLU1}.}
	\label{FIG:PPG-Versus-NIR_Spectroscopy}
\end{figure}

The change of blood volume with absorption of the light from tissue has been detected with PPG signal  \cite{habbu2019estimation}.  The change of the blood volume has been measured using pressure pulse with help of light detector \cite{monte2011non}. The change in volume of blood would result as the change of light intensity hence it may not be occur due to glucose molecule.  This may result as inaccurate glucose measurement. The difference of NIR against PPG has been shown in Fig \ref{FIG:PPG-Versus-NIR_Spectroscopy}. The intelligent glucose measurement device iGLU is mainly based on principle of NIR spectroscopy which helps to have precise glucose measurement.  There have been several work for glucose detection based on PPG signal \cite{paul2012design}. The data from patient body has logged to estimate the presence of glucose using PPG. Subsequently, various machine leaning models have been used for prediction of body glucose value \cite{philip2017continous}. The different parameters from total 70 subjects of healthy and diabetes have been considered for the prediction using Auto-Regressive Moving Average (ARMA) models \cite{karimipour2009diabetic}. There have been also several other smart solutions for glucose estimation using PPG signal with intelligent algorithms  \cite{cruz2019application, zhang2019non, yamakoshi2017side}.

One of the optical based techniques is Photo-plethysmography (PPG) which is used in advanced health care. It is non-invasive glucose measurement technique. In NIR spectrum a sensor similar to a pulse oximeter is used to record the PPG signal \cite{habbu2019estimation}. Photo transmitter and receiver is used to build the sensor which will operate in near infrared region at 920nm. At wavelength 920nm, by measuring changes in the absorption of light, a PPG signal can be obtained. The veins in the finger grow and contract with every heartbeat.

A method of measuring blood glucose using pulse oximeter and transmission of the PPG glucose monitoring system is available \cite{paul2012design}. As the glucose concentration increases, there is decrease in the light absorbance in the blood. The obtained signal is in the form of photo current, and for the filtering of this signal is then changed into the measurable voltage values. For the processing of filtered signal, lab view is used to estimate the blood glucose level.

A system using machine learning techniques and PPG system for the measurement of blood glucose level non-invasively has been prototyped \cite{monte2011non}. In this model, a PPG sensor, an activity detector, and a signal processing module is used to extract the features of PPG waveform. It finds the shape of the PPG waveform and the blood pressure glucose levels, the functional relationship between these two can be obtained then.

In PPG, the change in light intensity will be varied according to changes in blood volume. PPG signal analysis is not based on the principle of glucose molecule detection. Hence, the system has limited accuracy \cite{Joshi_TCE_iGLU2_TCE.2020.3011966, Jain_IEEE-MCE_2020-Jan_iGLU1}. Fig. \ref{FIG:PPG-Versus-NIR_Spectroscopy} illustrates the differences.

\subsection{Impedance Spectroscopy}

\begin{figure}[htbp]
	\begin{center}
		\includegraphics[width=0.75\textwidth]{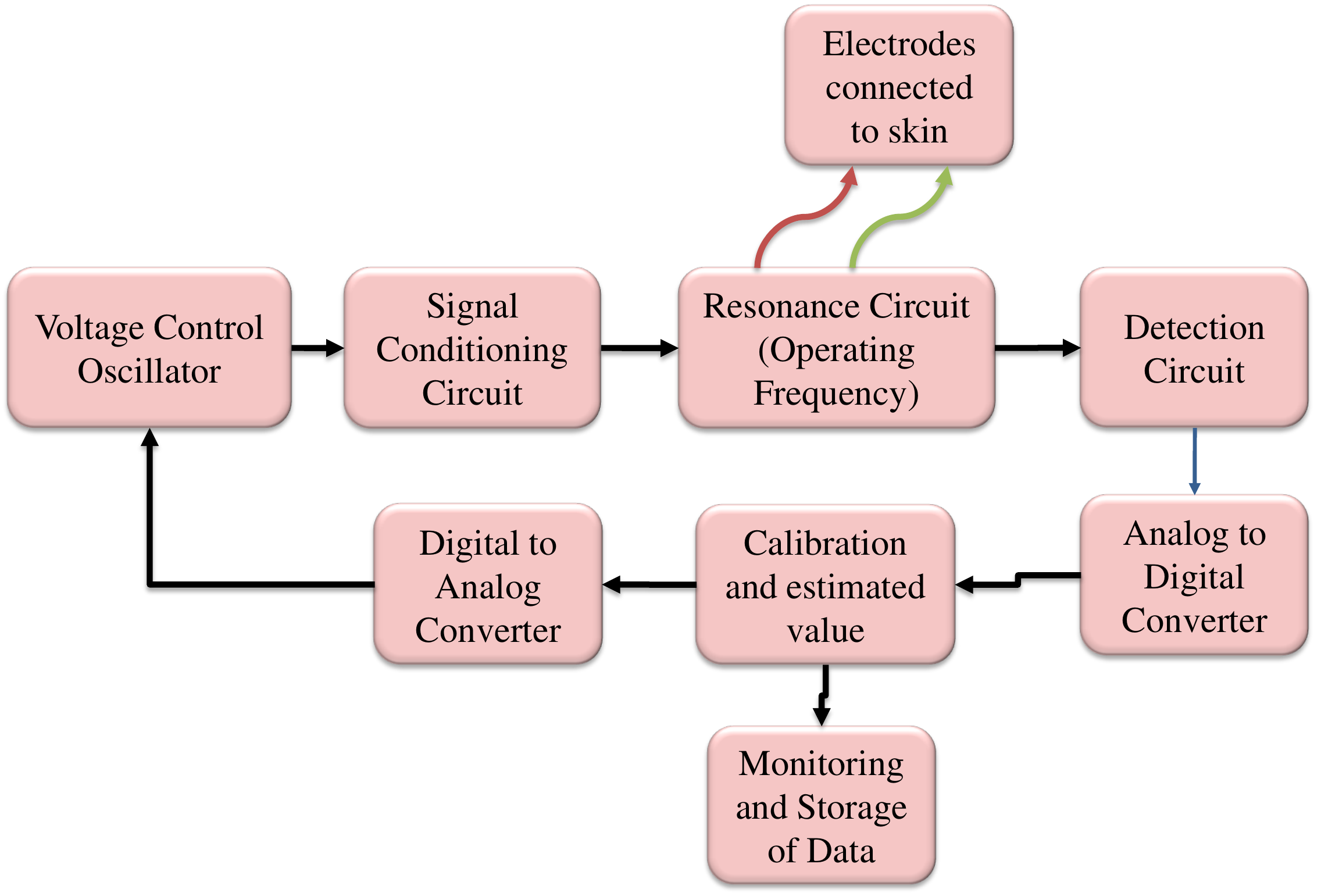}
		\caption{The Steps of Impedance Spectroscopy (IMPS).}
		\label{FIG:IMPS_Spectroscopy}
	\end{center}
\end{figure}

Impedance spectroscopy (IMPS) refers to the dielectric spectroscopy \cite{Olarte2013}. The steps of impedance spectroscopy (IMPS) is shown in Fig. \ref{FIG:IMPS_Spectroscopy}. This technique finds the dielectric properties of skin \cite{Dhar2013}. The current is directed through the skin \cite{Khawam2013}. Due to directed small current at multiple wavelengths, the impedance range is obtained \cite{Anas2013}. The range lies between 100 Hz to 100 MHz \cite{Amaral2007, Jain2017}. Change in glucose concentration will reflect the change in sodium ions and potassium ions concentration \cite{Paul2012}. So, the cell membrane potential difference will be changed \cite{Hofmann2012}. Thus, the dielectric value will be changed which predicts the glucose value of human body \cite{Liu2016}.

An enzyme sensor in a flow cell has been explored for glucose measurement in saliva \cite{677170}. Polypyrrole (PPy) supported with copper (Cu) nanoparticles on alkali anodized steel (AS) electrode for glucose detection in human saliva is available in \cite{8347021}.
The high precision level cannot be possible through these methods as sweat and saliva properties vary according to person. Hence, this approach is not suitable for glucose measurement in smart healthcare.

\subsection{Raman Spectroscopy}

\begin{figure}[htbp]
	\begin{center}
		\includegraphics[width=0.75\textwidth]{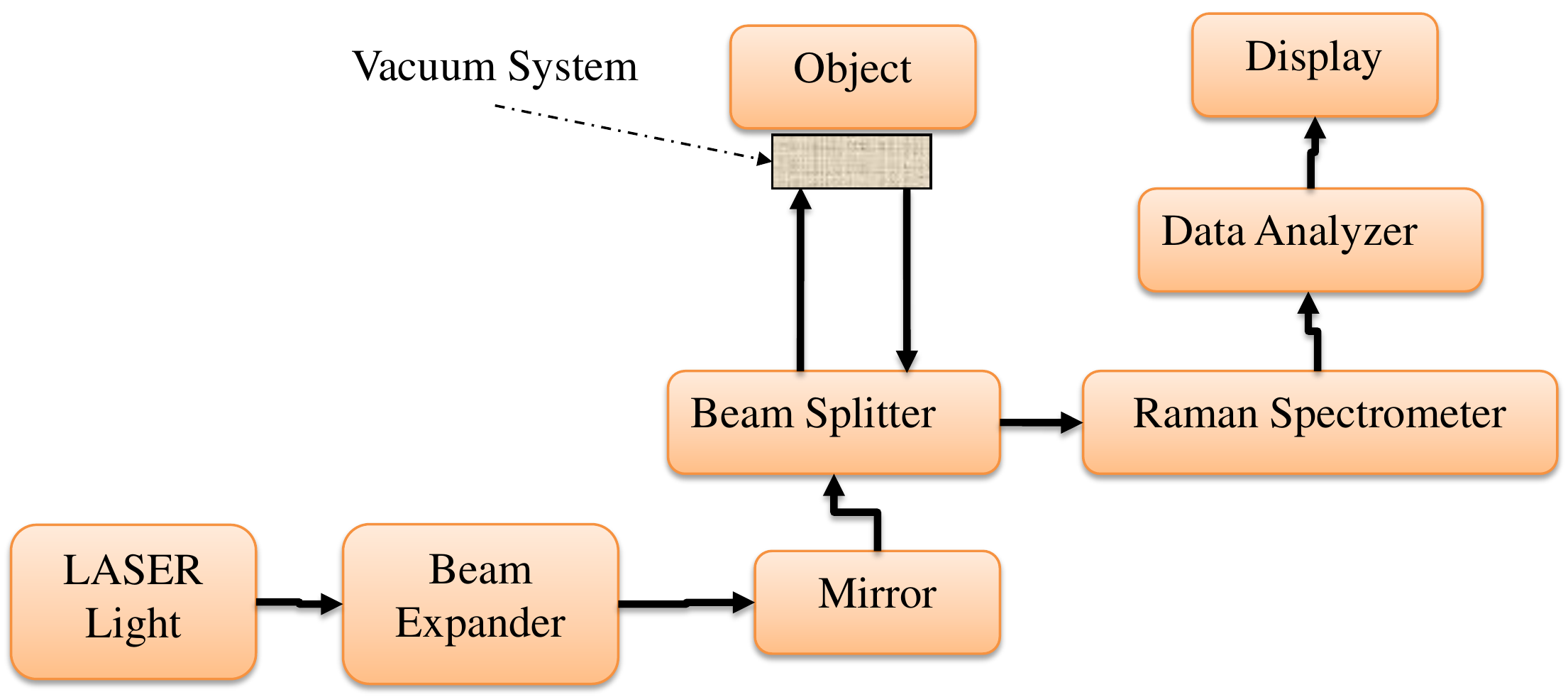}
		\caption{Building blocks of Raman spectroscopy.}
		\label{FIG:Raman_Spectroscopy_Basics}
	\end{center}
\end{figure}

\begin{figure}[htbp]
	\begin{center}
		\includegraphics[width=0.75\textwidth]{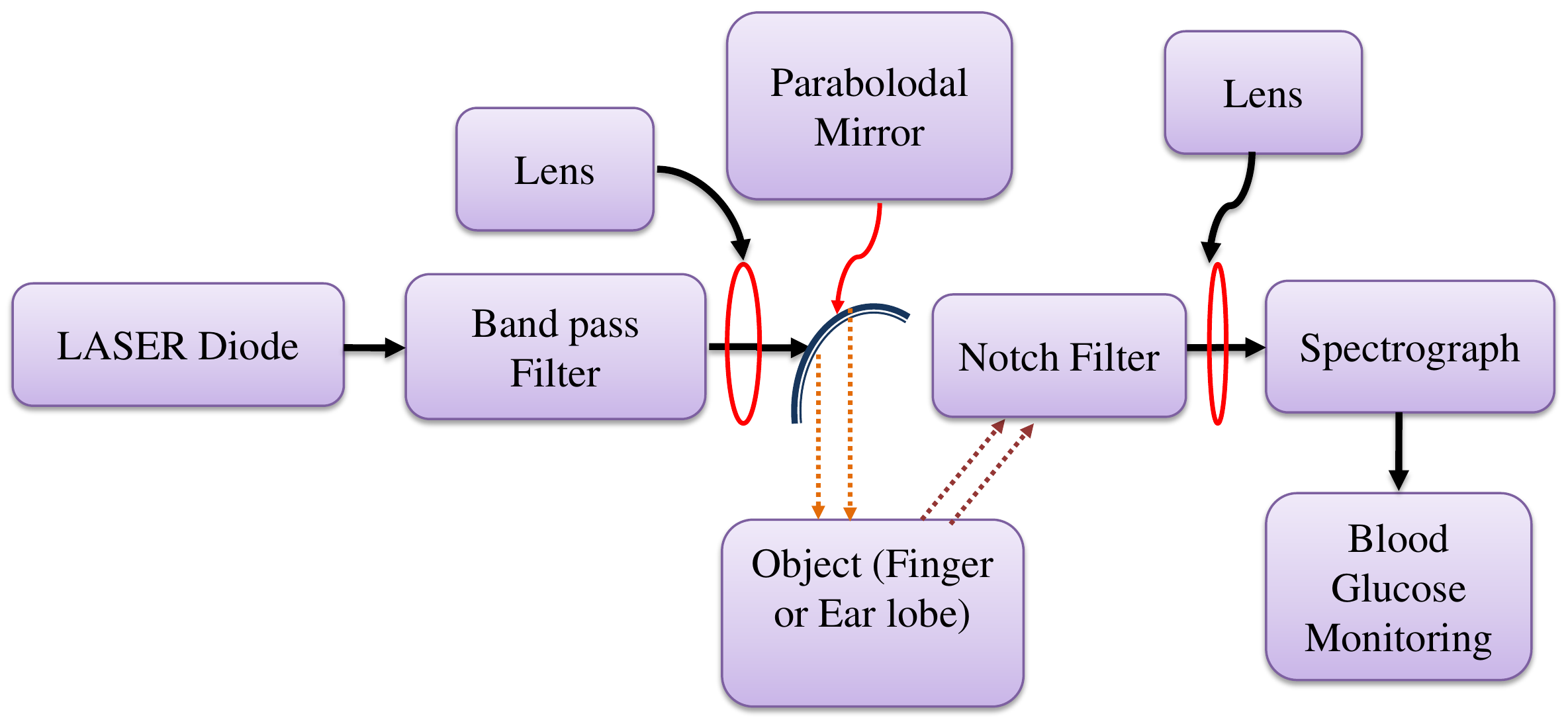}
		\caption{Noninvasive glucose measurement using Raman spectroscopy.}
		\label{FIG:Raman_Spectroscopy}
	\end{center}
\end{figure}

Due to the interaction of light with a glucose molecule, the polarization of the detected molecule will change \cite{Yoon1999}. In this technique, oscillation and rotation of molecules of the solution are possible through the incident of LASER light \cite{Yamakoshi2007}. The vibration of the molecule affects the emission of scattered light \cite{Yoon1998}. Due to this principle, blood glucose concentration can be predicted as \cite{Ishizawa2008}. This technique provides more accuracy with compared to infra-red spectroscopy technique \cite{Harada2007}.
There has been several research based on Raman spectroscopy to have precise glucose measurement. The validation has been also carried out on using vivo testing.
%It has been presented the development of existing Raman spectroscopy system and analyzed the algorithm for precise glucose sensing. The setup and algorithm were validated in a preclinical work in which a dog model has been taken.
Fig. \ref{FIG:Raman_Spectroscopy_Basics} presents basic framework of Raman spectroscopy, whereas 
Fig. \ref{FIG:Raman_Spectroscopy} presents its usage for noninvasive glucose measurement.

%Shih et al. explored the blood glucose measurement using Raman spectroscopy \cite{shih2015noninvasive}. The experimental setup for Raman spectroscopy required a large area and will not be portable. Raman spectroscopy technique is applicable for a laboratory test.

\subsection{Time of Flight and THz Domain}

The blood glucose estimation is adopted though Time of Flight (TOF) measurements for vitro testing \cite{JSTQE.2011.2175202}. The short pulse of laser light is inserted in the sample for photon migration measurement. This photon will experience scattering and absorption phenomenon while traveling from the sample. The optical analysis of the photons would be useful for precise glucose measurement.

\subsection{Photo Acoustic Spectroscopy}

\begin{figure}[htbp]
	\begin{center}
		\includegraphics[width=0.80\textwidth]{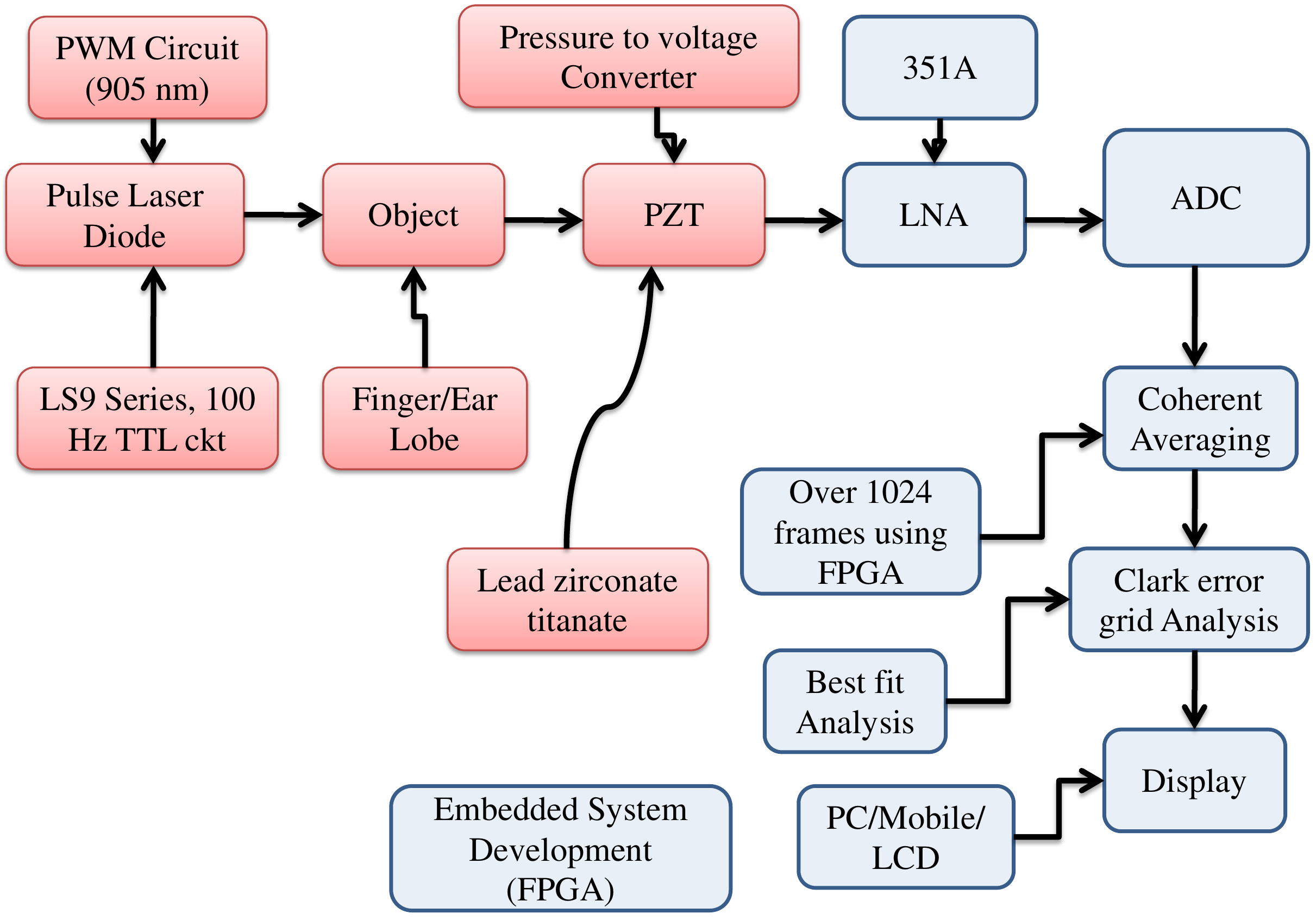}
		\caption{Photo acoustic spectroscopy.}
		\label{FIG:Photo-Acoustic_Spectroscopy}
	\end{center}
\end{figure}

Photoacoustic spectroscopy refers to the photoacoustic effect for the generation of the acoustic pressure wave from an object (refer Fig. \ref{FIG:Photo-Acoustic_Spectroscopy}) \cite{Sim2016}. In this spectroscopy technique, the absorption of modulated optical input provides the estimation of blood glucose detection \cite{Pai2015}. High intense optical light is absorbed by an object according to its optical conditions \cite{Pai2015b}. This process provided excitations of particular molecules according to its resonant frequency \cite{Tanaka2015}. The absorbed light is considered as heat which provides rising in localized temperature and thermal expansion of the sample \cite{Xiaoli2015}. The expansion in volume generates pressure in acoustic form \cite{Naam2015}. The generated photoacoustic wave can be used to predict the glucose concentration through specific excited wavelengths which are resonant for the vibration of glucose molecules \cite{Camou2011a}. At the specific resonance frequency, the glucose molecule changes own characteristic. This change is in the acoustic waveform \cite{Wadamori2008}. In previous work, 905 nm wavelength optical light is used for excitation \cite{Koyama2010, Domachuk2008}. 

\subsection{Capacitance Spectroscopy}

\begin{figure}[htbp]
	\begin{center}
		\includegraphics[width=0.80\textwidth]{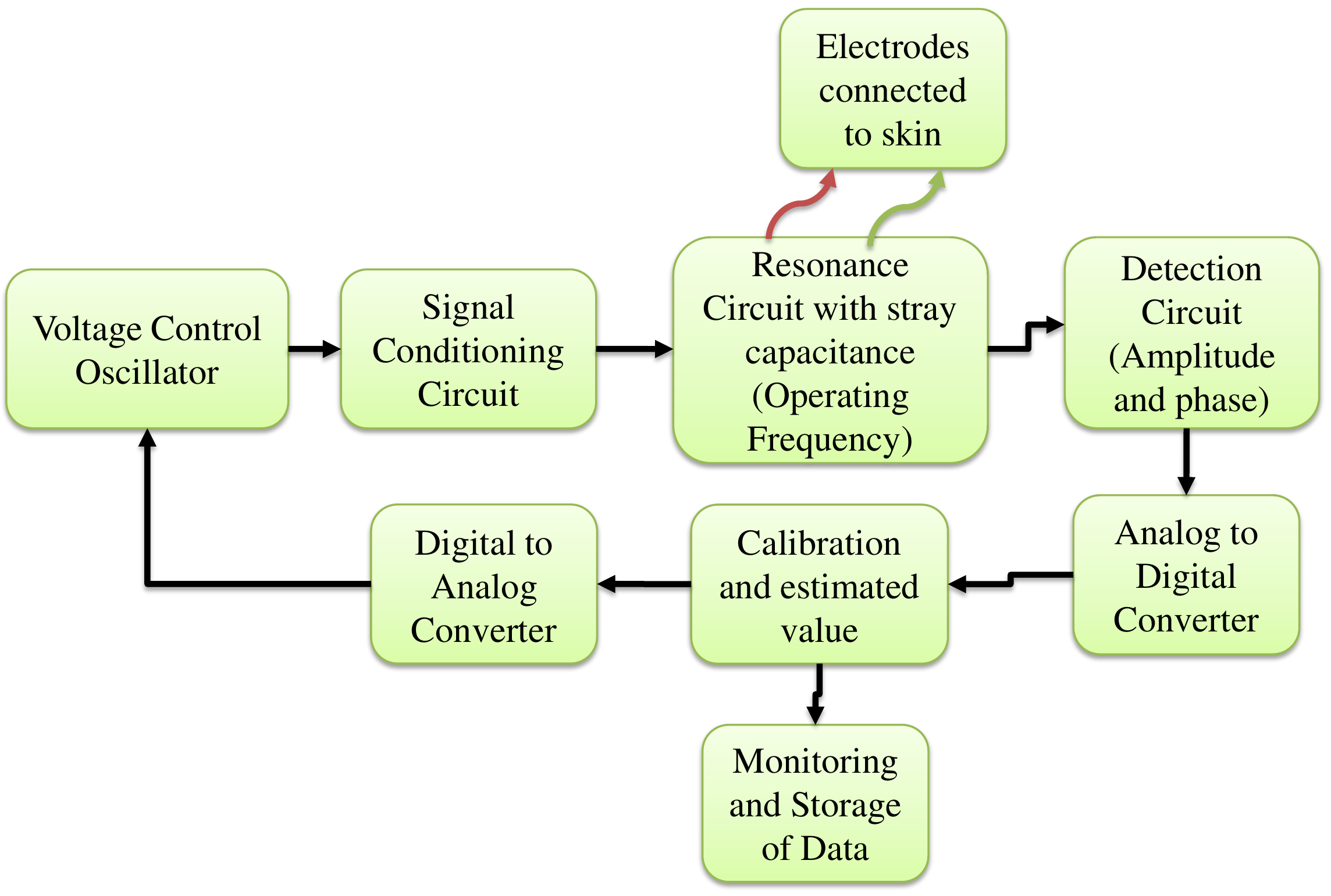}
		\caption{The typical steps of capacitance spectroscopy.}
		\label{FIG:Capacitive_Spectroscopy}
	\end{center}
\end{figure}

In the capacitance spectroscopy technique, inductor stray capacitance varies according to body capacitance (Fig. \ref{FIG:Capacitive_Spectroscopy}) \cite{Periyasamy2016}. The body capacitance is used to estimate body glucose concentration \cite{Yilmaz2014}. Flexible inductor based sensor follows the coupling capacitance principle for body glucose detection. In this technique, there is not any interaction between the inductive sensor and body skin through the current \cite{Gourzi2003}. This is the advantage of the impedance spectroscopy technique. The stray capacitance of the inductive sensor will vary according to body glucose. In this technique, the effect of fat and muscles will be negligible with respect to body glucose \cite{Turgul2015}.                                 

\subsection{Surface Plasmon Resonance (SPR)}

The Surface Plasmon Resonance (SPR) utilizes electron oscillation approach at dielectric and metal interface for glucose sensing \cite{LI201558}. It detects mainly the change in refractive index before as well as after the analytes interaction. The optical fiber based SPR has been used for point of care measurement for glucose due to its portability.

\subsection{Radio Frequency (RF) Technique and Microwave Sensing}

\begin{figure}[htbp]
	\begin{center}
		\includegraphics[width=0.70\textwidth]{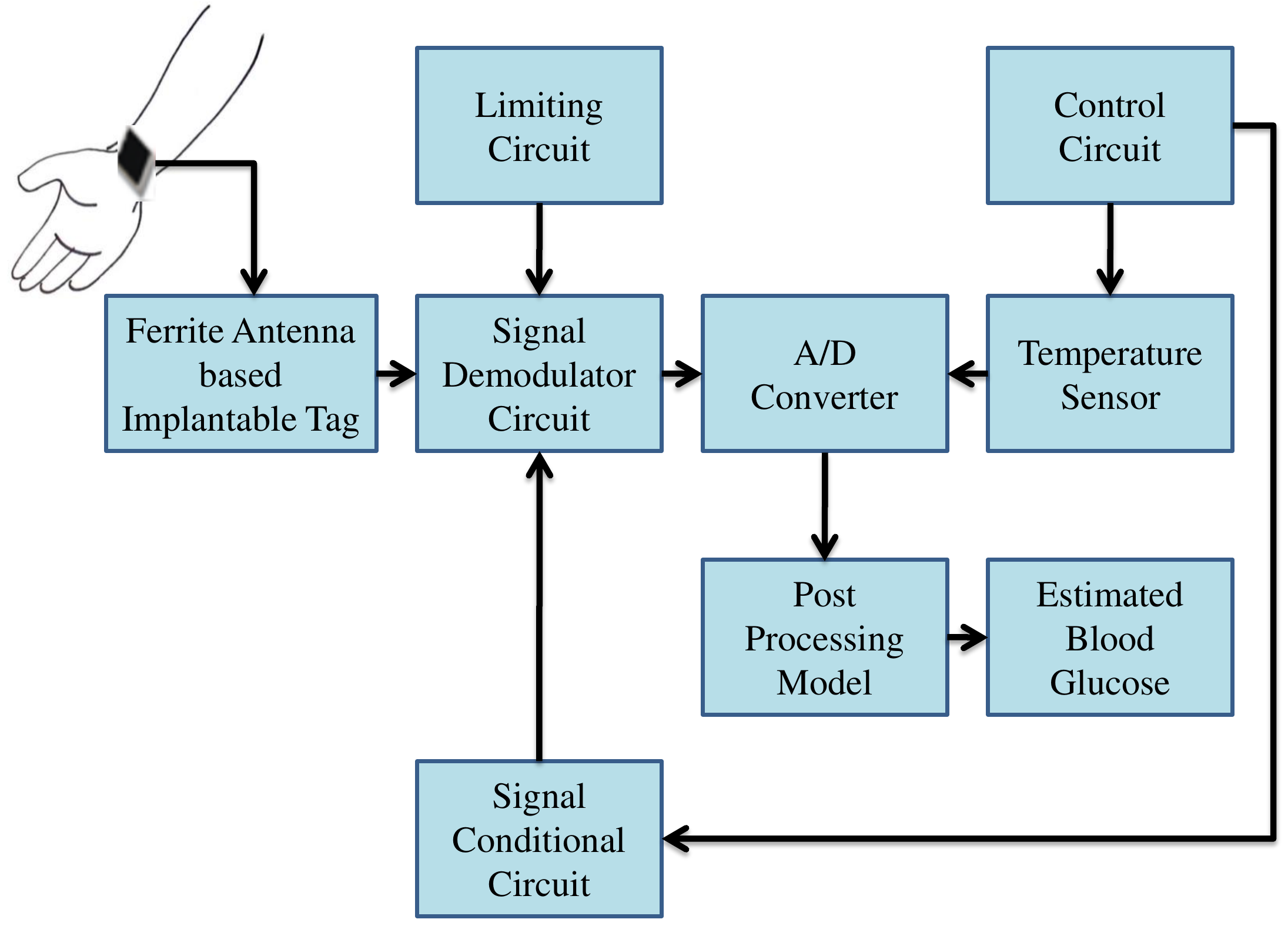}
		\caption{Glucose measurement using RF sensing technique.}
		\label{FIG:RF-Based_Measurement_Technique}
	\end{center}
\end{figure}

In the RF technique, the variation in the s-parameters response reflects the change in blood glucose \cite{Kaul2016, Turgul2017}. Fig. \ref{FIG:RF-Based_Measurement_Technique} shows typical steps of this technique. The response is determined through the antenna or resonator \cite{Cano-Garcia2016, Saha2016}. They follow the changes in dielectric constant value through the transmission \cite{Turgul2016}. The change in dielectric constant can be found as the change in resonance frequency spectrum through the antenna or resonator \cite{Ali2016, Choi2014}. The dielectric of blood varies according to blood glucose concentration. The human finger is an appropriate measurement object but there are many factors that play a cardinal and dominant role in the accuracy of measurement and repeatability. These are; the skin thickness, fingerprints, the applied pressure by the fingertip during measurement and positioning of a finger on the sensor \cite{Turgul2017a}.

\subsection{Ocular Spectroscopy}
In the Ocular Spectroscopy technique, glucose concentration is measured through the tears.
A specific lens is used to predict the body glucose concentration \cite{Miyauchi2011}. 
A hydrogel wafer is deposited to the lens. This wafer is prepared by boronic acid with 7 $\mu m$ thickness. The wafer is deposited on lens and then optical rays are inserted on the lens. 
%7 $\micro$m thickness. The wafer is deposited on lens and then optical rays are inserted on the lens. 
Then reflected light will change its wavelength. Change in wavelength will refer to a change in glucose concentration in tears.

\subsection{Iontophoresis}

In the Iontophoresis or Ionization technique, a small electric current passes through the skin diffusively. Three electrodes are used for the same \cite{Kim2004}. A small potential is applied through the electrodes to the different behaviour electrodes. During this process, glucose is transferred towards the cathode. The working electrode can have the bio-sensing function by the generation of current during applied potential through electrodes. This biosensor determines passively body glucose. The measurement is possible through wrist frequently \cite{Hofmann2011}.

\subsection{Optical Coherence Tomography}

The Optical Coherence Tomography technique is based on the principle followed by reflectance spectroscopy. In this technique, low coherent light is excited through the sample (sample is placed in an interferometer). In an interferometer, a moving mirror is placed in reference arc. A photodetector is placed on another side and it detects the interferometric signal. This signal contains backscattered and reflected light. Due to this process, we could get high-quality 2-D images. The glucose concentration increases with the increment of the refractive index in interstitial fluids. Change in the refractive index indicates the change in the scattering coefficient [107]. So, the scattering coefficient relates to glucose concentration indirectly.

\subsection{Polarimetry}

The Polarimetry technique is commonly used in a clinical laboratory with more accuracy. The optical linear polarization-based technique is used for glucose monitoring \cite{Mitsubayashi2014}. This technique is usually based on the rotation of vector due to thickness, temperature and concentration of blood glucose. Due to the process of prediction of glucose, the polarized light is transmitted through the medium containing glucose molecule. Due to high scattering through the skin, the depolarization of beam is possible. To overcome this drawback, a polarimetric test has been done through the eye. The light passes through the cornea. This technique is totally unaffected due to rotation of temperature and pH value of blood \cite{Cameron1996}.

\begin{figure}[htbp]
	\begin{center}
		\includegraphics[width=0.75\textwidth]{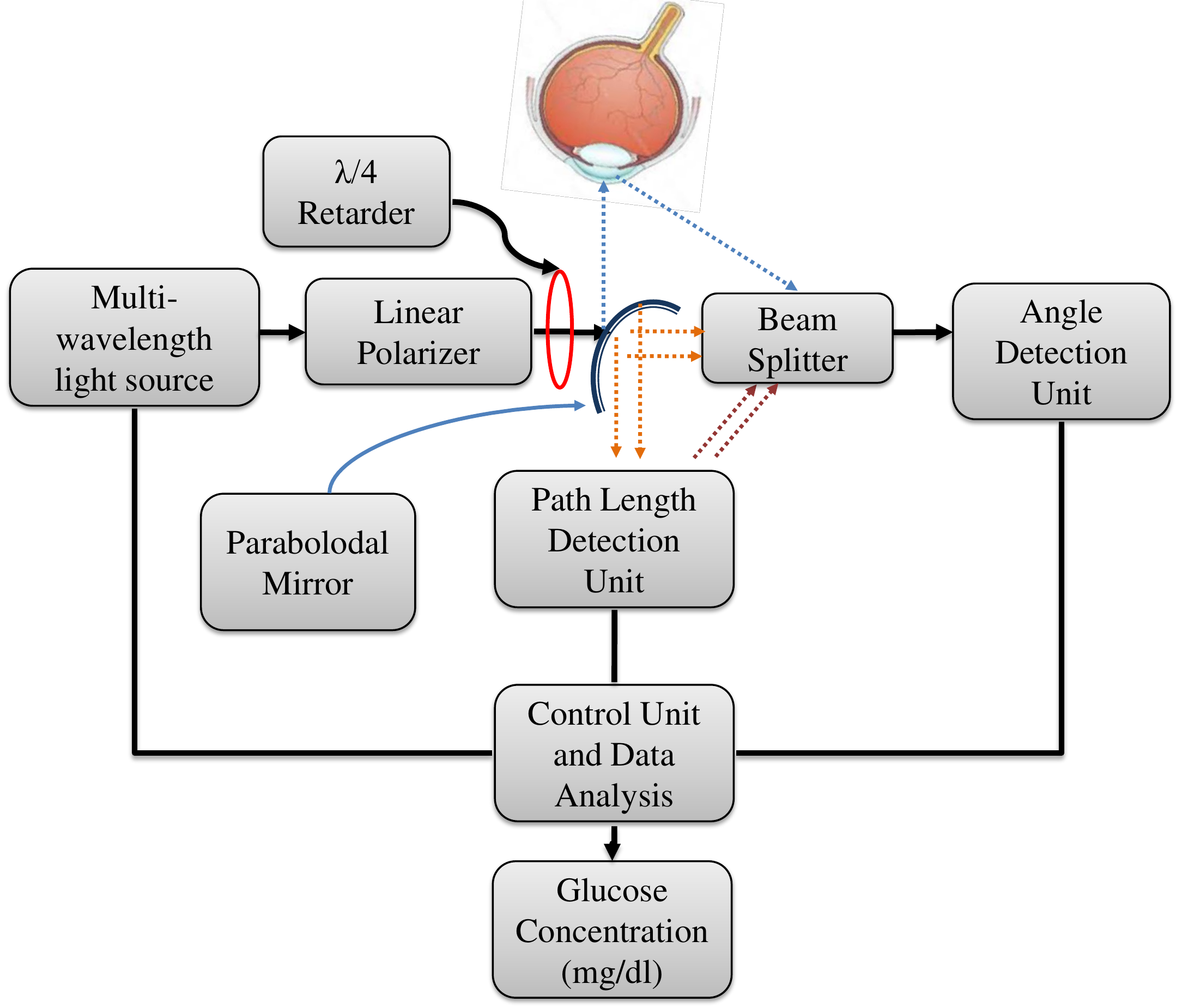}
		\caption{Non-invasive glucose measurement using Polarimetry.}
		\label{FIG:Polarimetry-Based_Measurement}
	\end{center}
\end{figure}

\subsection{Thermal Emission Spectroscopy}

The Thermal Emission Spectroscopy based technique is based on the naturally generated IR wave from the body. The emitted IR waves will vary according to body glucose concentration. 
The usual mid-IR emission from tympanic membrane of human body is modulated with tissue emitting. The selectivity  of this technique is same as the absorption spectroscopy. 
Due to this technique, glucose can be determined through the skin, fingers and earlobe. This technique is highly precise and accurate for glucose measurement \cite{Buford2008}.
It could provide the useful solution which is precise and  acceptable at clinical with measurement of thermal emission from tympanic
membrane.

\subsection{Ultrasound}

The Ultrasound method is based on low frequency components to extract the molecules from skin similar as reverse iontophoresis method \cite{j.ultrasmedbio.2005.04.004}. It is also alike sonophoresis and has larger skin permeability than reverse iontophoresis. Few or several tens of minutes of ultrasound exposure are required to pull glucose outward through the skin.
There are few attempts for such technology and there is not any commercial device with such type of technology.

%Low-frequency ultrasound enhances transdermal transport of molecules. Thus a technique has been utilized to extract the molecules across living skin as well as reverse iontophoresis. This process is also known as sonophoresis. Its effect on skin permeability is stronger than that of reverse iontophoresis.53 Although a clinically acceptable result for patients with diabetes53 and some encouraging results for animals have been demonstrated,54,55 no device for practical use with this technique has been developed so far. Advantages: Strong effect on skin permeability. Limitations: Interference from some biological compounds, temperature changes, and pressure variation. A few or several tens of minutes of ultrasound exposure are required to pull glucose outward through the skin.

\subsection{Metabolic Heat Conformation (MHC)}

The Metabolic Heat Conformation (MHC) method helps to measure the glucose value with metabolic heat and oxygen level along various physiological parameters considerations \cite{clinchem.2004.036954}. The mathematical model for metabolic energy conservation has been modified by several physiological parameters consideration such as pulse rate, oxyhemoglobin saturation, heat metabolic rate and the blood flow volume.  This method has shown good reproducibility and decent accuracy in humans.

\subsection{Fluorescence}

The Fluorescence technique is based on the excitation of blood vessels by UV rays at particular specified frequency ranges \cite{dia.2011.0041}. This is followed through the detection of fluorescence at a specified wavelength. The sensing of glucose using fluorescence through tear has been done by the diffraction of visible light. At 380 nm, an ultraviolet LASER was taken for excitation through the glucose solution medium. Fluorescence was estimated which is directly related to glucose concentration. In this technique, the signal is not affected by variation in light intensity through the environment.

\subsection{Kromoscopy}

The Kromoscopy technique uses the response from various spectroscopic of NIR light with four different detectors over different wavelength \cite{12.468318}. It employs the multi-channel approach with overlapped band-pass series filters to determine the glucose molecule. In this method, the radiation of IR are imparted on the sample and this will be divided among four detectors with band-pass filter. Each detector will detect the light of the similar structure of the tissue. Subsequently, the complex vector analysis has been utilised to measure the glucose concentration.

%Kromoscopy assesses the relative intensities of overlapping
%spectroscopic responses from four detectors recording spectra
%over different wavelengths of NIR light. Complex vector analysis
%is used to enhance differentiation between the target analyte
%and interferents.
%To date, urea and glucose have been successfully differentiated
%in vitro
%in binary solutions. It is a multi-channel, real time correlated method with a series of overlapped broad band-pass filters for the determination of selective quantification of analyte, such as glucose[127]. Selectivity of a four-channel kromoscopic signal is demonstrated by the resolution of glucose information collected over 800-1300nm NIR spectra [128]. In this technique, IR radiations are passed through the sample and transmitted light evenly divided into four detectors having band pass filters as shown in figure 15. These four detectors are arranged in such a way that the light reaching each detector has examined the same structures in the tissue. To evaluate target analyte such as glucose from interferents, a complex vector analysis is used. In vitro glucose and urea is successfully differentiated in a binary mixture .

\subsection{Electromagnetic Sensing}

In the Electromagnetic Sensing method, the variations in blood sample conductivity is observed by change in blood glucose concentration \cite{zhang2019noninvasive}. The alternation of electric field would be measured by electromagnetic sensor whenever there will be change in blood glucose concentration. This method utilizes the dielectric parameter of the blood samples. The frequency range for electromagnetic sensing is in the range of 2.4 to 2.9 MHz. The glucose molecule has maximum sensitivity at particular optimal frequency for given temperature of the medium.

%A change in blood glucose concentration causes a
%variation in the blood’s dielectric parameters such
%as its conductivity. An electromagnetic sensor based
%on Eddy currents due to changes in the electromagnetic
%field can determine the concentration
%of glucose through the aforementioned variations42.
%The principle of the sensor is based on two coils,
%with a separation of 40 mm between them. The
%conductivity fluctuations as a result of the change
%in glucose concentration involve the variation of
%the coils’output signal. The technique consists on
%applying a signal with a frequency of approximately
%4 MHz to the primary coil, while the output signal is
%measured on the secondary coil.
%Similar to bioimpedance spectroscopy, this technology assesses dielectric parameters of blood. The difference between them is that an electric current is used in bioimpedance spectroscopy, while the electromagnetic coupling between two inductors is used in electromagnetic sensing.16,17 The sensor uses electric currents to detect variation of the dielectric parameters of the blood, which may be caused by glucose concentration changes.18 The frequency range used in this technique is 2.4–2.9 MHz. However, depending on the temperature of the investigated medium, there is an optimal frequency, where sensitivity to glucose changes reaches its maximum. 

\subsection{Bioimpedance Spectroscopy and Dielectric Spectroscopy}

It is useful to measure the variation of the blood glucose  with help of conductivity and permittivity from red blood cells membrane \cite{Bertemes-Filho_ICEB_2016}. The spectrum of bioimpedance spectrum is measured from 0.1 to 100 MHz frequency range. 
It help to find the resistance
with passing through electric current which is flowing from human biological tissue. 
The change of plasma glucose would allow the changes in potassium and sodium to have the change in conductivity of the membrane of the red blood cell.  The multisensor approach is usually incorporated with this spectroscopy in order to measure sweat, moisture, movement and temperature for precise glucose measurement.

\subsection{Reverse Ionospheresis}

The small DC current is passed from anode to  cathode on the skin surface to have small interstitial fluid (ISF). 
Iontophoresis is employed for ionized molecules penetration at skin surface by such low current \cite{10.1002/dmrr.210}. The electric potential is passed from anode and cathode to
electroosmotic flow across the skin. This would allow to extract the molecules through skin whereas the the  molecules of glucose are moved towards the cathode. The enzyme method helps to sense the concentration of glucose molecules through oxidation process.  the method has very widely accepted and has good potential to measure accurate glucose value.

\subsection{Sonophoresis}

The Sonophoresis technique is based on the cutaneous
permittivity of the interstitial fluid (ISF) \cite{10.1211/jpp.61.06.0001}. 
It also uses enzyme method for glucose measurement. The low frequency ultrasound wave has been applied in order to have glucose molcules at the skin surface. The cutaneous permittivity of the ISF is increased to enable glucose at the epidermis surface. 
The contraction and expansion occurs in stratum corneum that subsequently opens the ISF pathway. There has been some attempts with this method for glucose detection but it has been observed that it could be helpful in drug delivery in stead of glucose measurement.

\subsection{Occlusion Spectroscopy}

The Occlusion spectroscopy based methods depend on the concept of light scattering which is of inverse proportion of glucose concentration \cite{10.1177/193229680700100403}. The flow is ceased for few seconds by applying pressure with pneumatic cuff. The volume of blood would change due to pulse generated from the pressure excursion. The light is transmitted through the sample and the variation of the intensity of in a received light defines the glucose concentration. 
The momentary blood flow cessation helps to get higher SNR value of the received signal.  Hence, the sensitivity for glucose detection would be increases with good robustness for accurate glucose measurement.

\subsection{Skin Suction Bluster Technique}

The Skin Suction Bluster technique uses the concept of blister generation through vacuum suction over limited skin area \cite{10.3109/00365519509104982}. The glucose measurement is performed on fluid which is collected from the blister. It has lower glucose molecules than plasma but it is well enough to have the glucose measurement. This method has low risk of infection, painless and well-tolerated.  It is actually useful to measure HbA1c value which represents three month average glucose value.

\subsection{Multimodal approach based measurement}

A two modal spectroscopy combining IMPS and mNIR spectroscopy is explored for high-level reproducibility of non-invasive blood glucose measurement \cite{Kossowski2017}. These two techniques are combined to overcome the limitation of individual employed technique \cite{Ficorella2015}. Impedance spectroscopy based circuit measures the dielectric constant value of skin or tissue through RLC resonant frequency and impedance to predict glucose level \cite{Song2014}. To improve the accuracy of NIR spectroscopy, mNIR spectroscopy technique is used. In this technique, three wavelengths 850 nm, 950 nm and 1300 nm are used \cite{Litinskaia2017}. For precise and accurate measurement, IMPS and mNIR are joined by an ANN (Artificial Neural Network) through DSP processor \cite{Song2015}. Therefore multimodel approaches have been explored for precise glucose measurement in the literature  \cite{joo2020vivo,feng2019multi}.

%Multiple techniques are used concurrently to reduce noise from the signal \cite{Jain2016}. Optimized signal conditioning circuits are used for removing garbage part of the received signal for final estimated value. Somewhere, optimized differential amplifiers are designed for noise cancellation. These are useful for biomedical measurement applications \cite{PRATEEK2016}. These signal conditioning circuits are used as post-processing approaches for glucose measurement \cite{Jain2014, Jain2013}. This device explored high-level accuracy in clinical trials. Based on the Clarke error grid analysis, the measured values are in A zone. This device has USB connectivity, battery charge and alerts for various stages of glucose level.

\begin{figure}[htbp]
	\begin{center}
\includegraphics[width=0.65\textwidth]{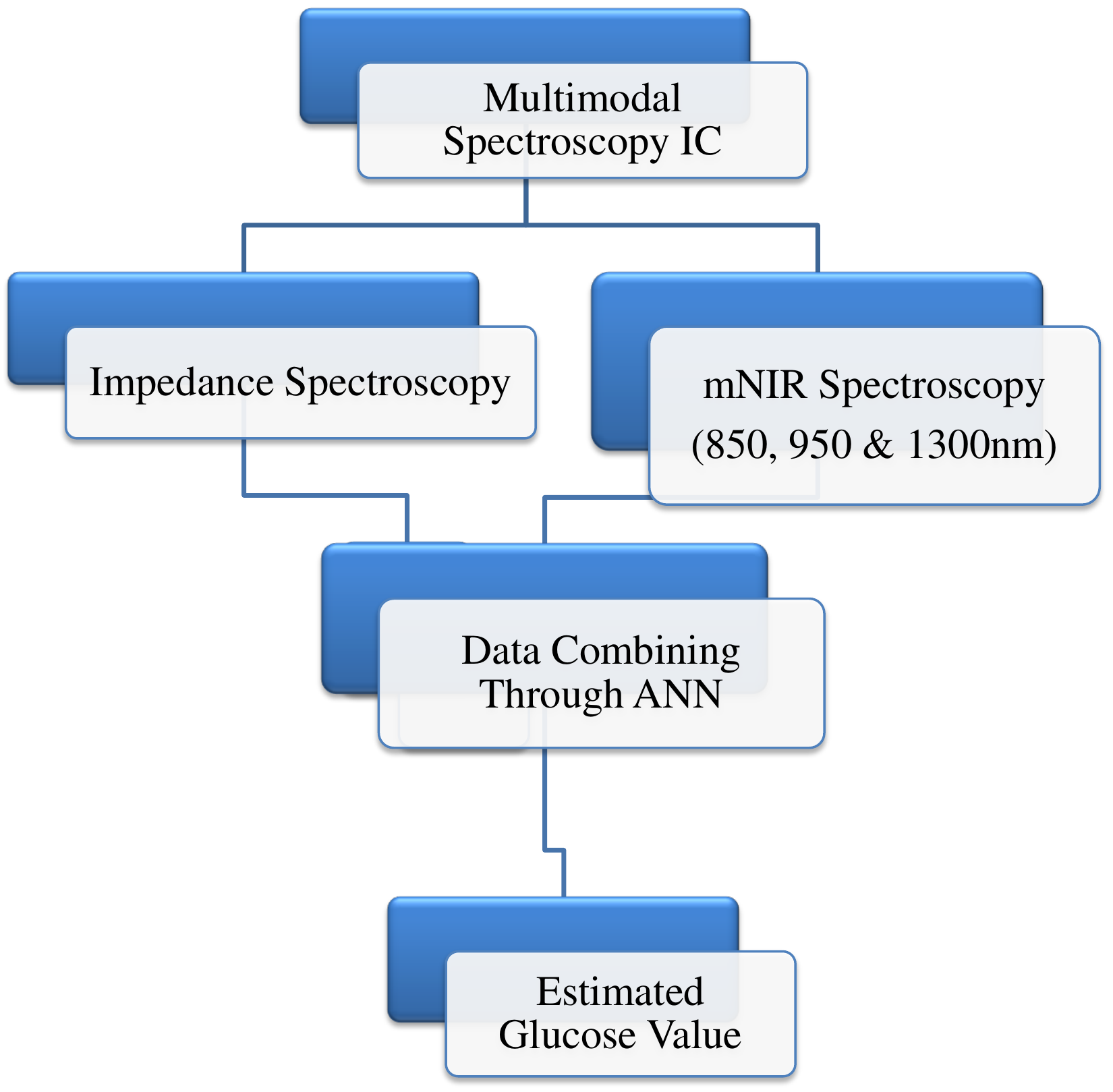}
		\caption{Multimodal IC based non invasive glucose measurement.}
		\label{FIG:Multimodal-IC-Based_Noninvasive_Glucose_Measurement}
	\end{center}
\end{figure}

\begin{table*}[htbp]
	\caption{Approaches Comparison with Noninvasive Works \cite{Joshi_TCE_iGLU2_TCE.2020.3011966, Jain_IEEE-MCE_2020-Jan_iGLU1}.}
	\label{table_example}
	\centering
	\begin{tabular}{p{3cm}ccccc}
		\hline
		\textbf{Works}&\textbf{Spectroscopy}&\textbf{Spectra}&\textbf{Specific}&\textbf{Measurement}&\textbf{Linearity}\\
		\textbf{}&\textbf{technique}&\textbf{}&\textbf{wavelength}&\textbf{range}&\textbf{(\%)}\\
		\hline		 \hline
		Singh, et al. \cite{8727488}&Optical&-&-&32-516 mg/dl&80\\
		\hline
		Song, et al.	\cite{Song2015}&Impedance and Reflectance&NIR&850-1300 nm&80-180 mg/dl&-\\
		\hline
		Pai, et al.	\cite{Pai2018}&Photoacoustic&NIR&905 nm&upto 500 mg/dl&-\\
		\hline
		Dai, et al.	\cite{Dai2009}&Bioimpedance&-&-&-&-\\
		\hline
		Beach, et al.	\cite{Beach2005}&Biosensing&-&-&-&-\\
		\hline
		Ali, et al.	\cite{Ali2017}&Transmittance and Refraction&NIR&650 nm&upto 450 mg/dl&-\\
		\hline
		Haxha, et al. \cite{haxha2016optical}&Transmission&NIR&940 nm&70-120 mg/dl&96\\
	\hline
		Jain, et al. \cite{jain2019iomt}&Absorption and Reflectance&NIR&940 nm&80-350 mg/dl&90\\
	\hline
Jain, et al. (iGLU 1.0) \cite{Jain_IEEE-MCE_2020-Jan_iGLU1, Jain_arXiv_2019-Nov30-1911-04471_iGLU1}&Absorption and Reflectance&NIR&940 and 1300 nm&80-420 mg/dl&95\\
		\hline
Jain et al. (iGLU 2.0) \cite{Mohanty_arXiv_2020-Jan-28-2001-09182_iGLU2, Joshi_TCE_iGLU2_TCE.2020.3011966} & Absorption and Reflectance &NIR& 940 and 1300 nm & 80-420 mg/dl & 97\\
		\hline
	\end{tabular}
\end{table*}

\begin{table*}[htbp]
	\caption{Statistical and Parametrical Comparison with Noninvasive Works \cite{Joshi_TCE_iGLU2_TCE.2020.3011966, Jain_IEEE-MCE_2020-Jan_iGLU1}.}
	\label{table_example1}
	\centering
	\begin{tabular}{p{3cm}ccccccccc}
		\hline
		\textbf{Works}&\textbf{R}&\textbf{MARD}&\textbf{AvgE}&\textbf{MAD}&\textbf{RMSE}&\textbf{Samples}&\textbf{Used}&\textbf{Measurement}&\textbf{Device}\\
		\textbf{}&\textbf{value}&\textbf{(\%)}&\textbf{(\%)}&\textbf{(mg/dl)}&\textbf{(mg/dl)} &\textbf{(100\%)}&\textbf{model}&\textbf{sample}&\textbf{cost}\\
		\hline		 \hline
		Singh, et al. \cite{8727488}&0.80&-&-&-&-&A\&B&Human&Saliva&Cheaper\\
		\hline
		Song, et al.	\cite{Song2015}&-&8.3&19&-&-&A\&B&Human&Blood&Cheaper\\
		\hline
		Pai, et al.	\cite{Pai2018}&-&7.01&-&5.23&7.64&A\&B&in-vitro&Blood&Costly\\
		\hline
		Dai, et al.	\cite{Dai2009}&-&5.99&5.58&-&-&-&in-vivo&Blood&Cheaper\\
		\hline
		Beach, et al.	\cite{Beach2005}&-&-&7.33&-&-&-&in-vitro&Solution&-\\
		\hline
		Ali, et al.	\cite{Ali2017}&-&8.0&-&-&-&A\&B&Human&Blood&Cheaper\\
		\hline
		Haxha, et al. \cite{haxha2016optical}&0.96&-&-&-&33.49&A\&B&Human&Blood&Cheaper\\
		\hline
		Jain, et al. \cite{jain2019iomt}&0.90&5.20&5.14&5.82&7.5&A\&B&Human&Blood&Cheaper\\
		\hline
		Jain, et al. (iGLU 1.0) \cite{Jain_IEEE-MCE_2020-Jan_iGLU1}&0.95&6.65&7.30&12.67&21.95&A\&B&Human&Blood&Cheaper\\
		\hline
Jain et al. (iGLU 2.0) \cite{Mohanty_arXiv_2020-Jan-28-2001-09182_iGLU2} & 0.97 &\ 4.86 & 4.88 & 9.42 & 13.57 & Zone A & Human & Serum & Cheaper\\
		\hline
	\end{tabular}
\end{table*}

%%%%%%%%%%%%%%%%%%%%%%%%%%%%%%%%%%%%%%%%%%%%%%%
%\section{Post processing and calibration techniques for non-invasive measurement}
%\label{sec4}
%%%%%%%%%%%%%%%%%%%%%%%%%%%%%%%%%%%%%%%%%%%%%%%
\section{Post-processing and calibration techniques for non-invasive Glucose-Level measurement}
\label{SEC:Postprocessing_and_Calibration_Techniques}

This Section presents various post-processing and calibration techniques which are deployed in various systems or frameworks for noninvasive glucose-level monitoring.

\subsection{Post processing and calibration techniques}

Various calibration processes have been applied for a high level of accuracy and noise reduction from received signal. These post-processing techniques are used to design the model for errorless continuous monitoring \cite{Malik2016,Yotha2016}.

\subsubsection{Noise Minimization and Signal Conditioning}
The coherent averaging technique has been adapted to minimize the variance of random noise \cite{Sarangi2014a}. The impact of noise is minimized with averaging of N number of individual samples coming from the continuous frames \cite{Naqvi2008}. Frames in the maximum count have been chosen for averaging to have SNR improvement \cite{Yamakoshi2009}. This proposed coherent averaging has been used frequently through MATLAB and coherent averaged signal acquired. Golay code has been proposed as calibration of measured data. The filtering or cancellation of unusual measured data has been achieved through the implementation of Golay code \cite{Ming2009,Sarangi2014,Heise1996}. 

\subsubsection{Computation Models for glucose Estimation}
The regression model of regularized least square is proposed by several researchers for measurement \cite{Pai2015a}. The estimated value is computed from photoacoustic signals. These photoacoustic signals are used to calibrate for estimation of glucose concentration \cite{Stemmann2010}. This can be possible through multi-variable linear regression model \cite{Rollins2010}. With the objective of high-level accuracy, a post-processing SVM technique is proposed \cite{Malik2015}. Support vector machine is a better option of correct measurement in glucose monitoring system \cite{Ogawa2007}. Artificial Neural Network (ANN) has also been proposed for data combining \cite{Dag2011}. The measured data from multiple techniques are combined through the proposed neural network model \cite{Olarte2011}. This artificial neural network has been implemented in DSP processor \cite{Savage2000, pancholi2018portable}. This proposed data interpretation model has been used for combining and calibrating of data for final estimated glucose concentration \cite{Baghbani2015}.

\subsection{Metrics for Model Validation}

\begin{figure}[htbp]
	\begin{center}
		\includegraphics[width=0.85\textwidth]{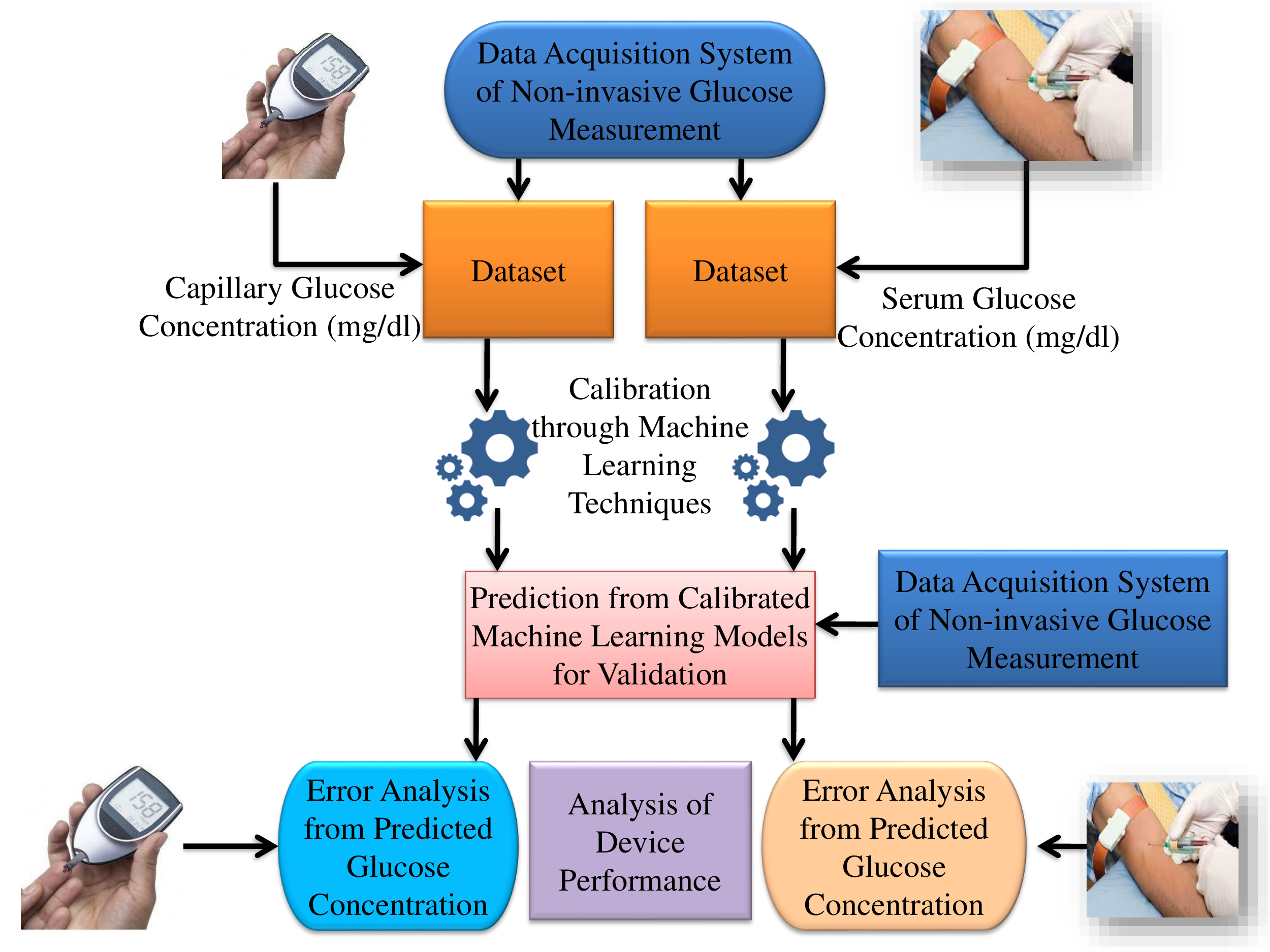}
		\caption{Representation of Metrics for model Validation.}
		\label{FIG:metrics}
	\end{center}
\end{figure}

The calibration method is used to have precise blood glucose estimation for measurements \cite{Parab2016}. The obtained glucose concentration values are used to compare with conventionally measured glucose concentrations \cite{Lekha2015a}. The Clarke error grid analysis  has been considered maximum measurement for  analysis which is used to check the performance of any device for accuracy measurement \cite{Olarte2012}. The process flow is represented in Fig. \ref{FIG:metrics}.

\subsection{Clinical Accuracy Evaluations using Clarke error grid analysis}

The Clarke Error Grid has been analysed as benchmark tool to examine the clinical precision for biomedical application. It has prediction of point as well as rate accuracy, and it amends for physiologic time lags inherent for measurement of body glucose. The exploitation of the Clarke error grid modelling will significantly make easy the development and refinement of a precise biomedical device. In 1970, this technique was developed by C.G. Clark  to identify the accuracy of  the clinical trials which helps to find the precision of estimated blood glucose with blood glucose value through the conventional method. A description of the Error Grid Analysis came into view in diabetes care in 1987. The grid is divided with five different zones mainly as zone A, zone B, zone C, zone D, and zone E. If the values residing in either zone A or zone B then it signifies satisfactory or accurate prediction of glucose results according to Beckman analyzer. The zone C values may prompt gratuitous corrections which may lead to a poor outcome. If the values are under zone D which actually defines a hazardous failure to sense. Zone E reflects the ``erroneous treatment'' \cite{Clarke2005}.

\begin{figure}[htbp]
	\begin{center}
		\includegraphics[width=0.65\textwidth]{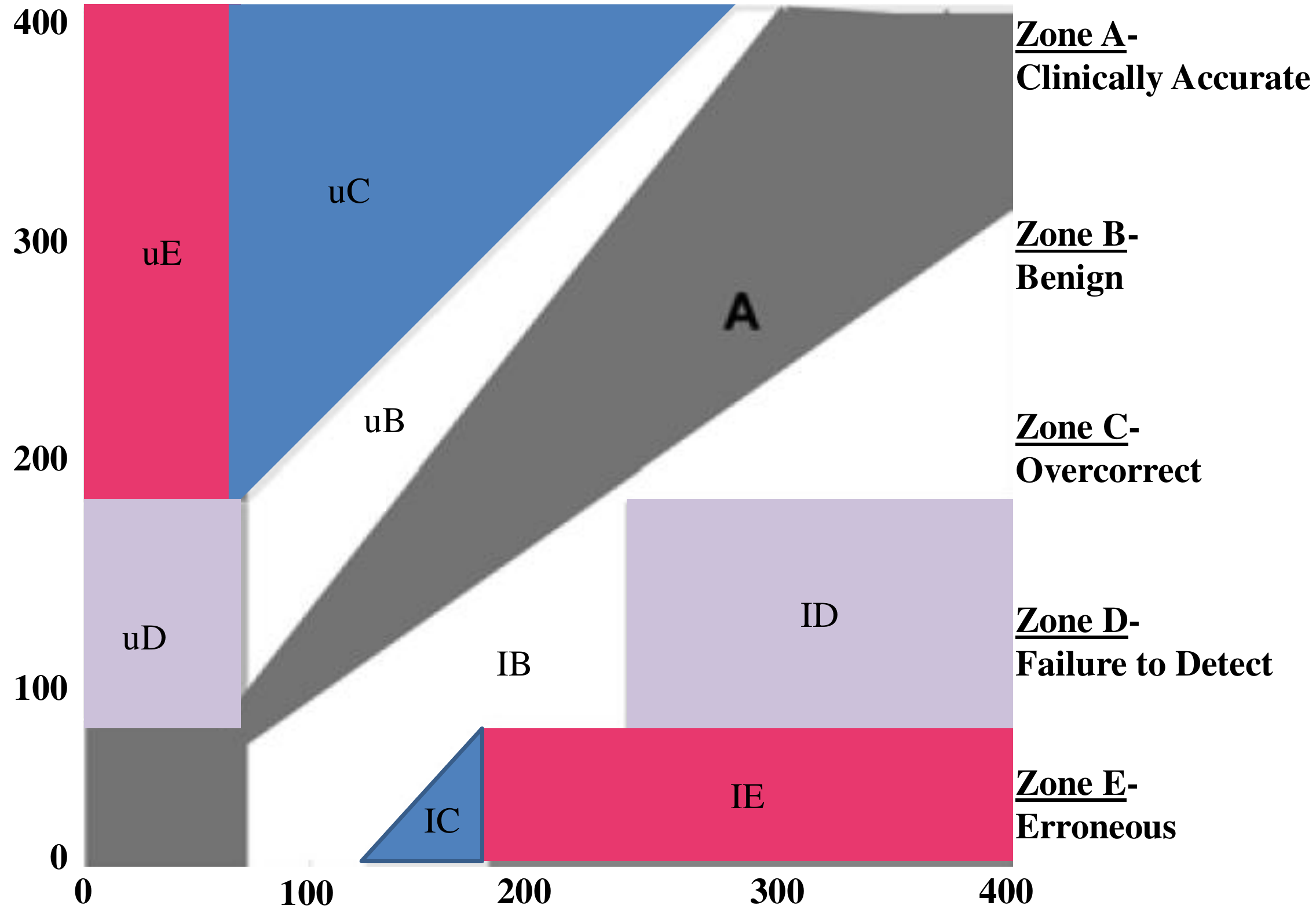}
		\caption{Clarke error grid analysis.}
		\label{FIG:Clarke_Error_Grid_Analysis}
	\end{center}
\end{figure}

%%%%%%%%%%%%%%%%%%%%%%%%%%%%%%%%%%%%%%%%%%%%%%%%%
%\section{Selected glucose measurement consumer products}
%%%%%%%%%%%%%%%%%%%%%%%%%%%%%%%%%%%%%%%%%%%%%%%%%
\section{Consumer Products for Glucose-Level Measurement}
\label{SEC:Consumer_Products_for_Measurement}

There have been several non-invasive glucometer at market(such as Freestyle Libre sensor, SugarBEAT from Nemaura medical) which used for proper medication. They would be like skin-patch with daily disposable feature and adhesive to have the continuous glucose monitoring. Most of the consumer products fail to provide precise glucose measurement and hence they are much popular for diabetes management.
There are some products as DiaMon Tech, glucowise, glucotrack, glutrac and CNOGA medical device. Glutrac is smart healthcare device but it has accuracy  issues for the blood glucose measurement. It has higher cost while precision is still not acceptable. The non-invasive stripless device known as Omelon B-2  has been used for the CGM. The fluorescent technique based Glucosense has been made for  contonuous monitoring of the glucose value. The flexible textile-based biosensor has developed from Texas University to measure the glucose level. All the available device have accuracy issues and considerable higher cost.

\subsection{Wearable versus Non-wearable for Glucose Monitoring}

The glucose monitoring have been attempted using  non-wearable and wearable solutions in the literature. Most of the non-wearable approaches are based on various spectroscopy such as photoacoustic spectroscopy, Raman spectroscopy  etc. The implantable devices are of semi-invasive type and are mainly of biosensors in nature. Sweat patches, Glucowatch and Smart contact lenses are of wearable devices category. LifePlus has developed non-invasive and wearable device for CGM purpose and it is under consideration for commercialization. Most of non-invasive device are wearable and helpful for frequent glucose measurement.
The continuous glucose monitoring would be more acceptable if they could measure the blood glucose values in day to day life. Therefore, the wearable devices are more state of art solutions then non wearable devices. 

\subsection{Noninvasive Glucose Measurement Consumer Products}

There are variety of products such as GlucoTrack\textsuperscript{\textregistered}, glucometer from Labiotech \cite{fernandez2018needle}, and similar available solutions have accuracy issues and cost is also high.  The glucowise is another non-invasive device for continuous glucose measurement from Medical Training Initiative (MTI) \textsuperscript{\texttrademark}. The Raman scattering spectroscopy based non-invasive solution is also developed by 2M Engineering \cite{schemmann2013blood}. These devices are not much popular because of their  cost and precision. Further for the high level of accuracy of glucose measurement, Glucotrack\textsuperscript{\texttrademark} has been developed by integrity applications Ltd.  \cite{Gal2011}. This non-invasive glucose monitoring device employed three consecutive ultrasonic spectroscopy, thermal emission and electromagnetic techniques. This device is highly precise and accurate because of a combination of three techniques \cite{Huber2007}. 
A comparative perspective of various consumer products for noninvasive glucose measurement has been summarized in Table \ref{TBL:Glucose_Measurement_Products}.

%\label{sec5}
%​\begin{figure}
%	\begin{center}
%		\includegraphics[width=1.0\textwidth]{tables-1.png}
%		\label{tables-1}
%	\end{center}
%\end{figure}
%​\begin{figure}
%	\begin{center}
%		\includegraphics[width=1.0\textwidth]{tables-2.png}
%		\label{tables-2}
%	\end{center}
%\end{figure}

%%%%%%%%%%%%%%%%%%%%%%%%%%%%%%%%%%%%%%%%%%%%%%%%%
\noindent
\begin{center}
	%
	%\tablefirsthead{First & Project Title  & Agency & Amount & Duration & Role \\
	%\hline}
	%
	\tablehead{
%\hline 
%\begin{longtable}{p{1.5cm}p{1.8cm}p{2.2cm}p{1.5cm}p{5cm}p{2cm}}
%\begin{table*}[htbp]
%\begin{table*}[htbp]
%	\caption{Various non invasive commercial glucose measurement devices}
%	\label{commercial}
	%\centering
%	\begin{tabular}{p{1.5cm}p{1.5cm}p{2.2cm}p{1cm}p{5cm}p{2cm}}
	\hline
		\textbf{Company}&\textbf{Device}&\textbf{Technology}&\textbf{Object}&\textbf{Summary}&\textbf{Snapshot}\\
		%\textbf{}&\textbf{}&\textbf{employed}&\textbf{}&\textbf{}&\textbf{}\\
		\hline		 \hline
}
\tabletail{\hline}
\tablecaption{A Comparative perspective of a selected consumer products for noninvasive glucose measurement.}
\label{TBL:Glucose_Measurement_Products}
\begin{supertabular}{p{1.5cm}p{1.8cm}p{2.2cm}p{1.5cm}p{5cm}p{2cm}}
        Cygus Inc. (USA) & 	GlucoWatch G2 Biographer  & Reverse iontophoresis & Wrist skin & It would be worn as watch is used with disposable component, autosensor which is to be attached at back of biographer that contact with the skin to provide frequent glucose monitoring &
        \raisebox{-\totalheight}{\includegraphics[height=0.6in,keepaspectratio]{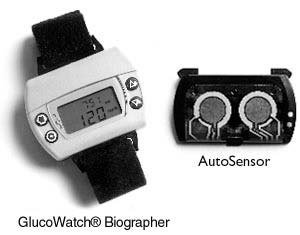}}\\
		\hline	
		CNOGA (Israel) &Combo glucometer&Tissue photography analysis&Finger&On basis of tissue photography analysis from fingertip capillaries, this device can analyze various bio parameters in very short time&	\raisebox{-\totalheight}{\includegraphics[height=0.6in,keepaspectratio]{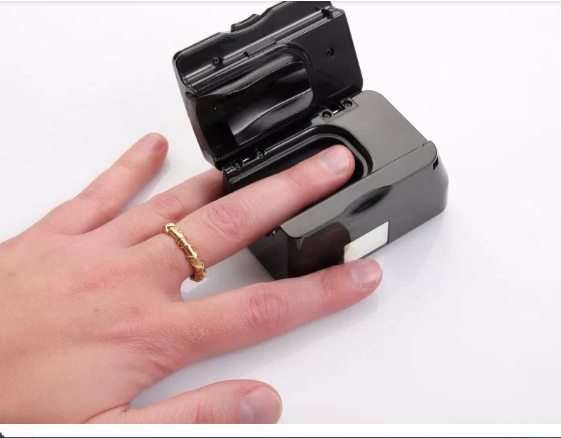}}\\
		\hline
		Pendragon Medical (Switzerland) & Pendra & Impedance Spectroscopy & Wrist Skin & It helps to measure the glucose with sodium transport of erythrocyte membrane,  The change of fluxes of transmembraneous sodium occur due to impedance field which is detected by device to generate final glucose value &
		\raisebox{-\totalheight}{\includegraphics[height=0.6in,keepaspectratio]{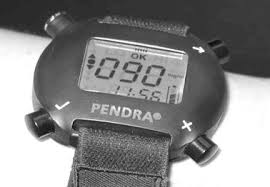}}\\
				\hline
			OrSense Ltd. (Israel) 	 &  OrSense NBM-200G & Occlusion Spectroscopy & Fingertip skin &  It is based on optical concept on
finger which is attached to a ring-shaped sensor probe. The probe has red/near-infrared
RNIR spectral region light source as well as detector. It has pneumatic cuffs which generates
systolic pressure to produce optical signal for glucose monitoring &
		\raisebox{-\totalheight}{\includegraphics[height=0.6in,keepaspectratio]{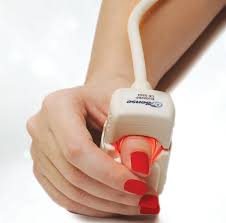}}\\
				\hline
				C8 Medisensors (USA) 	 &  C8 Medisensor  Glucose detector & Raman Spectroscopy & Fingertip skin &  This technique is based on monochromatic light source passes through skin where scattered light is detected. The generated colors from Raman spectra helps to exact chemical structure of glucose molecule &
		\raisebox{-\totalheight}{\includegraphics[height=0.6in,keepaspectratio]{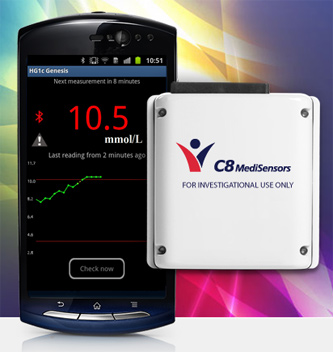}}\\
				\hline
						Integrity Applications (Israel) & Glucotrack & Combination of Electromagnetic, ultrasonic and Thermal
		& Ear lobe tissue & In this device, three different techniques are used concurrently to increase the accuracy and precision&	\raisebox{-\totalheight}{\includegraphics[height=0.6in,keepaspectratio]{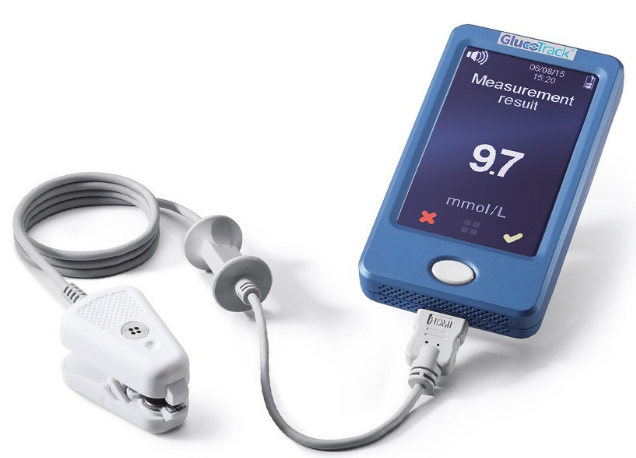}}\\
		\hline
		Tech4Life Enterprises (USA) & Non invasive glucometer & Infra red Spectroscopy & Finger & 
It is helpful for Hyperglycemia or Pre-Diabetic patients which allow for regular monitoring of precise blood glucose measurement at every 30 seconds		
		&	\raisebox{-\totalheight}{\includegraphics[height=0.75in,keepaspectratio]{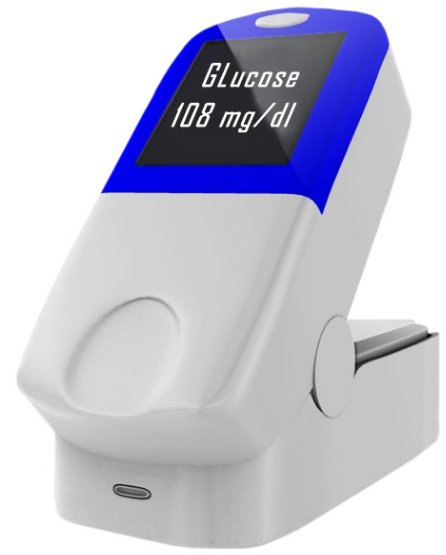}}\\
		\hline
		MediWise Ltd. (United Kingdom) & Glucowise & Radio Wave Spectroscopy & Forefinger skin/Earlobe & This non invasive wireless device can measure glucose concentration in very short time. It is based on electromagnetic waves of specific frequencies for blood glucose detection. It uses a thin-film layer of metamaterial which increases the penetration for precise glucose measurement	&	\raisebox{-\totalheight}{\includegraphics[height=0.5in,keepaspectratio]{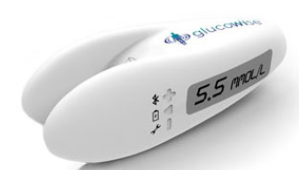}}\\
		\hline
		Nemaura Medical (united Kingdom) & SugarBeat & Reverse iontophoresis & Arm,Leg and adbomen & This has been proved accurate device, pain-free continuous blood glucose monitoring. 
		 SugarBEAT\textsuperscript{\textregistered} provides real-time, needle-free glucose measurement. Generally, it needs one time finger-prick test for calibration. One time finger prick is used when new patch is required to insert&\raisebox{-\totalheight}{\includegraphics[height=0.75in,keepaspectratio]{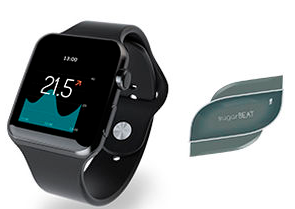}}\\
		\hline
		Abott Ltd. (USA)  & Free Style Libre & Glucose oxidase method & Fore-arm skin &
It uses enzyme glucose sensing technology for the detection of glucose levels through interstitial fluid		
		Glucose oxidase method is applied through sensor where electrical current proportional to the glucose concentration and glucose can be measured.&	\raisebox{-\totalheight}{\includegraphics[height=0.7in,keepaspectratio]{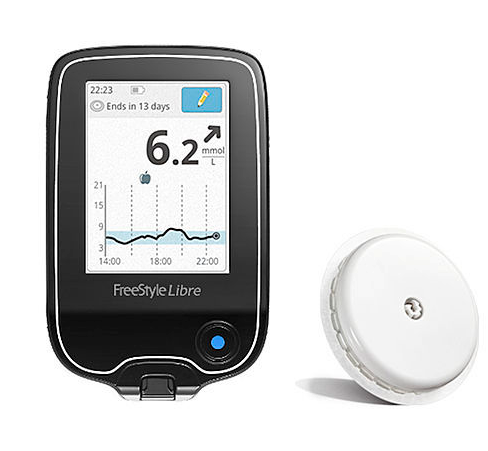}}\\
		\hline
		C8 Medisensor&Non invasive glucose monitor&Raman Spectroscopy&Fore arm skin&Raman spectroscopy technique based this device can detect glucose in blood through returning spectrum from the skin&	\raisebox{-\totalheight}{\includegraphics[height=0.6in,keepaspectratio]{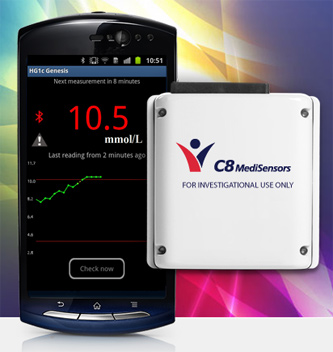}}\\
		\hline
%		BODYO&Non invasive glucometer&Optical Technology&Finger&NA&	\raisebox{-\totalheight}{\includegraphics[height=0.7in,keepaspectratio]{BODYO}}\\
%		\hline
%	\end{tabular}
%\end{longtable}
\end{supertabular}
\end{center}
%%%%%%%%%%%%%%%%%%%%%%%%%%%%%%%%%%%%%%%%%%%%%

%%%%%%%%%%%%%%%%%%%%%%%%%%%%%%%%%%%%%
%\section{Glucose Controls Approaches and Consumer Products}
%%%%%%%%%%%%%%%%%%%%%%%%%%%%%%%%%%%%%
\section{Glucose Level Controls Approaches and Consumer Products}
\label{SEC:Glucose_Controls_Approaches_and_Consumer-Products}

Various models have been developed for diet control using various parameters for glucose-insulin balance. The parameters are mainly includes net hepatic glucose balance, renal excretion rate, glucose absorption rate and peripheral glucose utilization for the glucose consumption prediction for the diabetic patients. These are useful parameters to calculate the  glucose level by proper insulin dosage along with scheduled diet plan. Therefore, the glucose-insulin control model was designed to balance glucose insulin level in the body for diabetes persons using proper medication.

\subsection{Glucose Controls Approaches}

The  mathematical models for insulin delivery have been presented to determine the coefficients of blood regulation. The model has been proposed for insulin secretion with glycemic profile for type 2 diabetic person \cite{bolie1961coefficients,bergman1981physiologic}.  The non-linear model is developed using differential equation with delay model with help of non-diabetic subjects \cite{de2000mathematical}. Most poplar “Uva/Padova Simulator” was also explored which was approved from FDA to have the proper clinical trials. The parameter are extracted with type 1 diabetic virtual patients \cite{cobelli2009diabetes}. The intravenous test for glucose tolerance with Hovorka maximal model has explored for non-diabetic subjects \cite{hovorka2002partitioning}. The samples from type 1 diabetic persons were collected to explain the model with help of time monitoring. The model is proposed mathematically for blood glucose value prediction in the postprandial mode for type 1 diabetes patients \cite{haidar2013stochastic,kirchsteiger2011estimating}. The mathematical model for glucose-insulin balance for longer period is explored using two days clinical information \cite{magdelaine2015long}. A  algorithm was developed for T1DM patient meal detection for the purpose of frequent glucose measurement. The work has integrated bolus meal mathematical model for glucose-insulin delivery model \cite{turksoy2015meal}. Diabetic and healthy people were considered  to acquire the values for the variable state dimension algorithm. The diet plan was examined in the absence of meal profile to have the glycemic profile balance, an intelligent PID controller (iPID) was developed to type 1 diabetic person \cite{xie2016variable,mohammadridha2017model}.

\subsection{Glucose Controls Consumer Products}

Type-1 diabetic patients aren't able to produce insulin. Insulin is a hormone that can balance body sugar (glucose) which is a prime source of energy that obtains from carbohydrates. If anybody has type 1 diabetes, it is necessary to be ready for insulin therapy. Insulin may be injected by self-injection, patients who take multiple injections daily of insulin may also think about use of an insulin pump. An insulin pump gives short-acting insulin all day long continuously. The insulin pump replaces the requirement of long-acting insulin. A pump also substitutes the requirement of multiple injections per day along with continuous insulin infusion and also serves to improve the glucose levels. Various types of insulin pumps are already available in the market as consumable product mainly as Animas, Medtronic, Roche, Tandem and Omnipod insulin pump are consumables. These insulin pumps are advanced to each other in terms of their upgraded features. A comparative perspective of a selected state of art approach for glucose measurement to have better glyncenic profile control is presented in Table \ref{TBL:Glucose_Control_Products}.

\noindent
\begin{center}
%\begin{center}
%\begin{longtable}{|m{0.25cm}|m{1.5cm}|m{0.5cm}||m{2.5cm}||m{2.5cm}|}
%\caption{State of Art research in Glucose Measurement} \label{tab:long} \\
%
%\hline \multicolumn{1}{|c|}{\textbf{Prior Work}} &
% \multicolumn{1}{c|}{\textbf{Technology}} & \multicolumn{1}{c|}{\textbf{Target}} &
%\multicolumn{1}{c|}{\textbf{Comments}}  & 
%\multicolumn{1}{c|}{\textbf{Results/Observations}}
%\\ \hline 
%\endfirsthead
%
%\multicolumn{5}{c}%
%{{\bfseries \tablename\ \thetable{} -- continued from previous page}} \\
%\hline \multicolumn{1}{|c|}{\textbf{Prior Work}} & \multicolumn{1}{c|}{\textbf{Technology}} & \multicolumn{1}{c|}{\textbf{Target}} &
%\multicolumn{1}{c|}{\textbf{Comments}}  & 
%\multicolumn{1}{c|}{\textbf{Results/Observations}} \\ \hline 
%\endhead
%
%\hline \multicolumn{5}{|r|}{{Continued on next page}} \\ \hline
%\endfoot
%
%\hline \hline
%\endlastfoot
%\begin{longtable}{p{0.5cm}p{2.5cm}p{2.0cm}p{4.5cm}p{5cm}}
%\begin{table*}[htbp]
%\caption{State of Art Research}
%	\label{Recent_reserach}
%	\centering
%	\begin{tabular}{p{0.5cm}p{3.5cm}p{2.0cm}p{3.5cm}p{3.5cm}}
%		\hline
\tablehead{\hline
	\textbf{Work} & \textbf{~Technology} & \textbf{Object} & \textbf{Findings} & \textbf{Observation} 
\\ \hline \hline}
\tabletail{\hline}
\tablecaption{A comparative perspective of a selected state of art approaches for glucose measurement}
\label{TBL:Glucose_Control_Products}
\begin{supertabular}{p{0.5cm}p{2.5cm}p{2.0cm}p{4.5cm}p{5cm}}
\cite{monte2011non} & photoplethy-\newline -smography (PPG)  & Finger & It helps to extract the features of PPG signal through machine
learning models to estimate Systolic and diastolic blood pressure and blood glucose values & machine learning models applied where random forest technique has best prediction results as $R^{2}_{SBP}$ = 0.91, $R^{2}_{DBP}$ = 0.89 and 
$R^{2}_{BGL}$ = 0.90. CEG has 87.7\% observation in Zone A, 10.3 \% in Zone B, and 1.9\% in Zone D \\
	\hline
\cite{kino2016hollow} & mid-infrared attenuated total reflection (ATR) spectroscopy and  trapezoidal multi-reflection ATR prism & oral mucosa inner lips & 
 Using a
multi-reflection prism brought about higher sensitivity, and the flat and
wide contact surface of the prism resulted in higher measurement
reproducibility \& spectra around 1155 cm$-1$
 for different blood glucose levels for fasting and before fasting &
 the coefficient of determination $R^2$
 is
0.75. The standard error is 12 mg/dl, and all the measured values are in
Region A\\
	\hline
\cite{chen2018noninvasive} & Optical Coherence Tomography &  Fingertip & It measures the optical rotation angle and depolarization index of aqueous glucose solutions with low and high scattering, respectively.  The value of angle increases while depolarization index decreases with glucose value increases & The correlation factor  has a value of $R^2$ 0.9101, the average deviation is found
around 0.027. \\
	\hline
\cite{park2018soft} & Contactlenses fluoresence & Tears & The
fabrication of a soft, smart contact lens in which glucose sensors, wireless power transfer circuits, and display pixels to
visualize sensing signals in real time are fully integrated using transparent and stretchable nanostructures & The usage of smart and soft lens would provide the wireless operation at real-time for glucose monitoring in tears  \\
	\hline
\cite{singh2019fabrication} &  transmission
spectroscopy &  Sliva &  After completely absorbing the sufficient amount of saliva on
the strip, the sample would reach detection zone via paper
microfluidic movement and enzymatic reaction between GOx
and salivary glucose would initiate a pH change, resulting in a
change in strip color that was recorded by using RGB detector
on the handheld instrument which helps for glucose detection & The developed biosensor had a wide detection range
of detection between 32- and 516-mg/dL glucose concentration
while the sensitivity of detection was 1.0 mg/dL/count at a limit
of detection (LOD) of 32 mg/dL within a response time of 15 s \\
	\hline
\cite{song2015impedance} & impedance
spectroscopy (IMPS) and multi-wavelength near-infrared spectroscopy (mNIRS)  & Left Handand wrist Hand & IMPS and mNIRS use the indirect dielectric
characteristics of the surrounding tissue around blood and the
optical scattering characteristics of glucose itself in blood,
respectively, the proposed IC can remove various systemic
noises to enhance the glucose level estimation accuracy&
mean absolute relative differences (mARD) to
8.3\% from 15.0\% of the IMPS and 15.0–20.0\% of the mNIRS
in the blood glucose level range of 80–180 mg/dL. From
the Clarke grid error (CGE) analysis, all of the measurement
results are clinically acceptable and 90\% of total samples can
be used for clinical treatment \\
	\hline
\cite{jain2019iglu} & NIR Spectroscopy & Fingertip & short NIR waves with absorption and reflectance of light using specific wavelengths (940 and 1,300 nm) has been introduced & The Pearson's correlation coefficient (R) is 0.953 and MAD is 09.89 which is RMSE 11.56 \\
	\hline
\cite{li2019absorption} & Microwave Detection & earlobe &  The absorption spectrum of microwave signal helps to measure using two antenna.  The sine wave of 500 MHz is for blood glucose measurement. & It can measure blood glucose from 0 to 500 mg/dl with step size of 200 mg/dl used for the experiment for testing the resolution. It obtained 0.5226 mean standard deviation while the minimum value of standard deviation is 0.04119.\\
\hline
\cite{zhang2019non}  &  PPG  & Finger & The  prediction of blood glucose was with machine-learning
using a smartphone camera. First the invalid data was separated  and The system did not require any type of calibration & The device was able to measure glucose only 70-130 mg/dl range. The results show accuracy of 
98.2\% for invalid single-period classification and
and the overall accuracy is 86.2\%. \\
\hline
\cite{wang2017wearable} &  MEMS  & Finger & 
It is minimally invasive technique known as e-Mosquito which extracts blood sample with shape memory alloy (SMA)-based
microactuator. It considered as 
first ever wearable device which performs the automatic situ blood extraction and performs the glucose analysis. & The method provided linear correlation ($R^2$ = 0.9733) between standard
measurements and the e-Mosquito prototype. \\
\hline
\cite{rachim2019wearable} & Visible NIR & Wrist  & The  paper developed biosensor which helps to exploit pulsation of arterial blood volume from the wrist tissue. The visible NIR spectroscopy was used for reflected optical signal to estimate blood glucose. &  The correlation coefficient (Rp) value after averaging all observation is 0.86, whereas the standard prediction error is around 6.16 mg/dl. \\
\hline
\cite{kino2016hollow} & mid-infrared attenuated total reflection (ATR) spectroscopy & inner lip
mucosa & Novel optical fiber probe was introduced  using multireflection
prism with ATR spectroscopy. The sensitivity increases with the number of reflections while
measurement reproducibility was higher due to prism's wide and flat \& wide contact surface. & The experimental results reveals the glucose signature at various spectra between fasting state and after the  glucose injection. The plot for calibration defines peak for absorption at 1155 $cm^{-1}$  which has glucose measurement error less than 20\%\\
%$cm^{−1}$ that originates from the pyranose ring structure of glucose gave measurement errors less than 20\%.  \\
\hline
\cite{chowdhury2016noninvasive} & modulated ultrasound and infrared technique &  Finger & the  MATLAB toolbox is used with Fast Fourier
Transform (FFT) for blood glucose extraction. The random blood glucose level test and oral glucose tolerance
test was done for the human subjects for performance measurement & The RMSE value of
noninvasive and invasive measurement from both tests
28.20 mg/dl and 23.76 mg/dl. The pearson correlation
coefficient was 0.85 and 0.76, respectively.
At the same time MSE was
17.76 mg/dl and 15.92 mg/dl.\\
\hline
%\cite{govind2020design} &  Microwave sensing &  Pig and goar samples were used & In this context, the design of a reusable microfluidic sensor for monitoring blood glucose at microwave
%frequencies is presented. The sensor is inspired by the metamaterial structure of complementary electric LC
%resonator. The central arm of the resonator is modified to form a cavity by carving in a groove deep into the
%substrate and then coating metal on the sidewalls for enhancing the capacitance. \\
\hline
%\end{tabular}
%\end{table*}
\end{supertabular}
\end{center}

%%%%%%%%%%%%%%%%%%%%%%%%%%%%%%%%%%%%%
%\section{Glucose-Level Measurement and Controls - An IoMT Perspective}
%%%%%%%%%%%%%%%%%%%%%%%%%%%%%%%%%%%%%
\section{Glucose-Level Measurement and Controls - IoMT Perspectives}
\label{SEC:IoMT_Perspectives}

The practical and sustainable mechanisms are the prime factors of smart and automated healthcare system. These are being optimized to support the population migration and quality of life in smart cities and smart villages \cite{Mohanty_CEM_2016-July, 9153927}. The features of smart healthcare system are continuous monitoring for critical care, ambient intelligence and quality of service for proper point of care mechanism \cite{9109415, 8977815}. The non-invasive and precise glucose measurement is requirement for diabetic person and would also needed to store the information using IoMT for proper treatment \cite{Mohanty_POT_2006-Mar}. The traditional method for glucose measurement has limited capability and is not able to assist the remotely located healthcare provider. The diabetic person would like to monitor their glycemic profile frequently in a day with support of storing at cloud server. The smart health care system would allow the point of care treatment for diabetes person with frequent monitoring.

The internet of Medical Things (IoMT) has allowed to connect the patients with doctors remotely for rapid treatment and special assistance using smart healthcare \cite{Mohanty_CEM_2016-July}. The continuous monitoring of vital parameters have provided to awareness about the diet plan and routine activity management with contemporary  healthcare consumers devices. With the active support of remote healthcare solution, the smart healthcare has potential to ameliorate the quality of service at reduced cost. The smart sensors would capture the patient data continuously and help to store the data on cloud data centre. It is also useful for the analysing the data and easy exchange of the information through mobile applications to doctors as well as patients. The healthcare Cyber-Physical System (H-CPS) has been used successfully to address the various challenges of healthcare sector with intelligent algorithms.

\begin{figure}[htbp]
	\centering
\includegraphics[width=0.75\textwidth]{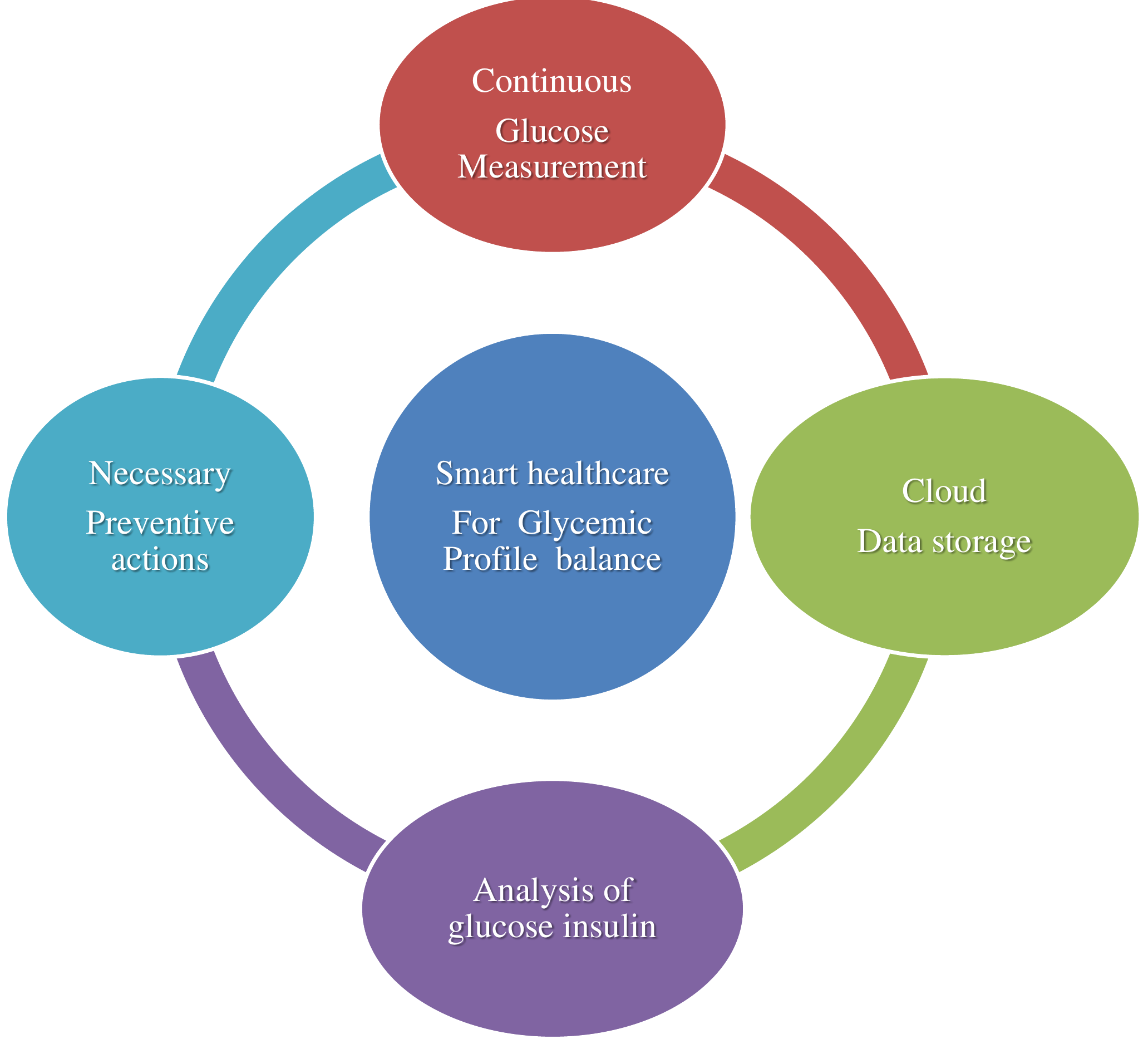}
	\caption{Blood glucose diagnosis and Control in smart health care system.}
	\label{FIG:IoMT_Glucose_Measurement_and_Control_Broad-View}
\end{figure}

The continuous glucose monitoring would certainly help the diabetic patients to plan their diet for the purpose of glucose control. The solution should be precise, low cost and easy to operate for rapid diagnosis \cite{Jain_IEEE-MCE_2020-Jan_iGLU1, Jain_arXiv_2019-Nov30-1911-04471_iGLU1}. The  serum glucose would always consider as accurate  than capillary measurement. Therefore, the rapid  serum glucose measurement solution with continuous monitoring is desired for the smart healthcare. The novel serum glucometer is portable device and is also integrated with IoMT to store the glucose values continuously at cloud. It would be useful for the healthcare provider to track the health records of remote located diabetes person. The smart healthcare management of continuous glucose measurement  is defined in Fig. \ref{FIG:IoMT_Glucose_Measurement_and_Control_Broad-View}.  

\begin{figure}[htbp]
	\centering
	\includegraphics[width=0.80\textwidth]{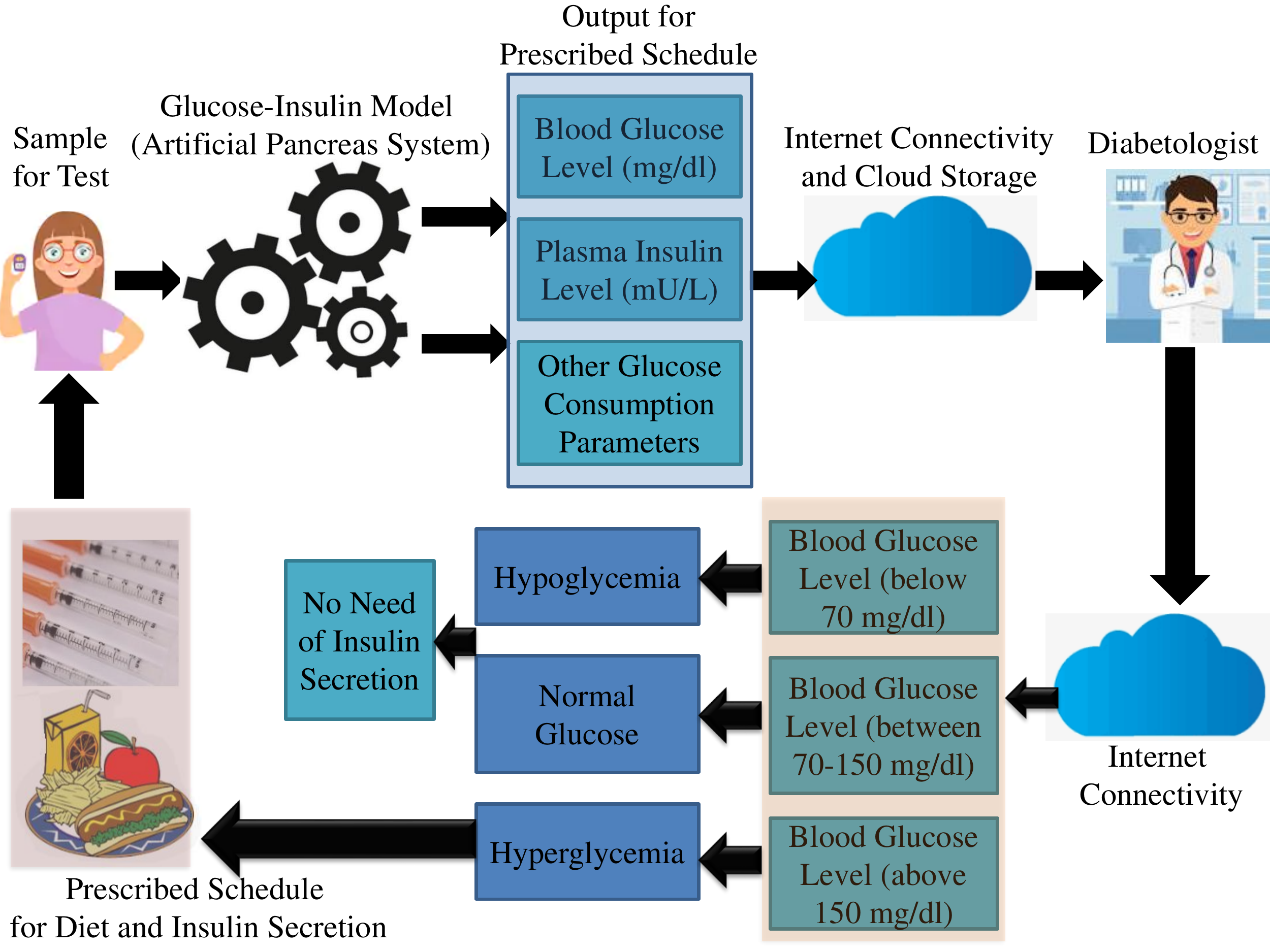}
\caption{A closed-loop automated insulin secretion diabetes control system in an IoMT framework \cite{jain2020iglu}.}
	\label{FIG:IoMT_Glucose_Measurement_and_Control_Detailed}
\end{figure}

A detailed example of a closed-loop system that presents glucose-level monitoring and insulin release to control it is illustrated in Fig. \ref{FIG:IoMT_Glucose_Measurement_and_Control_Detailed} \cite{jain2020iglu}.
This IoMT framework can provide a better solution for evaluation of insulin doses through
the closed-loop automated insulin secretion diabetes control. 
Such an integrated IoMT framework can be implemented to diagnose and
for the treatment of diabetic patients in terms of controlling
their blood glucose level in smart healthcare and be effective in smart village and smart cities for healthcare with limited medical personnel.

\begin{figure}[t]
	\centering
\includegraphics[width=0.80\textwidth]{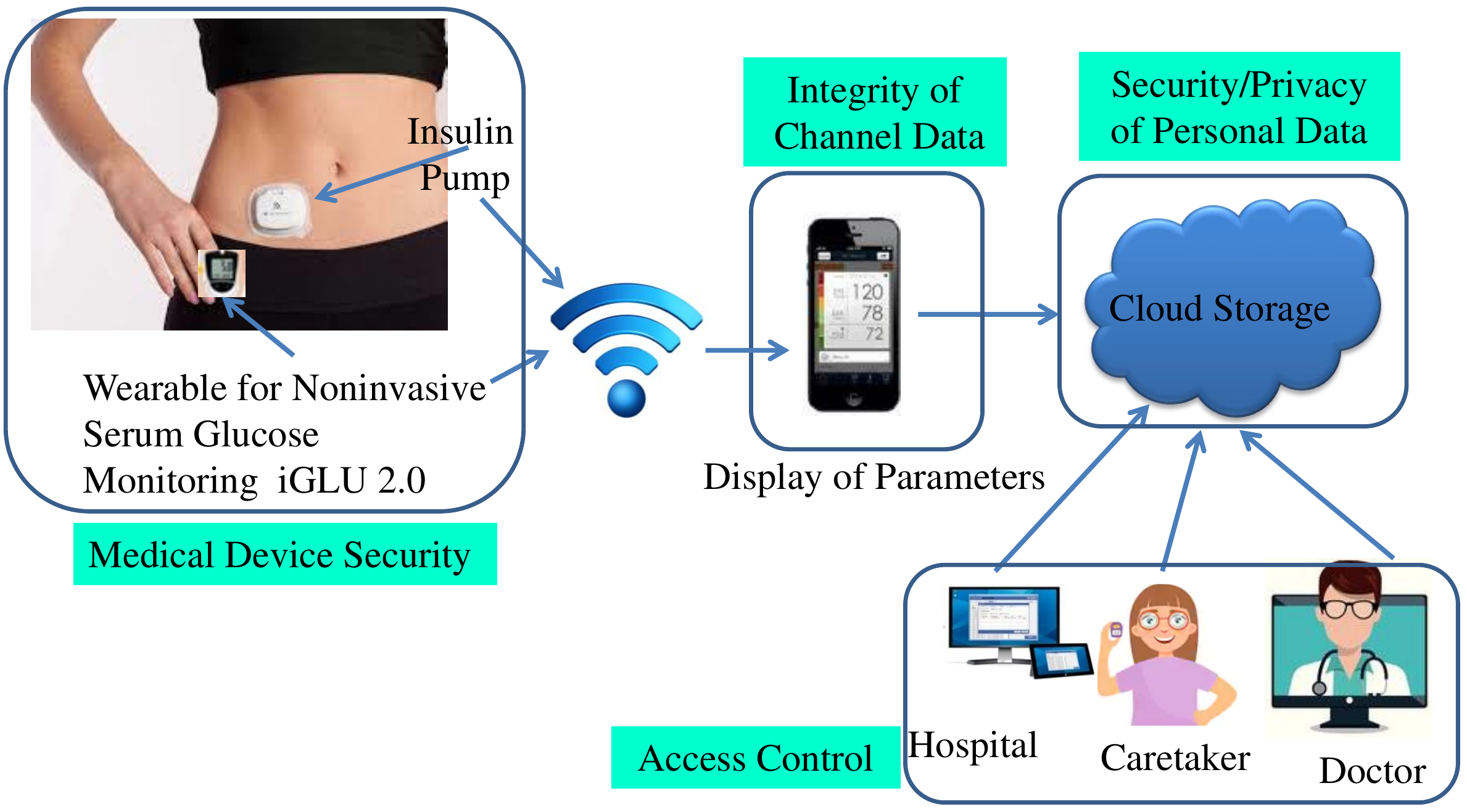}
	\caption{Our Long-Term Vision of Security-Assured Non-invasive Glucose-Level Measurement and Control through our Proposed iGLU.}
	\label{FIG:Secure-iGLU_Overview}
\end{figure}

The security and privacy issues of the medical devices are paramount aspect in any IoT network. 
The hardware security of wearable device is very crucial because control actions mainly occur in wireless media. The security vulnerabilities are defined for glucose measurement device and its control are shown in Fig. \ref{FIG:Secure-iGLU_Overview}. The devices security are important due to connected health system in an insecure and unreliable IoMT framework  \cite{joshi2020secure}. The integrity of useful  medical information is also crucial security aspect of smart healthcare. All patients medical records are stored over the server therefore the security of such data are also really important. The controlled access with proper authentication is required to have secure monitoring with proper patient treatment.

%%%%%%%%%%%%%%%%%%%%%%%%%%%%%%%%%%%%%
%\section{Short-Comings of Existing Works and Open Problems}
%%%%%%%%%%%%%%%%%%%%%%%%%%%%%%%%%%%%%
\section{Short-Comings of Existing Works and Open Problems}
\label{SEC:Shortcomings_and_Open-Problems}

This Section outlines the shortcomings and discusses some open problems of glucose level measurements and control.

\subsection{Limitations of the Existing Approaches and Products}

\begin{enumerate}
	
\item 
Photoacoustic spectroscopy has been implemented for glucose measurement. Real-time testing and validation have not been done from human blood. The artificial solution was prepared in the laboratory for glucose measurement. The prototype module with LASER and corresponding detector is costly and at the same time requires considerable bigger area and does not provide portable solution. Therefore, it is not much popular solution for continuous glucose monitoring.
	
\item 
Raman spectroscopy is a nonlinear scattering which occurs when monochromatic light interacts with a certain sample. Raman spectroscopy based solution is applicable for a laboratory test and also occupies the significant larger area. Hence, the system based on this approach will not be applicable for frequent glucose measurement.
	
\item 
The retina based glucose measurement is also one of the alternate non-invasive glucose detection approach, data has also been collected through retina for glucose measurement. Such technique is not useful for the glucose measurement all the time.
	
\item 
In case of bio-capacitance spectroscopy, the slight difference in placing the sensor at the same location might affect the output of the sensor. Effect of pressure on the sensor, body temperature and sweat on the skin may also affect the output of the sensor.
	
\item 
Glucose detection is performed with the impedance spectroscopy (IMPS) by electrodes connection to the skin which is affected with skin.
The accuracy is always an issue as the saliva and sweat could change for each individual and that may reflect to the precision of glucose.
	% The precise measurement may not be guaranteed by such methods as the saliva and sweat vary for each and individual. 
Therefore, this technique is not best for reliable glucose measurement in smart healthcare.
	
\item 
PPG signal has been used to extract features for blood glucose level prediction.  But the PPG may be precise blood glucose measurement technique where the output value would vary according the blood volume only.  
Therefore, the glucose molecule has not been detected precisely in the blood sample using this technique. 

%The optical technique has been observed as more accurate in comparison with PPG to measure the non-invasive blood glucose.

\end{enumerate}

%\begin{figure}
%	\begin{center}
%		\includegraphics[width=1.0\textwidth]{Capture14.png}
%		\caption{Graphical representation of Clarke error grid analysis \cite{Clarke2005a}}
%		\label{fig9}
%	\end{center}
%\end{figure}

\subsection{The Open Problems in Non-invasive Glucose Measurement}

\begin{figure}[htbp]
	\centering
	\includegraphics[width=0.55\textwidth]{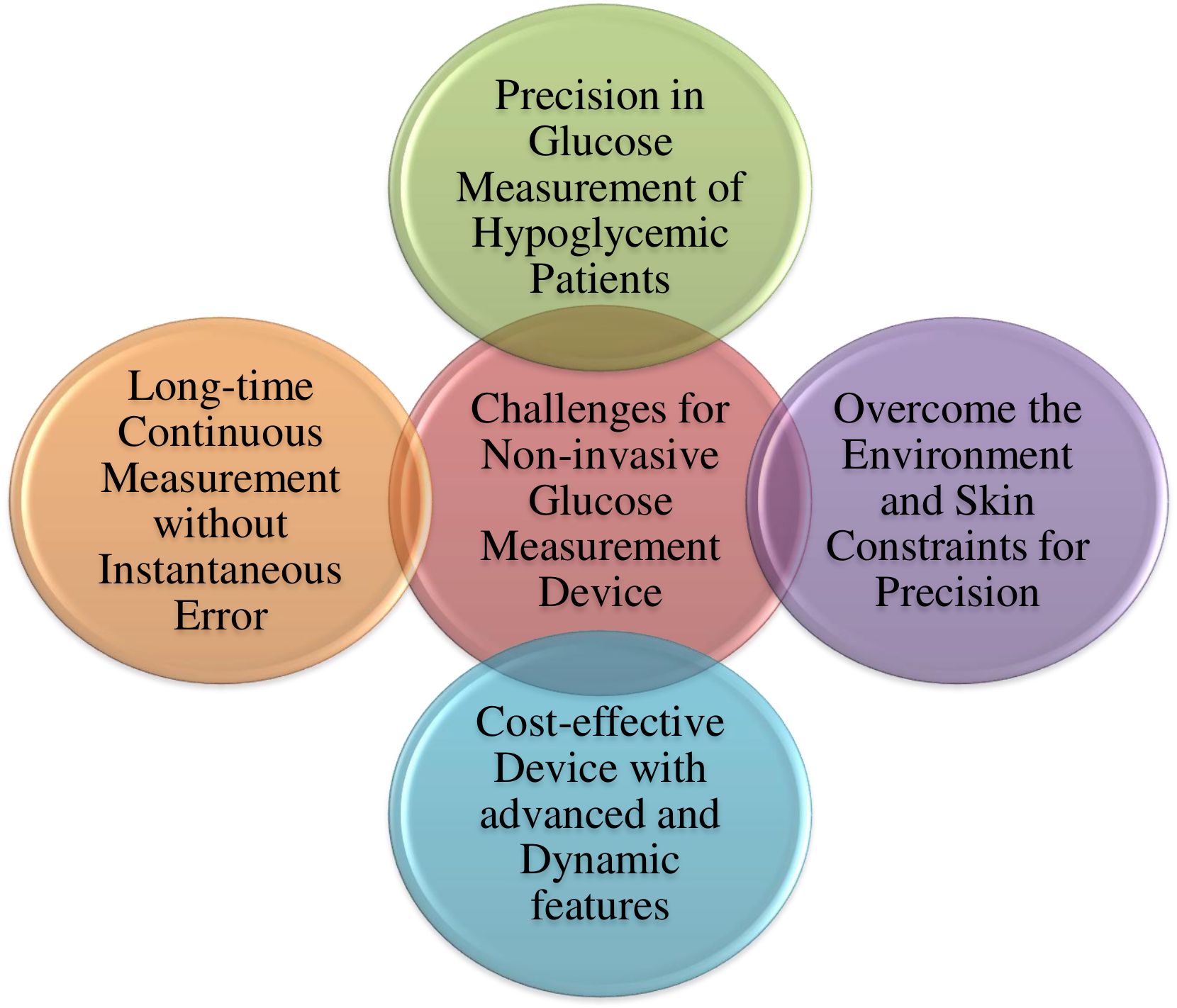}
	\caption{Open Challenges in Noninvasive Glucose-Level Measurement.}
	\label{FIG:Open_problem}
\end{figure}

There are lots of challenges for commercialization of non-invasive glucose measurement device. But, some open problems have been discussed which are prime challenges for precise non-invasive glucose measurement. These challenges have been represented in Fig. \ref{FIG:Open_problem}. The precise glucose measurement of hypoglycemic patient and long-time continuous glucose measurement without instantaneous error are the open problems which  are focussed by the researchers recently.

\begin{itemize}
\item The effect of blood pressure, body temperature and humidity have not been considered in the literature which affect the values of glucose measurement.
\item The cost effective and portable solution of continuous glucose measurement device has also not been addressed properly.
 \item The accurate glucose measurement has been also been open challenge for full rage from 40 mg/dl to 450 mg/dl.
  \item  The effective integration of glucometer with internet of medical things for continuously data logging to the cloud has still not potentially resolved.
  \item The mathematical model for automatic insulin secretion according to measured glucose value has to be address in better manner with internet framework.
     \item   The privacy and security issues of 
insulin and blood glucose measurement system is still not resolved yet.
  \item  The efficient power management mechanism
 has to be developed for continuous glucose measurement with insulin delivery system. 

\end{itemize}

%%%%%%%%%%%%%%%%%%%%%%%%%%%%%%%%%%%%%%%%%%
%\section{Conclusions}
%\label{SEC:Conclusions}
%%%%%%%%%%%%%%%%%%%%%%%%%%%%%%%%%%%%%%%%%%
\section{Conclusions and Future Research}
\label{SEC:Conclusions}

% FPGA architecture is proposed only for data acquisition, noise reduction and display only. The model for data interpretation has not implemented on FPGA for portable device. Non invasive devices available at present time are not more than efficient and accurate to measure low blood glucose concentration. The future non invasive device should be more than accurate compared to previous designed handheld device for hypoglycemic measurement. 
The paper presents survey of glucose measurement approaches along with overview of glucose control mechanism. 
Many techniques available in literature are only a proof of concept, showing good correlation between device estimated result and reference value of blood glucose. However, they are neither accurate nor cost effective solutions and not  available for  commercial purpose. The optical detection using short NIR has been potential solution to mitigate the drawbacks of all previous methods.
In future, the multi-model  approaches could be considered for precise glucose estimation. The device or prototype model should be more effective in different zones to support the continuous health monitoring. It should be implemented as a portable device for real time application with more frequently. This device should be developed as continuous health monitoring with minimum cost.

%\begin{figure}
%	\begin{center}
%		\includegraphics[width=1.0\textwidth]{fig10}
%		\caption{Block diagram representation of future aspect of non invasive glucose measurement}
%		\label{fig10}
%	\end{center}
%\end{figure}

%\subsection{Novelty}
%\label{subsec1}

%The proposed technique is advanced form of bio-impedance spectroscopy. The proposed approach can overcome the drawbacks of the conventional non-invasive glucose monitoring.
%Two techniques applied concurrently for estimation of glucose concentration and obtained values from both techniques are combined through artificial neural network (ANN). Final estimated data is more accurate with comparison of previous work.
%An FPGA based reconfigurable embedded architecture is proposed for high speed data acquisition, noise reduction and display of PA measurements. 

\begin{figure}[htbp]
	\centering
	\includegraphics[width=0.55\textwidth]{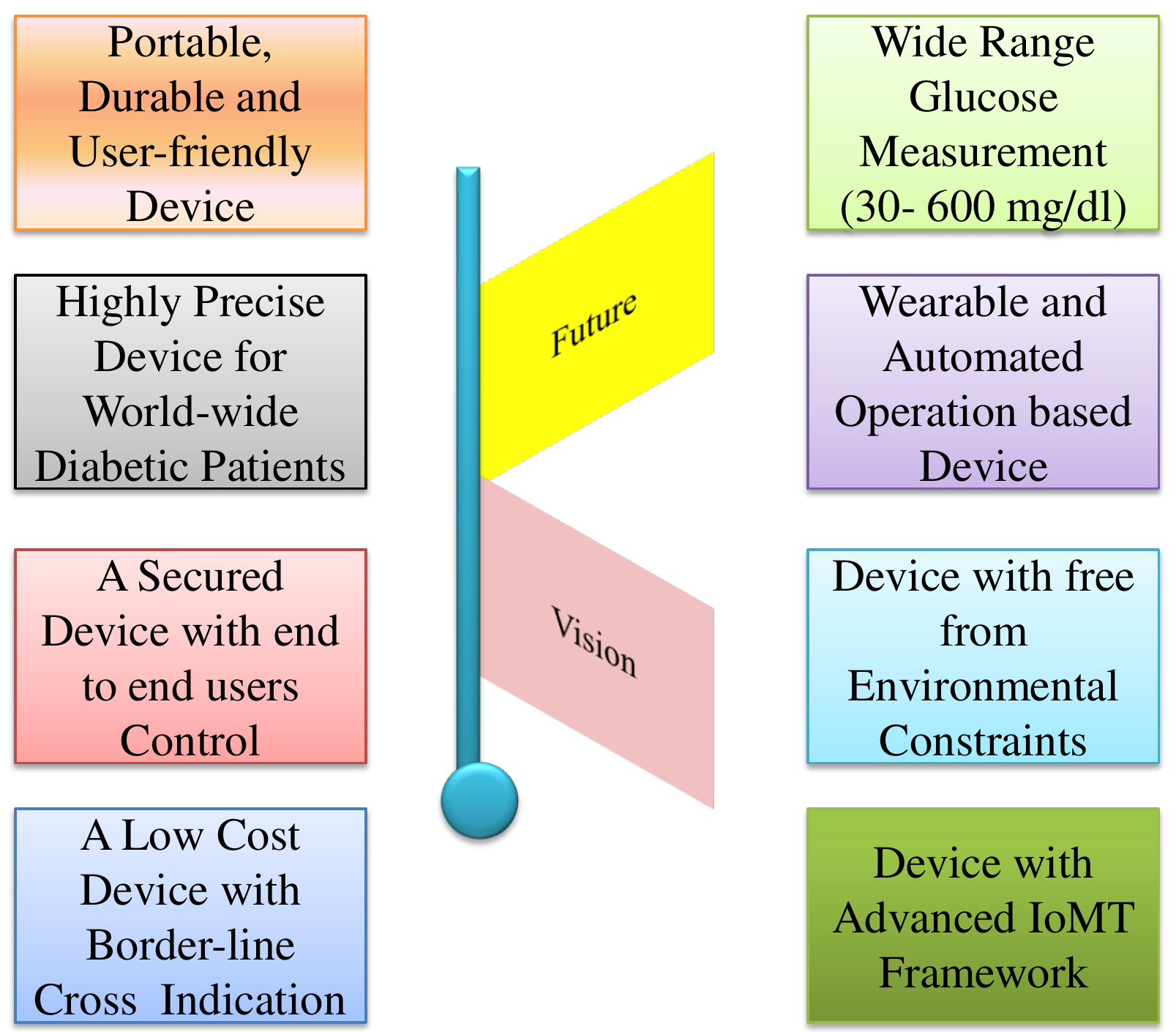}
	\caption{Our Future Vision for Non-invasive Glucose-Level Measurement.}
	\label{FIG:Future_vision}
\end{figure}

The future research for upcoming noninvasive glucose monitoring device is mentioned in Fig. \ref{FIG:Future_vision}. The device is required to be integrated with advanced IoMT framework. This advanced IoMT framework will alow to connect the device with all nearest diabetic centers to get best treatment. Unification of glucose-level measurement and automatic diet quantification can have strong impact on smart healthcare domain \cite{9011599}. The durability, portability and user-friendly device is also the future vision in this era. The device should have the feature of border-line cross indication. Because of this feature, any person will be aware to take own blood glucose level. A secured device with end to end users control and authentication is also necessary for future advancement. Physical Unclonable Function (PUF) based security of IoMT-devices can be effective for IoMT-devices which are intrinsically resource and battery constrained \cite{joshi2020secure, 8752409}. Unified healthcare Cyber-Physical System (H-CPS) with blockchain based data and device management can be effective and needs research \cite{9288682, 8977825}.

%%%%%%%%%%%%%%%%%%%%%%%%%%%%%%%%%%%%%%%%%%%%%%%%%%
\bibliographystyle{IEEEtran}
\bibliography{Bibliography_Glucose-Measurement-Survey}

\begin{thebibliography}{100}
\providecommand{\url}[1]{#1}
\csname url@samestyle\endcsname
\providecommand{\newblock}{\relax}
\providecommand{\bibinfo}[2]{#2}
\providecommand{\BIBentrySTDinterwordspacing}{\spaceskip=0pt\relax}
\providecommand{\BIBentryALTinterwordstretchfactor}{4}
\providecommand{\BIBentryALTinterwordspacing}{\spaceskip=\fontdimen2\font plus
\BIBentryALTinterwordstretchfactor\fontdimen3\font minus
  \fontdimen4\font\relax}
\providecommand{\BIBforeignlanguage}[2]{{%
\expandafter\ifx\csname l@#1\endcsname\relax
\typeout{** WARNING: IEEEtran.bst: No hyphenation pattern has been}%
\typeout{** loaded for the language `#1'. Using the pattern for}%
\typeout{** the default language instead.}%
\else
\language=\csname l@#1\endcsname
\fi
#2}}
\providecommand{\BIBdecl}{\relax}
\BIBdecl

\bibitem{Diabetestalk_URL_2018}
\BIBentryALTinterwordspacing
Diabetestalk, ``What is the normal fasting blood sugar range for adults,''
  2018, last Accessed on 18 Jan 2021. [Online]. Available:
  \url{https://diabetestalk.net/blood-sugar/what-is-the-normal-fasting-blood-sugar-range-for-adults}
\BIBentrySTDinterwordspacing

\bibitem{jain2020iglu}
P.~Jain, A.~M. Joshi, and S.~P. Mohanty, ``{iGLU 1.1}: Towards a
  glucose-insulin model based closed loop iomt framework for automatic insulin
  control of diabetic patients,'' in \emph{2020 IEEE 6th World Forum on
  Internet of Things (WF-IoT)}.\hskip 1em plus 0.5em minus 0.4em\relax IEEE,
  2020, pp. 1--6.

\bibitem{9174644}
A.~M. Joshi, U.~P. Shukla, and S.~P. Mohanty, ``Smart healthcare for diabetes
  during {COVID-19},'' \emph{IEEE Consumer Electronics Magazine}, vol.~10,
  no.~1, pp. 66--71, January 2020.

\bibitem{joshi2020smartA}
\vspace{0.2mm}Amit M~Joshi, U.~P. Shukla, and S.~P. Mohanty, ``Smart healthcare
  for diabetes: A {COVID-19} perspective,'' \emph{arXiv preprint
  arXiv:2008.11153}, 2020.

\bibitem{alsamman2020transcriptomic}
A.~M. Alsamman and H.~Zayed, ``{The transcriptomic profiling of COVID-19
  compared to SARS, MERS, Ebola, and H1N1},'' \emph{bioRxiv}, 2020.

\bibitem{diabetesatlas_URL_2020}
\BIBentryALTinterwordspacing
I.~D. Federation, ``{IDF Diabetes Atlas - Diabetes is rising worldwide... and
  is set to rise even further},'' 2019, last Accessed on 21 March 2020.
  [Online]. Available:
  \url{https://diabetesatlas.org/en/sections/worldwide-toll-of-diabetes.html}
\BIBentrySTDinterwordspacing

\bibitem{Cho_DRCP_2018-Apr}
N.~H. Cho, J.~E. Shaw, S.~Karuranga, Y.~Huang, J.~D. da~Rocha~Fernandes, A.~W.
  Ohlrogge, and B.~Malanda, ``{IDF} diabetes atlas: Global estimates of
  diabetes prevalence for 2017 and projections for 2045,'' \emph{Diabetes
  Research and Clinical Practice}, vol. 138, pp. 271--281, April 2018.

\bibitem{saeedi2019global}
P.~Saeedi, I.~Petersohn, P.~Salpea, B.~Malanda, S.~Karuranga, N.~Unwin,
  S.~Colagiuri, L.~Guariguata, A.~A. Motala, K.~Ogurtsova \emph{et~al.},
  ``Global and regional diabetes prevalence estimates for 2019 and projections
  for 2030 and 2045: Results from the international diabetes federation
  diabetes atlas,'' \emph{Diabetes research and clinical practice}, vol. 157,
  p. 107843, 2019.

\bibitem{clevelandclinic_URL_2020}
\BIBentryALTinterwordspacing
Clevelandclinic, ``{Diabetes Mellitus: An Overview},'' 2020, last Accessed on
  18 Jan 2021. [Online]. Available:
  \url{https://my.clevelandclinic.org/health/diseases/7104-diabetes-mellitus-an-overview}
\BIBentrySTDinterwordspacing

\bibitem{drugs_URL_2020}
\BIBentryALTinterwordspacing
Drugs, ``Type 1 diabetes mellitus,'' 2020, last Accessed on 18 Jan 2021.
  [Online]. Available:
  \url{https://www.drugs.com/health-guide/type-1-diabetes-mellitus.html}
\BIBentrySTDinterwordspacing

\bibitem{endocrineweb_URL_2019}
\BIBentryALTinterwordspacing
L.~M. Leontis and A.~Hess-Fischl, ``Diabetes mellitus: An overview,'' 2019,
  last Accessed on 18 Jan 2021. [Online]. Available:
  \url{https://www.endocrineweb.com/conditions/type-2-diabetes/type-2-diabetes-symptoms}
\BIBentrySTDinterwordspacing

\bibitem{healthline_URL_2018}
\BIBentryALTinterwordspacing
A.~Pietrangelo, ``What are the different types of diabetes?'' 2018, last
  Accessed on 18 Jan 2021. [Online]. Available:
  \url{https://www.healthline.com/health/diabetes/types-of-diabetes}
\BIBentrySTDinterwordspacing

\bibitem{Fowler42}
\BIBentryALTinterwordspacing
M.~J. Fowler, ``Diabetes: Magnitude and mechanisms,'' \emph{Clinical Diabetes},
  vol.~28, no.~1, pp. 42--46, 2010. [Online]. Available:
  \url{https://clinical.diabetesjournals.org/content/28/1/42}
\BIBentrySTDinterwordspacing

\bibitem{Yin_TETC_2019-2958946}
H.~{Yin}, B.~{Mukadam}, X.~{Dai}, and N.~{Jha}, ``{DiabDeep: Pervasive Diabetes
  Diagnosis based on Wearable Medical Sensors and Efficient Neural Networks},''
  \emph{IEEE Transactions on Emerging Topics in Computing}, pp. 1--1, 2019.

\bibitem{Zhang2011}
P.~Zhang, ``Global healthcare expenditure on diabetes for 2010 and 2030.''
  \emph{Diabetes Research and Clinical Practice}, 2011.

\bibitem{Venkataraman2011}
J.~Venkataraman and B.~Freer, ``Feasibility of non-invasive blood glucose
  monitoring: In-vitro measurements and phantom models,'' in \emph{2011 IEEE
  International Symposium on Antennas and Propagation (APSURSI)}, July 2011,
  pp. 603--606.

\bibitem{Wild2569}
\BIBentryALTinterwordspacing
S.~H. Wild, G.~Roglic, A.~Green, R.~Sicree, and H.~King, ``Global prevalence of
  diabetes: Estimates for the year 2000 and projections for 2030,''
  \emph{Diabetes Care}, vol.~27, no.~10, pp. 2569--2569, 2004. [Online].
  Available: \url{http://care.diabetesjournals.org/content/27/10/2569.2}
\BIBentrySTDinterwordspacing

\bibitem{WHITING2011311}
D.~R. Whiting, L.~Guariguata, C.~Weil, and J.~Shaw, ``Idf diabetes atlas:
  Global estimates of the prevalence of diabetes for 2011 and 2030,''
  \emph{Diabetes Research and Clinical Practice}, vol.~94, no.~3, pp. 311 --
  321, 2011.

\bibitem{Siegel2015}
P.~H. Siegel, A.~Tang, G.~Virbila, Y.~Kim, M.~C.~F. Chang, and V.~Pikov,
  ``Compact non-invasive millimeter-wave glucose sensor,'' in \emph{2015 40th
  International Conference on Infrared, Millimeter, and Terahertz waves
  (IRMMW-THz)}, Aug 2015, pp. 1--3.

\bibitem{Alavi2001}
S.~M. Alavi, M.~Gourzi, A.~Rouane, and M.~Nadi, ``An original method for
  non-invasive glucose measurement: preliminary results,'' in \emph{2001
  Conference Proceedings of the 23rd Annual International Conference of the
  IEEE Engineering in Medicine and Biology Society}, vol.~4, 2001, pp.
  3318--3320 vol.4.

\bibitem{Li2015}
X.~Li and C.~Li, ``Study on the application of wavelet transform to
  non-invasive glucose concentration measurement by nirs,'' in \emph{2015 Fifth
  International Conference on Instrumentation and Measurement, Computer,
  Communication and Control (IMCCC)}, Sept 2015, pp. 1294--1297.

\bibitem{Pai2017}
P.~P. Pai, P.~K. Sanki, S.~K. Sahoo, A.~De, S.~Bhattacharya, and S.~Banerjee,
  ``Cloud computing-based non-invasive glucose monitoring for diabetic care,''
  \emph{IEEE Transactions on Circuits and Systems I: Regular Papers}, vol.~PP,
  no.~99, pp. 1--14, 2017.

\bibitem{Reddy2017}
P.~S. Reddy and K.~Jyostna, ``Development of smart insulin device for non
  invasive blood glucose level monitoring,'' in \emph{2017 IEEE 7th
  International Advance Computing Conference (IACC)}, Jan 2017, pp. 516--519.

\bibitem{Jain2020}
P.~Jain, A.~M. Joshi, and S.~P. Mohanty, ``{iGLU: An Intelligent Device for
  Accurate Non-Invasive Blood Glucose-Level Monitoring in Smart Healthcare},''
  \emph{IEEE Consumer Electronics Magazine}, vol.~9, no.~1, p. Accepted,
  January 2020.

\bibitem{Joshi_TCE_iGLU2_TCE.2020.3011966}
A.~M. {Joshi}, P.~{Jain}, S.~P. {Mohanty}, and N.~{Agrawal}, ``{iGLU 2.0}: A
  new wearable for accurate non-invasive continuous serum glucose measurement
  in {IoMT} framework,'' \emph{IEEE Transactions on Consumer Electronics}, no.
  10.1109/TCE.2020.3011966, p. in Press, 2020.

\bibitem{Mohanty_POT_2006-Mar}
S.~P. Mohanty and E.~Kougianos, ``{Biosensors: A Tutorial Review},'' \emph{IEEE
  Potentials}, vol.~25, no.~2, pp. 35--40, March 2006.

\bibitem{Salam2016TheEO}
N.~A. Salam, W.~H.~M. Saad, Z.~Manap, and F.~Salehuddin, ``The evolution of
  non-invasive blood glucose monitoring system for personal application,''
  \emph{Journal of Telecommunication, Electronic and Computer Engineering},
  vol.~8, pp. 59--65, 2016.

\bibitem{Jain_arXiv_2019-Nov30-1911-04471_iGLU1}
\vspace{0.2mm}P. Jain, A.~M. Joshi, and S.~P. Mohanty, ``{iGLU 1.0: An Accurate
  Non-Invasive Near-Infrared Dual Short Wavelengths Spectroscopy based
  Glucometer for Smart Healthcare},'' \emph{arXiv Electrical Engineering and
  Systems Science}, no. arXiv:1911.04471, November 2019.

\bibitem{Mohanty_arXiv_2020-Jan-28-2001-09182_iGLU2}
\BIBentryALTinterwordspacing
\vspace{0mm}P. Jain, A.~M. Joshi, N.~Agrawal, and S.~P. Mohanty, ``{iGLU 2.0: A
  New Non-invasive, Accurate Serum Glucometer for Smart Healthcare},''
  \emph{arXiv Electrical Engineering and Systems Science}, vol. abs/2001.09182,
  2020. [Online]. Available: \url{http://arxiv.org/abs/2001.09182}
\BIBentrySTDinterwordspacing

\bibitem{acsnano7b06823}
Q.~Liu, Y.~Liu, F.~Wu, X.~Cao, Z.~Li, M.~Alharbi, A.~N. Abbas, M.~R. Amer, and
  C.~Zhou, ``{Highly Sensitive and Wearable In2O3 Nanoribbon Transistor
  Biosensors with Integrated On-Chip Gate for Glucose Monitoring in Body
  Fluids},'' \emph{ACS Nano}, vol.~12, no.~2, pp. 1170--1178, 2018, pMID:
  29338249.

\bibitem{Siegel2016}
P.~H. Siegel, W.~Dai, R.~A. Kloner, M.~Csete, and V.~Pikov, ``First
  millimeter-wave animal in vivo measurements of l-glucose and d-glucose:
  Further steps towards a non-invasive glucometer,'' in \emph{2016 41st
  International Conference on Infrared, Millimeter, and Terahertz waves
  (IRMMW-THz)}, Sept 2016, pp. 1--3.

\bibitem{Jain_IEEE-MCE_2020-Jan_iGLU1}
P.~Jain, A.~M. Joshi, and S.~P. Mohanty, ``{iGLU: An Intelligent Device for
  Accurate Non-Invasive Blood Glucose-Level Monitoring in Smart Healthcare},''
  \emph{IEEE Consumer Electronics Magazine}, vol.~9, no.~1, pp. 35--42, January
  2020.

\bibitem{Zhilo2017}
N.~M. Zhilo, P.~A. Rudenko, and A.~N. Zhigaylo, ``Development of
  hardware-software test bench for optical non-invasive glucometer
  improvement,'' in \emph{2017 IEEE Conference of Russian Young Researchers in
  Electrical and Electronic Engineering (EIConRus)}, Feb 2017, pp. 89--90.

\bibitem{Gusev2017}
S.~I. Gusev, A.~A. Simonova, P.~S. Demchenko, M.~K. Khodzitsky, and O.~P.
  Cherkasova, ``Blood glucose concentration sensing using biological molecules
  relaxation times determination,'' in \emph{2017 IEEE International Symposium
  on Medical Measurements and Applications (MeMeA)}, May 2017, pp. 458--463.

\bibitem{Sari2016}
M.~W. Sari and M.~Luthfi, ``Design and analysis of non-invasive blood glucose
  levels monitoring,'' in \emph{2016 International Seminar on Application for
  Technology of Information and Communication (ISemantic)}, Aug 2016, pp.
  134--137.

\bibitem{Lekha2015}
S.~Lekha and M.~Suchetha, ``Non- invasive diabetes detection and classification
  using breath analysis,'' in \emph{2015 International Conference on
  Communications and Signal Processing (ICCSP)}, April 2015, pp. 0955--0958.

\bibitem{li2016fine}
J.~Li, P.~Koinkar, Y.~Fuchiwaki, and M.~Yasuzawa, ``A fine pointed glucose
  oxidase immobilized electrode for low-invasive amperometric glucose
  monitoring,'' \emph{Biosensors and Bioelectronics}, vol.~86, pp. 90--94,
  2016.

\bibitem{demitri2017measuring}
N.~Demitri and A.~M. Zoubir, ``Measuring blood glucose concentrations in
  photometric glucometers requiring very small sample volumes,'' \emph{IEEE
  Transactions on Biomedical Engineering}, vol.~64, no.~1, pp. 28--39, 2017.

\bibitem{lucisano2017glucose}
J.~Y. Lucisano, T.~L. Routh, J.~T. Lin, and D.~A. Gough, ``Glucose monitoring
  in individuals with diabetes using a long-term implanted sensor/telemetry
  system and model,'' \emph{IEEE Transactions on Biomedical Engineering},
  vol.~64, no.~9, pp. 1982--1993, 2017.

\bibitem{7576627}
A.~{Sun}, A.~G. {Venkatesh}, and D.~A. {Hall}, ``A multi-technique
  reconfigurable electrochemical biosensor: Enabling personal health monitoring
  in mobile devices,'' \emph{IEEE Transactions on Biomedical Circuits and
  Systems}, vol.~10, no.~5, pp. 945--954, Oct 2016.

\bibitem{5291722}
A.~{Gani}, A.~V. {Gribok}, Y.~{Lu}, W.~K. {Ward}, R.~A. {Vigersky}, and
  J.~{Reifman}, ``Universal glucose models for predicting subcutaneous glucose
  concentration in humans,'' \emph{IEEE Transactions on Information Technology
  in Biomedicine}, vol.~14, no.~1, pp. 157--165, Jan 2010.

\bibitem{7933990}
G.~{Wang}, M.~D. {Poscente}, S.~S. {Park}, C.~N. {Andrews}, O.~{Yadid-Pecht},
  and M.~P. {Mintchev}, ``Wearable microsystem for minimally invasive,
  pseudo-continuous blood glucose monitoring: The e-mosquito,'' \emph{IEEE
  Transactions on Biomedical Circuits and Systems}, vol.~11, no.~5, pp.
  979--987, Oct 2017.

\bibitem{4956982}
M.~M. {Ahmadi} and G.~A. {Jullien}, ``A wireless-implantable microsystem for
  continuous blood glucose monitoring,'' \emph{IEEE Transactions on Biomedical
  Circuits and Systems}, vol.~3, no.~3, pp. 169--180, June 2009.

\bibitem{acciaroli2018reduction}
G.~Acciaroli, M.~Vettoretti, A.~Facchinetti, G.~Sparacino, and C.~Cobelli,
  ``Reduction of blood glucose measurements to calibrate subcutaneous glucose
  sensors: A bayesian multiday framework,'' \emph{IEEE Transactions on
  Biomedical Engineering}, vol.~65, no.~3, pp. 587--595, 2018.

\bibitem{6778812}
I.~{Pagkalos}, P.~{Herrero}, C.~{Toumazou}, and P.~{Georgiou}, ``Bio-inspired
  glucose control in diabetes based on an analogue implementation of a $\beta
  $-cell model,'' \emph{IEEE Transactions on Biomedical Circuits and Systems},
  vol.~8, no.~2, pp. 186--195, April 2014.

\bibitem{wang2017wearable}
G.~Wang, M.~D. Poscente, S.~S. Park, C.~N. Andrews, O.~Yadid-Pecht, and M.~P.
  Mintchev, ``Wearable microsystem for minimally invasive, pseudo-continuous
  blood glucose monitoring: The e-mosquito,'' \emph{IEEE Transactions on
  Biomedical Circuits and Systems}, vol.~11, no.~5, pp. 979--987, 2017.

\bibitem{Kossowski2016}
T.~Kossowski and R.~Stasinski, ``Robust ir attenuation measurement for
  non-invasive glucose level analysis,'' in \emph{2016 International Conference
  on Systems, Signals and Image Processing (IWSSIP)}, May 2016, pp. 1--4.

\bibitem{Pavlovich2013}
L.~P. Pavlovich and D.~Y. Mynziak, ``Noninvasive method for blood glucose
  measuring and monitoring,'' in \emph{2013 IEEE XXXIII International
  Scientific Conference Electronics and Nanotechnology (ELNANO)}, April 2013,
  pp. 255--257.

\bibitem{Liu2016b}
Y.~Liu, W.~Li, T.~Zheng, and W.~K. Ling, ``Overviews the methods of
  non-invasive blood glucose measurement,'' in \emph{2016 IEEE International
  Conference on Consumer Electronics-China (ICCE-China)}, Dec 2016, pp. 1--2.

\bibitem{Sharma2012}
N.~K. Sharma and S.~Singh, ``Designing a non invasive blood glucose measurement
  sensor,'' in \emph{2012 IEEE 7th International Conference on Industrial and
  Information Systems (ICIIS)}, Aug 2012, pp. 1--3.

\bibitem{Zhao2016}
X.~Zhao, Q.~Zheng, and Z.~M. Yang, ``Two types of photonic crystals applied to
  glucose sensor,'' in \emph{2016 IEEE International Nanoelectronics Conference
  (INEC)}, May 2016, pp. 1--2.

\bibitem{Tanaka2016}
Y.~Tanaka, C.~Purtill, T.~Tajima, M.~Seyama, and H.~Koizumi, ``Sensitivity
  improvement on cw dual-wavelength photoacoustic spectroscopy using acoustic
  resonant mode for noninvasive glucose monitor,'' in \emph{2016 IEEE SENSORS},
  Oct 2016, pp. 1--3.

\bibitem{Gouzouasis2016}
I.~Gouzouasis, H.~Cano-Garcia, I.~Sotiriou, S.~Saha, G.~Palikaras, P.~Kosmas,
  and E.~Kallos, ``Detection of varying glucose concentrations in water
  solutions using a prototype biomedical device for millimeter-wave
  non-invasive glucose sensing,'' in \emph{2016 10th European Conference on
  Antennas and Propagation (EuCAP)}, April 2016, pp. 1--4.

\bibitem{Nikawa2001}
Y.~Nikawa and D.~Someya, ``Non-invasive measurement of blood sugar level by
  millimeter waves,'' in \emph{2001 IEEE MTT-S International Microwave
  Sympsoium Digest (Cat. No.01CH37157)}, vol.~1, May 2001, pp. 171--174 vol.1.

\bibitem{Shao2016}
J.~Shao, F.~Yang, F.~Xia, Q.~Zhang, and Y.~Chen, ``A novel miniature spiral
  sensor for non-invasive blood glucose monitoring,'' in \emph{2016 10th
  European Conference on Antennas and Propagation (EuCAP)}, April 2016, pp.
  1--2.

\bibitem{Siegel2014}
P.~H. Siegel, Y.~Lee, and V.~Pikov, ``Millimeter-wave non-invasive monitoring
  of glucose in anesthetized rats,'' in \emph{2014 39th International
  Conference on Infrared, Millimeter, and Terahertz waves (IRMMW-THz)}, Sept
  2014, pp. 1--2.

\bibitem{Wang2014}
D.~Wang, ``An improved integration sensor of non-invasive blood glucose,'' in
  \emph{The 7th IEEE/International Conference on Advanced Infocomm Technology},
  Nov 2014, pp. 70--75.

\bibitem{Bayasi2013}
N.~Bayasi, H.~Saleh, B.~Mohammad, and M.~Ismail, ``The revolution of glucose
  monitoring methods and systems: A survey,'' in \emph{2013 IEEE 20th
  International Conference on Electronics, Circuits, and Systems (ICECS)}, Dec
  2013, pp. 92--93.

\bibitem{agrawal2013noninvasive}
R.~Agrawal, N.~Sharma, M.~Rathore, V.~Gupta, S.~Jain, V.~Agarwal, and S.~Goyal,
  ``Noninvasive method for glucose level estimation by saliva,'' \emph{J
  Diabetes Metab}, vol.~4, no.~5, pp. 2--5, 2013.

\bibitem{demitri2016measuring}
N.~Demitri and A.~M. Zoubir, ``Measuring blood glucose concentrations in
  photometric glucometers requiring very small sample volumes,'' \emph{IEEE
  Transactions on Biomedical Engineering}, vol.~64, no.~1, pp. 28--39, 2016.

\bibitem{10.3390/s20051251}
M.~Shokrekhodaei and S.~Quinones, ``{Review of Non-invasive Glucose Sensing
  Techniques: Optical, Electrical and Breath Acetone},'' \emph{Sensors},
  vol.~20, no.~5, p. 1251, 2020.

\bibitem{10.1007/s00216-018-1395-x}
S.~Delbeck, T.~Vahlsing, S.~Leonhardt, G.~S. G, and H.~M. Heise,
  ``{Non-invasive monitoring of blood glucose using optical methods for skin
  spectroscopy-opportunities and recent advances},'' \emph{Anal Bioanal Chem.},
  vol. 411, no.~1, pp. 63--77, 2019.

\bibitem{Madzhi2014}
N.~K. Madzhi, S.~A. Shamsuddin, and M.~F. Abdullah, ``Comparative investigation
  using gaas(950nm), gaaias (940nm) and ingaasp (1450nm) sensors for
  development of non-invasive optical blood glucose measurement system,'' in
  \emph{2014 IEEE International Conference on Smart Instrumentation,
  Measurement and Applications (ICSIMA)}, Nov 2014, pp. 1--6.

\bibitem{Aziz2014}
N.~A.~M. Aziz, N.~Arsad, P.~S. Menon, A.~R. Laili, M.~H. Laili, and A.~A.~A.
  Halim, ``Analysis of difference light sources for non-invasive aqueous
  glucose detection,'' in \emph{2014 IEEE 5th International Conference on
  Photonics (ICP)}, Sept 2014, pp. 150--152.

\bibitem{azolifesciences_URL_2020}
\BIBentryALTinterwordspacing
S.~Tommasone, ``Infrared spectroscopy: An overview,'' 2018, last Accessed on 18
  Jan 2021. [Online]. Available:
  \url{https://www.azolifesciences.com/article/Infrared-Spectroscopy-An-Overview.aspx}
\BIBentrySTDinterwordspacing

\bibitem{Muley2014}
A.~A. Muley and R.~B. Ghongade, ``Design and simulate an antenna for aqueous
  glucose measurement,'' in \emph{2014 Annual IEEE India Conference (INDICON)},
  Dec 2014, pp. 1--6.

\bibitem{menon2013voltage}
K.~U. Menon, D.~Hemachandran, and A.~T. Kunnath, ``Voltage intensity based
  non-invasive blood glucose monitoring,'' in \emph{2013 Fourth International
  Conference on Computing, Communications and Networking Technologies
  (ICCCNT)}.\hskip 1em plus 0.5em minus 0.4em\relax IEEE, 2013, pp. 1--5.

\bibitem{Lai2016}
J.~L. Lai, S.~Y. Huang, R.~S. Lin, and S.~C. Tsai, ``Design a non-invasive
  near-infrared led blood glucose sensor,'' in \emph{2016 International
  Conference on Applied System Innovation (ICASI)}, May 2016, pp. 1--4.

\bibitem{Tamilselvi2015}
M.~Tamilselvi and G.~Ramkumar, ``Non-invasive tracking and monitoring glucose
  content using near infrared spectroscopy,'' in \emph{2015 IEEE International
  Conference on Computational Intelligence and Computing Research (ICCIC)}, Dec
  2015, pp. 1--3.

\bibitem{Lawand2015}
K.~Lawand, M.~Parihar, and S.~N. Patil, ``Design and development of infrared
  led based non invasive blood glucometer,'' in \emph{2015 Annual IEEE India
  Conference (INDICON)}, Dec 2015, pp. 1--6.

\bibitem{Yadav2014}
J.~Yadav, A.~Rani, V.~Singh, and B.~M. Murari, ``Near-infrared led based
  non-invasive blood glucose sensor,'' in \emph{2014 International Conference
  on Signal Processing and Integrated Networks (SPIN)}, Feb 2014, pp. 591--594.

\bibitem{Abidin2013}
M.~T. B.~Z. Abidin, M.~K.~R. Rosli, S.~A. Shamsuddin, N.~K. Madzhi, and M.~F.
  Abdullah, ``Initial quantitative comparison of 940nm and 950nm infrared
  sensor performance for measuring glucose non-invasively,'' in \emph{2013 IEEE
  International Conference on Smart Instrumentation, Measurement and
  Applications (ICSIMA)}, Nov 2013, pp. 1--6.

\bibitem{Nikawa2006}
Y.~Nikawa and T.~Michiyama, ``Non-invasive measurement of blood-sugar level by
  reflection of millimeter-waves,'' in \emph{2006 Asia-Pacific Microwave
  Conference}, Dec 2006, pp. 47--50.

\bibitem{goodarzi2016selection}
M.~Goodarzi and W.~Saeys, ``Selection of the most informative near infrared
  spectroscopy wavebands for continuous glucose monitoring in human serum,''
  \emph{Talanta}, vol. 146, pp. 155--165, 2016.

\bibitem{sharma2013efficient}
S.~Sharma, M.~Goodarzi, L.~Wynants, H.~Ramon, and W.~Saeys, ``Efficient use of
  pure component and interferent spectra in multivariate calibration,''
  \emph{Analytica chimica acta}, vol. 778, pp. 15--23, 2013.

\bibitem{uwadaira2010factors}
Y.~Uwadaira, N.~Adachi, A.~Ikehata, and S.~Kawano, ``Factors affecting the
  accuracy of non-invasive blood glucose measurement by short-wavelength near
  infrared spectroscopy in the determination of the glycaemic index of foods,''
  \emph{Journal of Near Infrared Spectroscopy}, vol.~18, no.~5, pp. 291--300,
  2010.

\bibitem{haxha2016optical}
S.~Haxha and J.~Jhoja, ``Optical based noninvasive glucose monitoring sensor
  prototype,'' \emph{IEEE Photonics Journal}, vol.~8, no.~6, pp. 1--11, 2016.

\bibitem{zhang2013discussion}
W.~Zhang, R.~Liu, W.~Zhang, H.~Jia, and K.~Xu, ``Discussion on the validity of
  nir spectral data in non-invasive blood glucose sensing,'' \emph{Biomedical
  optics express}, vol.~4, no.~6, pp. 789--802, 2013.

\bibitem{golic2003short}
M.~Golic, K.~Walsh, and P.~Lawson, ``Short-wavelength near-infrared spectra of
  sucrose, glucose, and fructose with respect to sugar concentration and
  temperature,'' \emph{Applied spectroscopy}, vol.~57, no.~2, pp. 139--145,
  2003.

\bibitem{Song2015}
K.~Song, U.~Ha, S.~Park, J.~Bae, and H.~J. Yoo, ``An impedance and
  multi-wavelength near-infrared spectroscopy ic for non-invasive blood glucose
  estimation,'' \emph{IEEE Journal of Solid-State Circuits}, vol.~50, no.~4,
  pp. 1025--1037, April 2015.

\bibitem{ramasahayam2015noninvasive}
S.~Ramasahayam, K.~S. Haindavi, and S.~R. Chowdhury, ``Noninvasive estimation
  of blood glucose concentration using near infrared optodes,'' in
  \emph{Sensing Technology: Current Status and Future Trends IV}.\hskip 1em
  plus 0.5em minus 0.4em\relax Springer, 2015, pp. 67--82.

\bibitem{heller2005integrated}
A.~Heller, ``Integrated medical feedback systems for drug delivery,''
  \emph{AIChE journal}, vol.~51, no.~4, pp. 1054--1066, 2005.

\bibitem{pai2015nir}
P.~P. Pai, P.~K. Sanki, A.~De, and S.~Banerjee, ``{NIR} photoacoustic
  spectroscopy for non-invasive glucose measurement,'' in \emph{2015 37th
  Annual International Conference of the IEEE Engineering in Medicine and
  Biology Society (EMBC)}.\hskip 1em plus 0.5em minus 0.4em\relax IEEE, 2015,
  pp. 7978--7981.

\bibitem{jain2019iglu}
P.~Jain, A.~M. Joshi, and S.~P. Mohanty, ``{iGLU}: An intelligent device for
  accurate noninvasive blood glucose-level monitoring in smart healthcare,''
  \emph{IEEE Consumer Electronics Magazine}, vol.~9, no.~1, pp. 35--42, 2019.

\bibitem{jain2019precise}
P.~Jain, R.~Maddila, and A.~M. Joshi, ``{A precise non-invasive blood glucose
  measurement system using NIR spectroscopy and Huber’s regression model},''
  \emph{Optical and Quantum Electronics}, vol.~51, no.~2, p.~51, 2019.

\bibitem{monte2011non}
E.~Monte-Moreno, ``Non-invasive estimate of blood glucose and blood pressure
  from a photoplethysmograph by means of machine learning techniques,''
  \emph{Artificial intelligence in medicine}, vol.~53, no.~2, pp. 127--138,
  2011.

\bibitem{habbu2019estimation}
S.~Habbu, M.~Dale, and R.~Ghongade, ``Estimation of blood glucose by
  non-invasive method using photoplethysmography,'' \emph{S{\=a}dhan{\=a}},
  vol.~44, no.~6, p. 135, 2019.

\bibitem{Ali2017}
H.~{Ali}, F.~{Bensaali}, and F.~{Jaber}, ``Novel approach to non-invasive blood
  glucose monitoring based on transmittance and refraction of visible laser
  light,'' \emph{IEEE Access}, vol.~5, pp. 9163--9174, 2017.

\bibitem{C0AN00537A}
\BIBentryALTinterwordspacing
C.~Vrančić, A.~Fomichova, N.~Gretz, C.~Herrmann, S.~Neudecker, A.~Pucci, and
  W.~Petrich, ``Continuous glucose monitoring by means of mid-infrared
  transmission laser spectroscopyin vitro,'' \emph{Analyst}, vol. 136, pp.
  1192--1198, 2011. [Online]. Available:
  \url{http://dx.doi.org/10.1039/C0AN00537A}
\BIBentrySTDinterwordspacing

\bibitem{paul2012design}
B.~Paul, M.~P. Manuel, and Z.~C. Alex, ``Design and development of non invasive
  glucose measurement system,'' in \emph{2012 1st International Symposium on
  Physics and Technology of Sensors (ISPTS-1)}.\hskip 1em plus 0.5em minus
  0.4em\relax IEEE, 2012, pp. 43--46.

\bibitem{philip2017continous}
L.~A. Philip, K.~Rajasekaran, and E.~S.~J. Jothi, ``Continous monitoring of
  blood glucose using photophlythesmograph signal,'' in \emph{2017
  International Conference on Innovations in Electrical, Electronics,
  Instrumentation and Media Technology (ICEEIMT)}.\hskip 1em plus 0.5em minus
  0.4em\relax IEEE, 2017, pp. 187--191.

\bibitem{karimipour2009diabetic}
H.~Karimipour, H.~T. Shandiz, and E.~Zahedi, ``Diabetic diagnose test based on
  ppg signal and identification system,'' \emph{Journal of Biomedical Science
  and Engineering}, vol.~2, no.~06, p. 465, 2009.

\bibitem{cruz2019application}
F.~R.~G. Cruz, C.~C. Paglinawan, C.~N.~V. Catindig, J.~C.~B. Lamchek, D.~D.~C.
  Almiranez, and A.~F. Sanchez, ``Application of reflectance mode
  photoplethysmography for non-invasive monitoring of blood glucose level with
  moving average filter,'' in \emph{Proceedings of the 2019 9th International
  Conference on Biomedical Engineering and Technology}.\hskip 1em plus 0.5em
  minus 0.4em\relax ACM, 2019, pp. 22--26.

\bibitem{zhang2019non}
Y.~Zhang, Y.~Zhang, S.~A. Siddiqui, and A.~Kos, ``Non-invasive blood-glucose
  estimation using smartphone ppg signals and subspace knn classifier,''
  \emph{Elektrotehniski Vestnik}, vol.~86, no. 1/2, pp. 68--74, 2019.

\bibitem{yamakoshi2017side}
Y.~Yamakoshi, K.~Matsumura, T.~Yamakoshi, J.~Lee, P.~Rolfe, Y.~Kato,
  K.~Shimizu, and K.-i. Yamakoshi, ``Side-scattered
  finger-photoplethysmography: experimental investigations toward practical
  noninvasive measurement of blood glucose,'' \emph{Journal of biomedical
  optics}, vol.~22, no.~6, p. 067001, 2017.

\bibitem{Olarte2013}
O.~Olarte, W.~V. Moer, K.~Barbé, Y.~V. Ingelgem, and A.~Hubin, ``Influence of
  the type and position of the sensor on the precision of impedance glucose
  measurements,'' in \emph{2013 IEEE International Instrumentation and
  Measurement Technology Conference (I2MTC)}, May 2013, pp. 1750--1754.

\bibitem{Dhar2013}
S.~K. Dhar, P.~Biswas, and S.~Chakraborty, ``Dc impedance of human blood using
  eis: An appraoch to non-invasive blood glucose measurement,'' in \emph{2013
  International Conference on Informatics, Electronics and Vision (ICIEV)}, May
  2013, pp. 1--6.

\bibitem{Khawam2013}
Y.~Khawam, M.~Ali, H.~Shazada, S.~Kanan, and H.~Nashash, ``Non-invasive blood
  glucose measurement using transmission spectroscopy,'' in \emph{2013 1st
  International Conference on Communications, Signal Processing, and their
  Applications (ICCSPA)}, Feb 2013, pp. 1--4.

\bibitem{Anas2013}
M.~N. Anas and P.~K. Lim, ``A bio-impedance approach,'' in \emph{2013 IEEE
  International Conference on Smart Instrumentation, Measurement and
  Applications (ICSIMA)}, Nov 2013, pp. 1--5.

\bibitem{Amaral2007}
C.~E.~F. Amaral and B.~Wolf, ``Effects of glucose in blood and skin impedance
  spectroscopy,'' in \emph{AFRICON 2007}, Sept 2007, pp. 1--7.

\bibitem{Jain2017}
P.~Jain and A.~M. Joshi, ``Low leakage and high cmrr cmos differential
  amplifier for biomedical application,'' \emph{Analog Integrated Circuits and
  Signal Processing}, pp. 1--15, 2017.

\bibitem{Paul2012}
B.~Paul, M.~P. Manuel, and Z.~C. Alex, ``Design and development of non invasive
  glucose measurement system,'' in \emph{2012 1st International Symposium on
  Physics and Technology of Sensors (ISPTS-1)}, March 2012, pp. 43--46.

\bibitem{Hofmann2012}
M.~Hofmann, M.~Bloss, R.~Weigel, G.~Fischer, and D.~Kissinger, ``Non-invasive
  glucose monitoring using open electromagnetic waveguides,'' in \emph{2012
  42nd European Microwave Conference}, Oct 2012, pp. 546--549.

\bibitem{Liu2016}
Y.~Liu, M.~Xia, Z.~Nie, J.~Li, Y.~Zeng, and L.~Wang, ``In vivo wearable
  non-invasive glucose monitoring based on dielectric spectroscopy,'' in
  \emph{2016 IEEE 13th International Conference on Signal Processing (ICSP)},
  Nov 2016, pp. 1388--1391.

\bibitem{677170}
M.~{Yamaguchi}, M.~{Mitsumori}, and Y.~{Kano}, ``Noninvasively measuring blood
  glucose using saliva,'' \emph{IEEE Engineering in Medicine and Biology
  Magazine}, vol.~17, no.~3, pp. 59--63, May 1998.

\bibitem{8347021}
M.~S. {Prasad}, R.~{Chen}, Y.~{Li}, D.~{Rekha}, D.~{Li}, H.~{Ni}, and N.~Y.
  {Sreedhar}, ``Polypyrrole supported with copper nanoparticles modified alkali
  anodized steel electrode for probing of glucose in real samples,'' \emph{IEEE
  Sensors Journal}, vol.~18, no.~13, pp. 5203--5212, July 2018.

\bibitem{Yoon1999}
G.~Yoon, K.~J. Jeon, A.~K. Amerov, Y.-J. Kim, D.~Y. Hwang, J.~B. Kim, and H.~S.
  Kim, ``Non-invasive monitoring of blood glucose,'' in \emph{Lasers and
  Electro-Optics, 1999. CLEO/Pacific Rim '99. The Pacific Rim Conference on},
  vol.~4, Aug 1999, pp. 1233--1234 vol.4.

\bibitem{Yamakoshi2007}
Y.~Yamakoshi, M.~Ogawa, T.~Yamakoshi, M.~Satoh, M.~Nogawa, S.~Tanaka,
  T.~Tamura, P.~Rolfe, and K.~Yamakoshi, ``A new non-invasive method for
  measuring blood glucose using instantaneous differential near infrared
  spectrophotometry,'' in \emph{2007 29th Annual International Conference of
  the IEEE Engineering in Medicine and Biology Society}, Aug 2007, pp.
  2964--2967.

\bibitem{Yoon1998}
G.~Yoon, A.~K. Amerov, K.~J. Jeon, J.~B. Kim, and Y.-J. Kim, ``Optical
  measurement of glucose levels in scattering media,'' in \emph{Proceedings of
  the 20th Annual International Conference of the IEEE Engineering in Medicine
  and Biology Society. Vol.20 Biomedical Engineering Towards the Year 2000 and
  Beyond (Cat. No.98CH36286)}, vol.~4, Oct 1998, pp. 1897--1899 vol.4.

\bibitem{Ishizawa2008}
H.~Ishizawa, A.~Muro, T.~Takano, K.~Honda, and H.~Kanai, ``Non-invasive blood
  glucose measurement based on atr infrared spectroscopy,'' in \emph{2008 SICE
  Annual Conference}, Aug 2008, pp. 321--324.

\bibitem{Harada2007}
T.~Harada, K.~Yamamoto, M.~Kondo, K.~Gesho, and I.~Ishimaru, ``Spectroscpy
  optical coherence tomography of biomedical tissue,'' in \emph{SICE Annual
  Conference 2007}, Sept 2007, pp. 3056--3059.

\bibitem{JSTQE.2011.2175202}
A.~Popov, A.~Bykov, S.~Toppari, M.~Kinnunen, A.~Priezzhev, and R.~Myllylä,
  ``Glucose sensing in flowing blood and intralipid by laser pulse
  time-of-flight and optical coherence tomography techniques,'' \emph{IEEE
  Journal of Selected Topics in Quantum Electronics}, vol.~18, pp. 1335--1342,
  07 2012.

\bibitem{Sim2016}
J.~Y. Sim, C.~G. Ahn, E.~Jeong, and B.~K. Kim, ``Photoacoustic spectroscopy
  that uses a resonant characteristic of a microphone for in vitro measurements
  of glucose concentration,'' in \emph{2016 38th Annual International
  Conference of the IEEE Engineering in Medicine and Biology Society (EMBC)},
  Aug 2016, pp. 4861--4864.

\bibitem{Pai2015}
P.~P. Pai, P.~K. Sanki, and S.~Banerjee, ``A photoacoustics based continuous
  non-invasive blood glucose monitoring system,'' in \emph{2015 IEEE
  International Symposium on Medical Measurements and Applications (MeMeA)
  Proceedings}, May 2015, pp. 106--111.

\bibitem{Pai2015b}
P.~P. Pai, P.~K. Sanki, A.~De, and S.~Banerjee, ``Nir photoacoustic
  spectroscopy for non-invasive glucose measurement,'' in \emph{2015 37th
  Annual International Conference of the IEEE Engineering in Medicine and
  Biology Society (EMBC)}, Aug 2015, pp. 7978--7981.

\bibitem{Tanaka2015}
Y.~Tanaka, Y.~Higuchi, and S.~Camou, ``Noninvasive measurement of aqueous
  glucose solution at physiologically relevant blood concentration levels with
  differential continuous-wave laser photoacoustic technique,'' in \emph{2015
  IEEE SENSORS}, Nov 2015, pp. 1--4.

\bibitem{Xiaoli2015}
L.~Xiaoli and L.~Chengwei, ``Research on glucose concentration sensing with
  single wavelength laser,'' in \emph{2015 12th IEEE International Conference
  on Electronic Measurement Instruments (ICEMI)}, vol.~03, July 2015, pp.
  1547--1551.

\bibitem{Naam2015}
H.~A.~A. Naam, M.~O. Idrees, A.~Awad, O.~S. Abdalsalam, and F.~Mohamed, ``Non
  invasive blood glucose measurement based on photo-acoustic spectroscopy,'' in
  \emph{2015 International Conference on Computing, Control, Networking,
  Electronics and Embedded Systems Engineering (ICCNEEE)}, Sept 2015, pp. 1--4.

\bibitem{Camou2011a}
S.~Camou, Y.~Ueno, and E.~Tamechika, ``New cw-photoacoustic-based protocol for
  noninvasive and selective determination of aqueous glucose level: A potential
  alternative towards noninvasive blood sugar sensing,'' in \emph{2011 IEEE
  SENSORS Proceedings}, Oct 2011, pp. 798--801.

\bibitem{Wadamori2008}
N.~Wadamori, R.~Shinohara, and Y.~Ishihara, ``Photoacoustic depth profiling of
  a skin model for non-invasive glucose measurement,'' in \emph{2008 30th
  Annual International Conference of the IEEE Engineering in Medicine and
  Biology Society}, Aug 2008, pp. 5644--5647.

\bibitem{Koyama2010}
S.~Koyama, Y.~Miyauchi, T.~Horiguchi, and H.~Ishizawa, ``Non-invasive
  measurement of blood glucose of diabetic based on ir spectroscopy,'' in
  \emph{Proceedings of SICE Annual Conference 2010}, Aug 2010, pp. 3425--3426.

\bibitem{Domachuk2008}
P.~Domachuk, M.~Hunter, R.~Batorsky, M.~Cronin-Golomb, F.~G. Omenetto, A.~Wang,
  A.~K. George, and J.~C. Knight, ``A path for non-invasive glucose detection
  using mid-ir supercontinuum,'' in \emph{2008 Conference on Lasers and
  Electro-Optics and 2008 Conference on Quantum Electronics and Laser Science},
  May 2008, pp. 1--2.

\bibitem{Periyasamy2016}
R.~Periyasamy and S.~Anand, ``A study on non-invasive blood glucose estimation-
  an approach using capacitance measurement technique,'' in \emph{2016
  International Conference on Signal Processing, Communication, Power and
  Embedded System (SCOPES)}, Oct 2016, pp. 847--850.

\bibitem{Yilmaz2014}
T.~Yilmaz, R.~Foster, and Y.~Hao, ``Towards accurate dielectric property
  retrieval of biological tissues for blood glucose monitoring,'' \emph{IEEE
  Transactions on Microwave Theory and Techniques}, vol.~62, no.~12, pp.
  3193--3204, Dec 2014.

\bibitem{Gourzi2003}
M.~Gourzi, A.~Rouane, M.~B. McHugh, R.~Guelaz, and M.~Nadi, ``New biosensor for
  non-invasive glucose concentration measurement,'' in \emph{Proceedings of
  IEEE Sensors 2003 (IEEE Cat. No.03CH37498)}, vol.~2, Oct 2003, pp. 1343--1347
  Vol.2.

\bibitem{Turgul2015}
V.~Turgul and I.~Kale, ``On the accuracy of complex permittivity model of
  glucose/water solutions for non-invasive microwave blood glucose sensing,''
  in \emph{2015 E-Health and Bioengineering Conference (EHB)}, Nov 2015, pp.
  1--4.

\bibitem{LI201558}
\BIBentryALTinterwordspacing
D.~Li, D.~Yang, J.~Yang, Y.~Lin, Y.~Sun, H.~Yu, and K.~Xu, ``Glucose affinity
  measurement by surface plasmon resonance with borate polymer binding,''
  \emph{Sensors and Actuators A: Physical}, vol. 222, pp. 58 -- 66, 2015.
  [Online]. Available:
  \url{http://www.sciencedirect.com/science/article/pii/S0924424714004737}
\BIBentrySTDinterwordspacing

\bibitem{Kaul2016}
R.~Kaul and U.~P. Khot, ``Design of microstrip antennas for glucometer
  application,'' in \emph{2016 IEEE International Conference on Advances in
  Electronics, Communication and Computer Technology (ICAECCT)}, Dec 2016, pp.
  352--357.

\bibitem{Turgul2017}
V.~Turgul and I.~Kale, ``Influence of fingerprints and finger positioning on
  accuracy of rf blood glucose measurement from fingertips,'' \emph{Electronics
  Letters}, vol.~53, no.~4, pp. 218--220, 2017.

\bibitem{Cano-Garcia2016}
H.~Cano-Garcia, I.~Gouzouasis, I.~Sotiriou, S.~Saha, G.~Palikaras, P.~Kosmas,
  and E.~Kallos, ``Reflection and transmission measurements using 60 ghz patch
  antennas in the presence of animal tissue for non-invasive glucose sensing,''
  in \emph{2016 10th European Conference on Antennas and Propagation (EuCAP)},
  April 2016, pp. 1--3.

\bibitem{Saha2016}
S.~Saha, I.~Sotiriou, I.~Gouzouasis, H.~Cano-Garcia, G.~Palikaras, P.~Kosmas,
  and E.~Kallos, ``Evaluation of the sensitivity of transmission measurements
  at millimeter waves using patch antennas for non-invasive glucose sensing,''
  in \emph{2016 10th European Conference on Antennas and Propagation (EuCAP)},
  April 2016, pp. 1--4.

\bibitem{Turgul2016}
V.~Turgul and I.~Kale, ``A novel pressure sensing circuit for non-invasive
  rf/microwave blood glucose sensors,'' in \emph{2016 16th Mediterranean
  Microwave Symposium (MMS)}, Nov 2016, pp. 1--4.

\bibitem{Ali2016}
M.~S. Ali, N.~J. Shoumy, S.~Khatun, L.~M. Kamarudin, and V.~Vijayasarveswari,
  ``Non-invasive blood glucose measurement performance analysis through uwb
  imaging,'' in \emph{2016 3rd International Conference on Electronic Design
  (ICED)}, Aug 2016, pp. 513--516.

\bibitem{Choi2014}
H.~Choi, J.~Nylon, S.~Luzio, J.~Beutler, and A.~Porch, ``Design of continuous
  non-invasive blood glucose monitoring sensor based on a microwave split ring
  resonator,'' in \emph{2014 IEEE MTT-S International Microwave Workshop Series
  on RF and Wireless Technologies for Biomedical and Healthcare Applications
  (IMWS-Bio2014)}, Dec 2014, pp. 1--3.

\bibitem{Turgul2017a}
V.~Turgul and I.~Kale, ``Simulating the effects of skin thickness and
  fingerprints to highlight problems with non-invasive rf blood glucose sensing
  from fingertips,'' \emph{IEEE Sensors Journal}, vol.~17, no.~22, pp.
  7553--7560, Nov 2017.

\bibitem{Miyauchi2011}
Y.~Miyauchi, T.~Horiguchi, H.~Ishizawa, S.~i.~Tezuka, and H.~Hara, ``Blood
  glucose level measurement by confocal reflection photodetection system,'' in
  \emph{SICE Annual Conference 2011}, Sept 2011, pp. 2686--2689.

\bibitem{Kim2004}
D.~W. Kim, H.~S. Kim, D.~H. Lee, and H.~C. Kim, ``Importance of skin resistance
  in the reverse iontophoresis-based noninvasive glucose monitoring system,''
  in \emph{The 26th Annual International Conference of the IEEE Engineering in
  Medicine and Biology Society}, vol.~1, Sept 2004, pp. 2434--2437.

\bibitem{Hofmann2011}
M.~Hofmann, T.~Fersch, R.~Weigel, G.~Fischer, and D.~Kissinger, ``A novel
  approach to non-invasive blood glucose measurement based on rf
  transmission,'' in \emph{2011 IEEE International Symposium on Medical
  Measurements and Applications}, May 2011, pp. 39--42.

\bibitem{Mitsubayashi2014}
K.~Mitsubayashi, ``Novel biosensing devices for medical applications soft
  contact-lens sensors for monitoring tear sugar,'' in \emph{2014 International
  Conference on Simulation of Semiconductor Processes and Devices (SISPAD)},
  Sept 2014, pp. 349--352.

\bibitem{Cameron1996}
B.~D. Cameron and G.~L. Cote, ``Polarimetric glucose sensing in aqueous humor
  utilizing digital closed-loop control,'' in \emph{Proceedings of 18th Annual
  International Conference of the IEEE Engineering in Medicine and Biology
  Society}, vol.~1, Oct 1996, pp. 204--205 vol.1.

\bibitem{Buford2008}
R.~J. Buford, E.~C. Green, and M.~J. McClung, ``A microwave frequency sensor
  for non-invasive blood-glucose measurement,'' in \emph{2008 IEEE Sensors
  Applications Symposium}, Feb 2008, pp. 4--7.

\bibitem{j.ultrasmedbio.2005.04.004}
S.~Lee, V.~Nayak, J.~Dodds, M.~Pishkou, and N.~B. Smith, ``Glucose measurements
  with sensors and ultrasound,'' \emph{Ultrasound in Medicine and Biology},
  vol.~31, no.~7, pp. 971--977, 2005.

\bibitem{clinchem.2004.036954}
\BIBentryALTinterwordspacing
O.~K. Cho, Y.~O. Kim, H.~Mitsumaki, and K.~Kuwa, ``{Noninvasive Measurement of
  Glucose by Metabolic Heat Conformation Method},'' \emph{Clinical Chemistry},
  vol.~50, no.~10, pp. 1894--1898, 10 2004. [Online]. Available:
  \url{https://doi.org/10.1373/clinchem.2004.036954}
\BIBentrySTDinterwordspacing

\bibitem{dia.2011.0041}
A.~B. Blodgett, R.~K. Kothinti, I.~Kamyshko, D.~H. Petering, S.~Kumar, and
  N.~M. Tabatabai, ``A fluorescence method for measurement of glucose transport
  in kidney cells,'' \emph{Diabetes Technology \& Therapeutics}, vol.~13,
  no.~7, pp. 743--751, 2011, pMID: 21510766.

\bibitem{12.468318}
\BIBentryALTinterwordspacing
A.~K. Amerov, Y.~Sun, G.~W. Small, and M.~A. Arnold, ``{Kromoscopic measurement
  of glucose in the first overtone region of the near-infrared spectrum},'' in
  \emph{Optical Diagnostics and Sensing of Biological Fluids and Glucose and
  Cholesterol Monitoring II}, A.~V. Priezzhev and G.~L. Cote, Eds., vol. 4624,
  International Society for Optics and Photonics.\hskip 1em plus 0.5em minus
  0.4em\relax SPIE, 2002, pp. 11 -- 19. [Online]. Available:
  \url{https://doi.org/10.1117/12.468318}
\BIBentrySTDinterwordspacing

\bibitem{zhang2019noninvasive}
R.~Zhang, S.~Liu, H.~Jin, Y.~Luo, Z.~Zheng, F.~Gao, and Y.~Zheng, ``Noninvasive
  electromagnetic wave sensing of glucose,'' \emph{Sensors}, vol.~19, no.~5, p.
  1151, 2019.

\bibitem{Bertemes-Filho_ICEB_2016}
P.~Bertemes-Filho, R.~Weinert, T.~Barato, and T.~Baratto~de Albuquerque,
  ``Detection of glucose by using impedance spectroscopy,'' in \emph{Proc.
  International Conference on Electrical Bio-Impedance}, 06 2016.

\bibitem{10.1002/dmrr.210}
\BIBentryALTinterwordspacing
R.~O. Potts, J.~A.~Tamada, and M.~J.~Tierney, ``Glucose monitoring by reverse
  iontophoresis,'' \emph{Diabetes/Metabolism Research and Reviews}, vol.~18,
  no.~S1, pp. S49--S53, 2002. [Online]. Available:
  \url{https://onlinelibrary.wiley.com/doi/abs/10.1002/dmrr.210}
\BIBentrySTDinterwordspacing

\bibitem{10.1211/jpp.61.06.0001}
\BIBentryALTinterwordspacing
R.~Rao and S.~Nanda, ``Sonophoresis: recent advancements and future trends,''
  \emph{Journal of Pharmacy and Pharmacology}, vol.~61, no.~6, pp. 689--705,
  2009. [Online]. Available:
  \url{https://onlinelibrary.wiley.com/doi/abs/10.1211/jpp.61.06.0001}
\BIBentrySTDinterwordspacing

\bibitem{10.1177/193229680700100403}
O.~Amir, D.~Weinstein, S.~Zilberman, M.~Less, D.~Perl-Treves, H.~Primack,
  A.~Weinstein, E.~Gabis, B.~Fikhte, and A.~Karasik, ``Continuous noninvasive
  glucose monitoring technology based on ``occlusion spectroscopy'',''
  \emph{Journal of Diabetes Science and Technology}, vol.~1, no.~4, pp.
  463--469, 2007, pMID: 19885108.

\bibitem{10.3109/00365519509104982}
B.~M. Jensen, P.~Bjerring, J.~S. Christiansen, and H.~ørskov, ``Glucose
  content in human skin: relationship with blood glucose levels,''
  \emph{Scandinavian Journal of Clinical and Laboratory Investigation},
  vol.~55, no.~5, pp. 427--432, 1995.

\bibitem{Kossowski2017}
T.~Kossowski and R.~Stasiński, ``Multi-wavelength analysis of substances
  levels in human blood,'' in \emph{2017 International Conference on Systems,
  Signals and Image Processing (IWSSIP)}, May 2017, pp. 1--4.

\bibitem{Ficorella2015}
A.~Ficorella, A.~D'Amico, M.~Santonico, G.~Pennazza, S.~Grasso, and
  A.~Zompanti, ``A multi-frequency system for glucose detection with optical
  sensors,'' in \emph{2015 XVIII AISEM Annual Conference}, Feb 2015, pp. 1--3.

\bibitem{Song2014}
K.~Song, U.~Ha, S.~Park, and H.-J. Yoo, ``An impedance and multi-wavelength
  near-infrared spectroscopy ic for non-invasive blood glucose estimation,'' in
  \emph{2014 Symposium on VLSI Circuits Digest of Technical Papers}, June 2014,
  pp. 1--2.

\bibitem{Litinskaia2017}
E.~L. Litinskaia, N.~A. Bazaev, K.~V. Pozhar, and V.~M. Grinvald, ``Methods for
  improving accuracy of non-invasive blood glucose detection via optical
  glucometer,'' in \emph{2017 IEEE Conference of Russian Young Researchers in
  Electrical and Electronic Engineering (EIConRus)}, Feb 2017, pp. 47--49.

\bibitem{Jain2016}
P.~Jain and S.~Akashe, ``An innovative design: Mos based full-wave
  centre-tapped rectifier,'' \emph{Wireless Personal Communications}, vol.~90,
  no.~4, pp. 1673--1693, 2016.

\bibitem{PRATEEK2016}
P.~JAIN and S.~AKASHE, ``Performance analysis of analog to digital converter
  with augmented voltage swing boost logic cum schmitt trigger mos switches
  configuration.'' \emph{Journal of Active \& Passive Electronic Devices},
  vol.~11, 2016.

\bibitem{Jain2014}
P.~Jain and S.~Akashe, ``Analyzing the impact of bootstrapped adc with
  augmented nmos sleep transistors configuration on performance parameters,''
  \emph{Circuits, Systems, and Signal Processing}, vol.~33, no.~7, pp.
  2009--2025, 2014.

\bibitem{Jain2013}
P.Jain and S.~Akashe, ``Design and optimization of flash type analog to digital
  converter using augmented sleep transistors with current mode logic,''
  \emph{Radioelectronics and Communications Systems}, vol.~56, no.~10, pp.
  472--480, 2013.

\bibitem{8727488}
A.~K. {Singh} and S.~K. {Jha}, ``Fabrication and validation of a handheld
  non-invasive, optical biosensor for self-monitoring of glucose using
  saliva,'' \emph{IEEE Sensors Journal}, vol.~19, no.~18, pp. 8332--8339, Sep.
  2019.

\bibitem{Pai2018}
P.~P. Pai, A.~De, and S.~Banerjee, ``Accuracy enhancement for noninvasive
  glucose estimation using dual-wavelength photoacoustic measurements and
  kernel-based calibration,'' \emph{IEEE Transactions on Instrumentation and
  Measurement}, vol.~67, no.~1, pp. 126--136, 2018.

\bibitem{Dai2009}
T.~Dai and A.~Adler, ``In vivo blood characterization from bioimpedance
  spectroscopy of blood pooling,'' \emph{IEEE Transactions on Instrumentation
  and Measurement}, vol.~58, no.~11, p. 3831, 2009.

\bibitem{Beach2005}
R.~D. {Beach}, R.~W. {Conlan}, M.~C. {Godwin}, and F.~{Moussy}, ``Towards a
  miniature implantable in vivo telemetry monitoring system dynamically
  configurable as a potentiostat or galvanostat for two- and three-electrode
  biosensors,'' \emph{IEEE Transactions on Instrumentation and Measurement},
  vol.~54, no.~1, pp. 61--72, Feb 2005.

\bibitem{jain2019iomt}
P.~Jain, S.~Pancholi, and A.~M. Joshi, ``An iomt based non-invasive precise
  blood glucose measurement system,'' in \emph{2019 IEEE International
  Symposium on Smart Electronic Systems (iSES)(Formerly iNiS)}.\hskip 1em plus
  0.5em minus 0.4em\relax IEEE, 2019, pp. 111--116.

\bibitem{Malik2016}
B.~A. Malik, A.~Naqash, and G.~M. Bhat, ``Backpropagation artificial neural
  network for determination of glucose concentration from near-infrared
  spectra,'' in \emph{2016 International Conference on Advances in Computing,
  Communications and Informatics (ICACCI)}, Sept 2016, pp. 2688--2691.

\bibitem{Yotha2016}
D.~Yotha, C.~Pidthalek, S.~Yimman, and S.~Niramitmahapanya, ``Design and
  construction of the hypoglycemia monito wireless system for diabetic,'' in
  \emph{2016 9th Biomedical Engineering International Conference (BMEiCON)},
  Dec 2016, pp. 1--4.

\bibitem{Sarangi2014a}
S.~Sarangi, P.~P. Pai, P.~K. Sanki, and S.~Banerjee, ``Comparative analysis of
  golay code based excitation and coherent averaging for non-invasive glucose
  monitoring system,'' in \emph{2014 IEEE 27th International Symposium on
  Computer-Based Medical Systems}, May 2014, pp. 485--486.

\bibitem{Naqvi2008}
S.~R. Naqvi, N.~Z. Azeemi, A.~Hameed, R.~Baddar, and T.~Rasool, ``Improving
  accuracy of non-invasive glucose monitoring through non-local data
  denoising,'' in \emph{2008 Cairo International Biomedical Engineering
  Conference}, Dec 2008, pp. 1--4.

\bibitem{Yamakoshi2009}
Y.~Yamakoshi, M.~Ogawa, T.~Yamakoshi, T.~Tamura, and K.~i.~Yamakoshi,
  ``Multivariate regression and discreminant calibration models for a novel
  optical non-invasive blood glucose measurement method named pulse
  glucometry,'' in \emph{2009 Annual International Conference of the IEEE
  Engineering in Medicine and Biology Society}, Sept 2009, pp. 126--129.

\bibitem{Ming2009}
C.~Z. Ming, P.~Raveendran, and P.~S. Chew, ``A comparison analysis between
  partial least squares and neural network in non-invasive blood glucose
  concentration monitoring system,'' in \emph{2009 International Conference on
  Biomedical and Pharmaceutical Engineering}, Dec 2009, pp. 1--4.

\bibitem{Sarangi2014}
S.~Sarangi, P.~P. Pai, P.~K. Sanki, and S.~Banerjee, ``Comparative analysis of
  golay code based excitation and coherent averaging for non-invasive glucose
  monitoring system,'' in \emph{2014 IEEE 27th International Symposium on
  Computer-Based Medical Systems}, May 2014, pp. 485--486.

\bibitem{Heise1996}
H.~M. Heise, ``Technology for non-invasive monitoring of glucose,'' in
  \emph{Proceedings of 18th Annual International Conference of the IEEE
  Engineering in Medicine and Biology Society}, vol.~5, Oct 1996, pp.
  2159--2161 vol.5.

\bibitem{Pai2015a}
P.~P. Pai, S.~Bhattacharya, and S.~Banerjee, ``Regularized least squares
  regression for calibration of a photoacoustic spectroscopy based non-invasive
  glucose monitoring system,'' in \emph{2015 IEEE International Ultrasonics
  Symposium (IUS)}, Oct 2015, pp. 1--4.

\bibitem{Stemmann2010}
M.~Stemmann, F.~Ståhl, J.~Lallemand, E.~Renard, and R.~Johansson, ``Sensor
  calibration models for a non-invasive blood glucose measurement sensor,'' in
  \emph{2010 Annual International Conference of the IEEE Engineering in
  Medicine and Biology}, Aug 2010, pp. 4979--4982.

\bibitem{Rollins2010}
D.~K. Rollins, K.~Kotz, and C.~Stiehl, ``Non-invasive glucose monitoring from
  measured inputs,'' in \emph{UKACC International Conference on Control 2010},
  Sept 2010, pp. 1--5.

\bibitem{Malik2015}
B.~A. Malik, ``Determination of glucose concentration from near infrared
  spectra using least square support vector machine,'' in \emph{2015
  International Conference on Industrial Instrumentation and Control (ICIC)},
  May 2015, pp. 475--478.

\bibitem{Ogawa2007}
M.~Ogawa, Y.~Yamakoshi, M.~Satoh, M.~Nogawa, T.~Yamakoshi, S.~Tanaka, P.~Rolfe,
  T.~Tamura, and K.~i.~Yamakoshi, ``Support vector machines as multivariate
  calibration model for prediction of blood glucose concentration using a new
  non-invasive optical method named pulse glucometry,'' in \emph{2007 29th
  Annual International Conference of the IEEE Engineering in Medicine and
  Biology Society}, Aug 2007, pp. 4561--4563.

\bibitem{Dag2011}
Z.~T. Dag, E.~Koklukaya, F.~Temurtas, H.~M. Saraoglu, and S.~Altikat,
  ``Detection of the blood glucose and haemoglobin a1c with palm perspiration
  by using artificial neural networks,'' in \emph{2011 7th International
  Conference on Electrical and Electronics Engineering (ELECO)}, Dec 2011, pp.
  II--302--II--305.

\bibitem{Olarte2011}
O.~Olarte, W.~V. Moer, K.~Barbé, Y.~V. Ingelgem, and A.~Hubin, ``Using random
  phase multisines to perform non-invasive glucose measurements,'' in
  \emph{2011 IEEE International Symposium on Medical Measurements and
  Applications}, May 2011, pp. 300--304.

\bibitem{Savage2000}
M.~B. Savage, S.~Kun, H.~Harjunmaa, and R.~A. Peura, ``Development of a
  non-invasive blood glucose monitor: application of artificial neural networks
  for signal processing,'' in \emph{Proceedings of the IEEE 26th Annual
  Northeast Bioengineering Conference (Cat. No.00CH37114)}, 2000, pp. 29--30.

\bibitem{Baghbani2015}
R.~Baghbani, M.~A. Rad, and A.~Pourziad, ``Microwave sensor for non-invasive
  glucose measurements design and implementation of a novel linear,'' \emph{IET
  Wireless Sensor Systems}, vol.~5, no.~2, pp. 51--57, 2015.

\bibitem{Parab2016}
J.~S. Parab, R.~S. Gad, and G.~M. Naik, ``Influence of pca components on
  glucose prediction using non-invasive technique,'' in \emph{2016
  International Conference on Advances in Electrical, Electronic and Systems
  Engineering (ICAEES)}, Nov 2016, pp. 473--476.

\bibitem{Lekha2015a}
T.~R. J.~C. Lekha and C.~S. Kumar, ``Nir spectroscopic algorithm development
  for glucose detection,'' in \emph{2015 International Conference on
  Innovations in Information, Embedded and Communication Systems (ICIIECS)},
  March 2015, pp. 1--6.

\bibitem{Olarte2012}
O.~Olarte, W.~V. Moer, K.~Barbé, S.~Verguts, Y.~V. Ingelgem, and A.~Hubin,
  ``Using the best linear approximation as a first step to a new non-invasive
  glucose measurement,'' in \emph{2012 IEEE International Instrumentation and
  Measurement Technology Conference Proceedings}, May 2012, pp. 2747--2751.

\bibitem{Clarke2005}
W.~L. Clarke, ``{The original Clarke error grid analysis (EGA)},''
  \emph{Diabetes Technology \& Therapeutics}, vol.~7, no.~5, pp. 776--779,
  2005.

\bibitem{fernandez2018needle}
C.~Fernandez, ``Needle-free diabetes care: 7 devices that painlessly measure
  blood glucose,'' \emph{Labiotech}, vol.~23, 2018.

\bibitem{schemmann2013blood}
M.~F. Schemmann and T.~O'brien, ``Blood glucose sensor,'' Apr.~11 2013, uS
  Patent App. 13/646,721.

\bibitem{Gal2011}
A.~Gal, I.~Harman-Boehm, E.~Naidis, Y.~Mayzel, and L.~Trieman, ``Validity of
  glucotrack{\textregistered}, a non-invasive glucose monitor, for variety of
  diabetics,'' \emph{Age}, vol.~1, no. 295, p.~61, 2011.

\bibitem{Huber2007}
D.~Huber, L.~Falco-Jonasson, M.~Talary, F.~Dewarrat, A.~Caduff, W.~Stahel, and
  N.~Stadler, ``Multi-sensor data fusion for non-invasive continuous glucose
  monitoring,'' in \emph{2007 10th International Conference on Information
  Fusion}, July 2007, pp. 1--10.

\bibitem{bolie1961coefficients}
V.~W. Bolie, ``Coefficients of normal blood glucose regulation,'' \emph{Journal
  of applied physiology}, vol.~16, no.~5, pp. 783--788, 1961.

\bibitem{bergman1981physiologic}
R.~N. Bergman, L.~S. Phillips, and C.~Cobelli, ``Physiologic evaluation of
  factors controlling glucose tolerance in man: measurement of insulin
  sensitivity and beta-cell glucose sensitivity from the response to
  intravenous glucose.'' \emph{The Journal of clinical investigation}, vol.~68,
  no.~6, pp. 1456--1467, 1981.

\bibitem{de2000mathematical}
A.~De~Gaetano and O.~Arino, ``Mathematical modelling of the intravenous glucose
  tolerance test,'' \emph{Journal of mathematical biology}, vol.~40, no.~2, pp.
  136--168, 2000.

\bibitem{cobelli2009diabetes}
C.~Cobelli, C.~Dalla~Man, G.~Sparacino, L.~Magni, G.~De~Nicolao, and B.~P.
  Kovatchev, ``Diabetes: models, signals, and control,'' \emph{IEEE reviews in
  biomedical engineering}, vol.~2, pp. 54--96, 2009.

\bibitem{hovorka2002partitioning}
R.~Hovorka, F.~Shojaee-Moradie, P.~V. Carroll, L.~J. Chassin, I.~J. Gowrie,
  N.~C. Jackson, R.~S. Tudor, A.~M. Umpleby, and R.~H. Jones, ``Partitioning
  glucose distribution/transport, disposal, and endogenous production during
  ivgtt,'' \emph{American Journal of Physiology-Endocrinology and Metabolism},
  vol. 282, no.~5, pp. E992--E1007, 2002.

\bibitem{haidar2013stochastic}
A.~Haidar, M.~E. Wilinska, J.~A. Graveston, and R.~Hovorka, ``Stochastic
  virtual population of subjects with type 1 diabetes for the assessment of
  closed-loop glucose controllers,'' \emph{IEEE Transactions on Biomedical
  Engineering}, vol.~60, no.~12, pp. 3524--3533, 2013.

\bibitem{kirchsteiger2011estimating}
H.~Kirchsteiger, G.~C. Estrada, S.~P{\"o}lzer, E.~Renard, and L.~del Re,
  ``Estimating interval process models for type 1 diabetes for robust control
  design,'' \emph{IFAC Proc. Volumes}, vol.~44, no.~1, pp. 11\,761--11\,766,
  2011.

\bibitem{magdelaine2015long}
N.~Magdelaine, L.~Chaillous, I.~Guilhem, J.-Y. Poirier, M.~Krempf, C.~H. Moog,
  and E.~Le~Carpentier, ``A long-term model of the glucose--insulin dynamics of
  type 1 diabetes,'' \emph{IEEE Transactions on Biomedical Engineering},
  vol.~62, no.~6, pp. 1546--1552, 2015.

\bibitem{turksoy2015meal}
K.~Turksoy, S.~Samadi, J.~Feng, E.~Littlejohn, L.~Quinn, and A.~Cinar, ``Meal
  detection in patients with type 1 diabetes: a new module for the
  multivariable adaptive artificial pancreas control system,'' \emph{IEEE
  journal of biomedical and health informatics}, vol.~20, no.~1, pp. 47--54,
  2015.

\bibitem{xie2016variable}
J.~Xie and Q.~Wang, ``A variable state dimension approach to meal detection and
  meal size estimation: in silico evaluation through basal-bolus insulin
  therapy for type 1 diabetes,'' \emph{IEEE Transactions on Biomedical
  Engineering}, vol.~64, no.~6, pp. 1249--1260, 2016.

\bibitem{mohammadridha2017model}
T.~MohammadRidha, M.~A{\"\i}t-Ahmed, L.~Chaillous, M.~Krempf, I.~Guilhem, J.-Y.
  Poirier, and C.~H. Moog, ``Model free ipid control for glycemia regulation of
  type-1 diabetes,'' \emph{IEEE Transactions on Biomedical Engineering},
  vol.~65, no.~1, pp. 199--206, 2017.

\bibitem{kino2016hollow}
S.~Kino, S.~Omori, T.~Katagiri, and Y.~Matsuura, ``Hollow optical-fiber based
  infrared spectroscopy for measurement of blood glucose level by using
  multi-reflection prism,'' \emph{Biomedical optics express}, vol.~7, no.~2,
  pp. 701--708, 2016.

\bibitem{chen2018noninvasive}
T.-L. Chen, Y.-L. Lo, C.-C. Liao, and Q.-H. Phan, ``Noninvasive measurement of
  glucose concentration on human fingertip by optical coherence tomography,''
  \emph{Journal of biomedical optics}, vol.~23, no.~4, p. 047001, 2018.

\bibitem{park2018soft}
J.~Park, J.~Kim, S.-Y. Kim, W.~H. Cheong, J.~Jang, Y.-G. Park, K.~Na, Y.-T.
  Kim, J.~H. Heo, C.~Y. Lee \emph{et~al.}, ``Soft, smart contact lenses with
  integrations of wireless circuits, glucose sensors, and displays,''
  \emph{Science Advances}, vol.~4, no.~1, p. eaap9841, 2018.

\bibitem{singh2019fabrication}
A.~K. Singh and S.~K. Jha, ``Fabrication and validation of a handheld
  non-invasive, optical biosensor for self-monitoring of glucose using
  saliva,'' \emph{IEEE Sensors Journal}, vol.~19, no.~18, pp. 8332--8339, 2019.

\bibitem{song2015impedance}
K.~Song, U.~Ha, S.~Park, J.~Bae, and H.-J. Yoo, ``An impedance and
  multi-wavelength near-infrared spectroscopy ic for non-invasive blood glucose
  estimation,'' \emph{IEEE Journal of solid-state circuits}, vol.~50, no.~4,
  pp. 1025--1037, 2015.

\bibitem{li2019absorption}
Q.~Li, X.~Xiao, and T.~Kikkawa, ``Absorption spectrum for non-invasive blood
  glucose concentration detection by microwave signals,'' \emph{Journal of
  Electromagnetic Waves and Applications}, vol.~33, no.~9, pp. 1093--1106,
  2019.

\bibitem{rachim2019wearable}
V.~P. Rachim and W.-Y. Chung, ``Wearable-band type visible-near infrared
  optical biosensor for non-invasive blood glucose monitoring,'' \emph{Sensors
  and Actuators B: Chemical}, vol. 286, pp. 173--180, 2019.

\bibitem{Mohanty_CEM_2016-July}
S.~P. {Mohanty}, U.~{Choppali}, and E.~{Kougianos}, ``{Everything you wanted to
  know about smart cities: The Internet of things is the backbone},''
  \emph{IEEE Consumer Electronics Magazine}, vol.~5, no.~3, pp. 60--70, July
  2016.

\bibitem{9153927}
P.~{Chanak} and I.~{Banerjee}, ``Internet of things-enabled smart villages:
  Recent advances and challenges,'' \emph{IEEE Consumer Electronics Magazine},
  pp. 1--1, 2020.

\bibitem{9109415}
C.~P. {Antonopoulos}, G.~{Keramidas}, N.~S. {Voros}, M.~{Huebner},
  F.~{Schwiegelshohn}, D.~{Goehringer}, M.~{Dagioglou}, G.~{Stavrinos},
  S.~{Konstantopoulos}, and V.~{Karkaletsis}, ``Toward an ict-based service
  oriented health care paradigm,'' \emph{IEEE Consumer Electronics Magazine},
  vol.~9, no.~4, pp. 77--82, 2020.

\bibitem{8977815}
M.~{Aazam}, S.~{Zeadally}, and K.~A. {Harras}, ``Health fog for smart
  healthcare,'' \emph{IEEE Consumer Electronics Magazine}, vol.~9, no.~2, pp.
  96--102, 2020.

\bibitem{joshi2020secure}
A.~M. Joshi, P.~Jain, and S.~P. Mohanty, ``{Secure-iGLU}: A secure device for
  noninvasive glucose measurement and automatic insulin delivery in iomt
  framework,'' in \emph{2020 IEEE Computer Society Annual Symposium on VLSI
  (ISVLSI)}.\hskip 1em plus 0.5em minus 0.4em\relax IEEE, 2020, pp. 440--445.

\bibitem{9011599}
L.~{Rachakonda}, S.~P. {Mohanty}, and E.~{Kougianos}, ``{iLog}: An intelligent
  device for automatic food intake monitoring and stress detection in the
  iomt,'' \emph{IEEE Transactions on Consumer Electronics}, vol.~66, no.~2, pp.
  115--124, 2020.

\bibitem{8752409}
V.~P. {Yanambaka}, S.~P. {Mohanty}, E.~{Kougianos}, and D.~{Puthal}, ``{PMsec}:
  Physical unclonable function-based robust and lightweight authentication in
  the internet of medical things,'' \emph{IEEE Transactions on Consumer
  Electronics}, vol.~65, no.~3, pp. 388--397, 2019.

\bibitem{9288682}
L.~{Rachakonda}, A.~K. {Bapatla}, S.~P. {Mohanty}, and E.~{Kougianos},
  ``{SaYoPillow}: Blockchain-integrated privacy-assured iomt framework for
  stress management considering sleeping habits,'' \emph{IEEE Transactions on
  Consumer Electronics}, pp. 1--1, 2020.

\bibitem{8977825}
S.~P. {Mohanty}, V.~P. {Yanambaka}, E.~{Kougianos}, and D.~{Puthal},
  ``{PUFchain}: A hardware-assisted blockchain for sustainable simultaneous
  device and data security in the internet of everything ({IoE}),'' \emph{IEEE
  Consumer Electronics Magazine}, vol.~9, no.~2, pp. 8--16, 2020.

\end{thebibliography}

\pagebreak
%%%%%%%%%%%%%%%%%%%%%%%%%%%%%%%%%%%%%%%%%%%%
%%%%%%%%%%%%%% Authors' Bio
%%%%%%%%%%%%%%%%%%%%%%%%%%%%%%%%%%%%%%%%%%%%%%%%%%%%%%%
\section*{About the Authors}

%\textbf{Prateek Jain} is a Research Scholar in ECE department of MNIT, Jaipur, India. He can be contacted at: prtk.ieju@gmail.com.

%\textbf{Amit M. Joshi} is an Assistant Professor in Department of ECE, MNIT, Jaipur, India. He can be contacted at: amjoshi.ece@mnit.ac.in.

%\textbf{Saraju P. Mohanty} is the Editor in Chief of the IEEE Consumer Electronics Magazine and Professor in the Department of Computer Science and Engineering (CSE), University of North Texas (UNT), Denton, TX, USA. Contact him at Saraju.Mohanty@unt.edu.

\vspace{-0.8cm}

\begin{IEEEbiography}
	[{\includegraphics[height=1.25in,keepaspectratio]{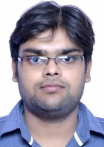}}] 
	{Prateek Jain} (Member, IEEE) earned his B.E. degree in Electronics Engineering from Jiwaji University, India in 2010 and Master degree from ITM University Gwalior. Currently, he is an Assistant Professor in SENSE, VIT University, Amaravati (A.P.). His current research interest includes VLSI design, Biomedical Systems and Instrumentation. He is an author of 14 peer-reviewed publications. He is a regular reviewer of 12 journals and 10 conferences.
\end{IEEEbiography}

\vspace{-0.8cm}

\begin{IEEEbiography}
	[{\includegraphics[height=1.25in,keepaspectratio]{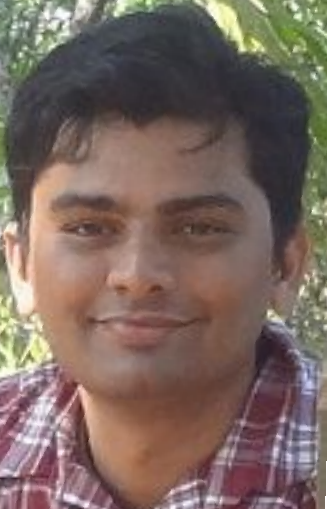}}]
	{Amit M. Joshi} (Member, IEEE) received the Ph.D. degree from the NIT, Surat, India. He is currently an Assistant Professor at National Institute of Technology, Jaipur. His area of specialization is Biomedical signal processing, Smart healthcare, VLSI DSP Systems and embedded system design. He has also published papers in international peer reviewed journals with high impact factors. He has published six book chapters and also published more than 70 research articles in excellent peer reviewed international journals/conferences. He has worked as a reviewer of technical journals such as IEEE Transactions/ IEEE Access, Springer, Elsevier and also served as Technical Programme Committee member for IEEE conferences which are related to biomedical field. He also received honour of UGC Travel fellowship, the award of SERB DST Travel grant and CSIR fellowship to attend well known IEEE Conferences TENCON, ISCAS, MENACOMM etc across the world. He has served as session chair at various IEEE Conferences like TENCON -2016, iSES-2018, iSES-2019, ICCIC-14 etc. He has also supervised 19 PG Dissertations and 16 UG projects. 
He has completed supervision of 4 Ph.D thesis and six more research scholars are also working.
\end{IEEEbiography}

\vspace{-0.8cm}

\begin{IEEEbiography}
	[{\includegraphics[height=1.25in,keepaspectratio]{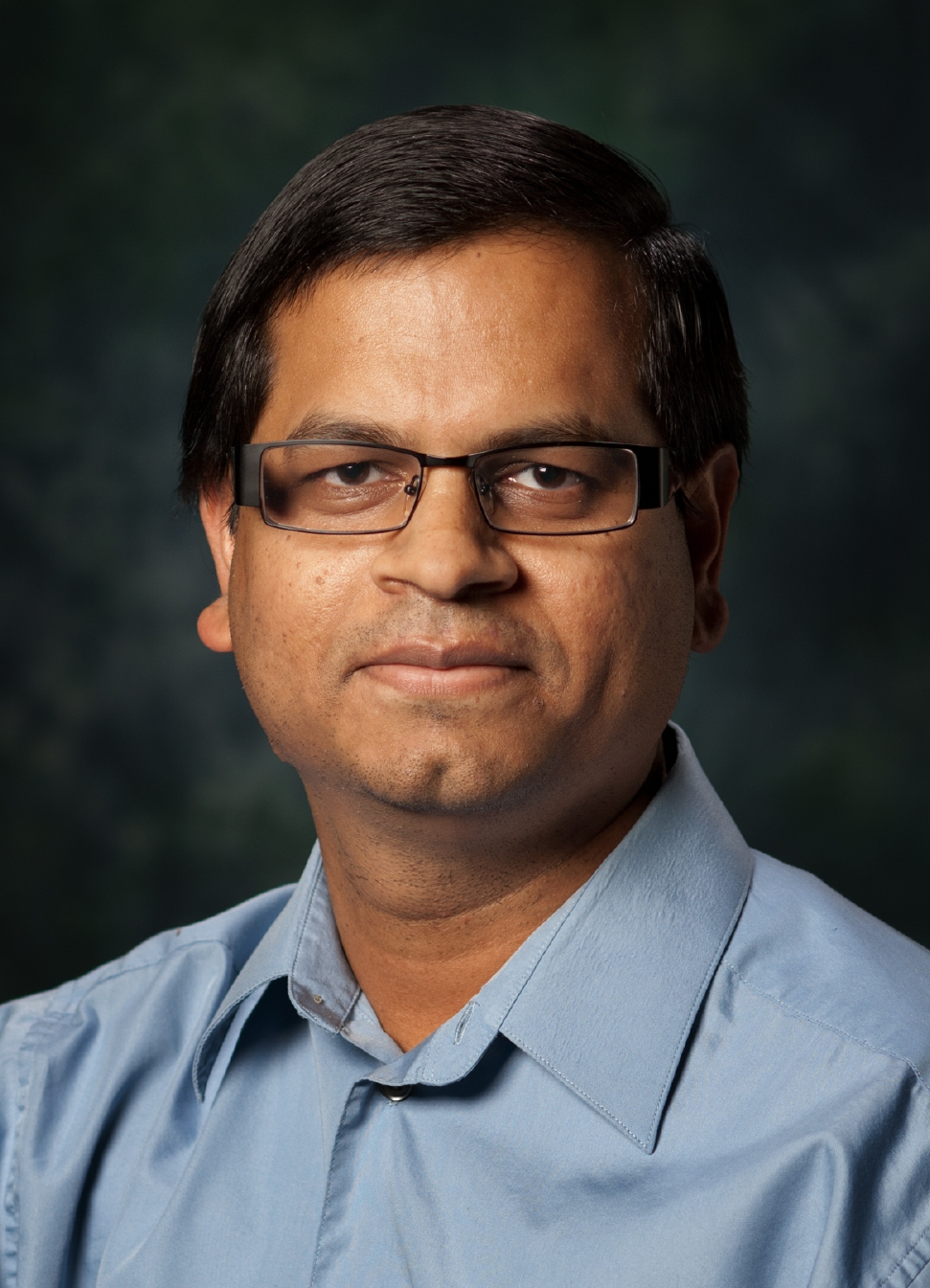}}] 
{Saraju P. Mohanty} (Senior Member, IEEE) received the bachelor's degree (Honors) in electrical engineering from the Orissa University of Agriculture and Technology, Bhubaneswar, in 1995, the master's degree in Systems Science and Automation from the Indian Institute of Science, Bengaluru, in 1999, and the Ph.D. degree in Computer Science and Engineering from the University of South Florida, Tampa, in 2003. He is a Professor with the University of North Texas. His research is in ``Smart Electronic Systems'' which has been funded by National Science Foundations (NSF), Semiconductor Research Corporation (SRC), U.S. Air Force, IUSSTF, and Mission Innovation. He has authored 350 research articles, 4 books, and invented 4 granted and 1 pending patents. His Google Scholar h-index is 39 and i10-index is 149 with 6600 citations. He is regarded as a visionary researcher on Smart Cities technology in which his research deals with security and energy aware, and AI/ML-integrated smart components. He introduced the Secure Digital Camera (SDC) in 2004 with built-in security features designed using Hardware-Assisted Security (HAS) or Security by Design (SbD) principle. He is widely credited as the designer for the first digital watermarking chip in 2004 and first the low-power digital watermarking chip in 2006. He is a recipient of 12 best paper awards, Fulbright Specialist Award in 2020, IEEE Consumer Technology Society Outstanding Service Award in 2020, the IEEE-CS-TCVLSI Distinguished Leadership Award in 2018, and the PROSE Award for Best Textbook in Physical Sciences and Mathematics category in 2016. He has delivered 10 keynotes and served on 11 panels at various International Conferences. He has been serving on the editorial board of several peer-reviewed international journals, including IEEE Transactions on Consumer Electronics (TCE), and IEEE Transactions on Big Data (TBD). He is the Editor-in-Chief (EiC) of the IEEE Consumer Electronics Magazine (MCE). He has been serving on the Board of Governors (BoG) of the IEEE Consumer Technology Society, and has served as the Chair of Technical Committee on Very Large Scale Integration (TCVLSI), IEEE Computer Society (IEEE-CS) during 2014-2018. He is the founding steering committee chair for the IEEE International Symposium on Smart Electronic Systems (iSES), steering committee vice-chair of the IEEE-CS Symposium on VLSI (ISVLSI), and steering committee vice-chair of the OITS International Conference on Information Technology (ICIT). He has mentored 2 post-doctoral researchers, and supervised 12 Ph.D. dissertations, 26 M.S. theses, and 12 undergraduate projects.
\end{IEEEbiography}

\end{document}